\documentclass[twocolumn,aps,pra,showpacs,amsmath,amssymb,superscriptaddress]{revtex4-1}

\usepackage{amsmath,amssymb}
\usepackage{graphics}
\usepackage{epsfig}

\usepackage{MnSymbol}
\usepackage{mathrsfs}
\usepackage[usenames]{color}

\begin{document}

%\title{Efimov physics in resonantly interacting ultracold quantum gases}
\title{Few-body physics in resonantly interacting ultracold quantum gases}

\author{Jos\'e P. D'Incao}
\address{JILA, University of Colorado and NIST, Boulder, Colorado, USA}
\address{Department of Physics, University of Colorado, Boulder, Colorado, USA}
\email{jpdincao@jila.colorado.edu}

\begin{abstract}
We provide a general discussion on the importance of three-body Efimov physics for strongly interacting ultracold quantum
gases. 
Using the adiabatic hyperspherical representation, we discuss a universal classification of three-body systems in terms 
of the attractive or repulsive character of the effective interactions. The hyperspherical representation leads to a simple and 
conceptually clear picture for the bound and scattering properties of three-body systems with strong $s$-wave interactions. 
Using our universal classification scheme, 
we present a detailed discussion of all relevant ultracold three-body scattering processes using a pathway analysis that makes 
evident the importance of Efimov physics in determining the energy and scattering length dependence of such processes. 
%JPD
%\textcolor{red}
{This article provides a general overview of the current status of the field and a discussion
of various issues relevant to the lifetime and stability of ultracold quantum gases along with universal properties of ultracold, resonantly interacting, 
few-body systems.}
\end{abstract}


\maketitle

{\small\tableofcontents}
%\newpage

% !TEX root = ./TutorialJPB.tex

%%%%%%%%%%%%%%%%%%%%%%%%%%%%%%%%%%%%%%%%%%%%%%%%%%%%
\section{Introduction}

%{\bf Use BO to explain the origin of the Efimov effect...}

%JPD
%\textcolor{red}
{Along with the recent theoretical and experimental progress in the exploration of the collective behavior of 
ultracold quantum gases, the relevance of few-body physics, i.e., the physics of systems with three or more particles, 
has increasingly been recognized for its fundamental importance in determining the stability and dynamics of ultracold gases
\cite{fletcher2013PRL,rem2013PRL,makotyn2014NTP,eismann2016PRX,fletcher2017Sci,sykes2014PRA,laurent2014PRL,
smith2014PRL,piatecki2014NTC,barth2015PRA,corson2015PRA,jiang2016PRA,chevy2016JPB,ding2017PRA,colussi2017ARX}. }
The key ingredient that sets apart ultracold gases 
from most other physical systems is that the strength of the interatomic interactions can be controlled.
By applying an external magnetic field, the intricate fine and hyperfine structure of alkali atoms allow for the occurrence of Feshbach 
resonances in the collision threshold for two atoms \cite{chin2010RMP}, causing the two-body $s$-wave scattering length, $a$,
to diverge. 
This ability to control the scattering length near Feshbach resonances is the pivotal element that enables driving  
ultracold quantum gases into the {\em strongly interacting regime}, a regime in which $|a|\gg r_0$ (where $r_0$ is the characteristic range of the interatomic 
interactions) and drastically different collective regimes are expected \cite{fletcher2013PRL,rem2013PRL,makotyn2014NTP,eismann2016PRX,fletcher2017Sci,sykes2014PRA,laurent2014PRL,
smith2014PRL,piatecki2014NTC,barth2015PRA,jiang2016PRA,chevy2016JPB,
dalfovo1999RMP,stringari2003,pethick2008,giorgini2008RMP,bloch2008RMP}. 
In practice, however, it was quickly realized \cite{anderson1995Sci,davis1995PRL,bradley1997PRL,burt1997PRL,
inouye1998NT,courteille1998PRL,stenger1999PRL,roberts2000PRL,marte2002PRL,weber2003Sci,weber2003PRL}
that three-body scattering processes within the gas lead to strong atomic losses, consequently limiting the lifetime and stability of the 
system. These observations were soon recognized \cite{esry1996JPB,esry1999PRL,nielsen1999PRLb,bedaque2000PRL,braaten2001PRL} 
as a window for accessing a much more fundamental effect governing strongly interacting few-body systems: the 
{\em Efimov effect}. 

The Efimov effect \cite{efimov1970PLB,efimov1971SJNP,efimov1972JETPL,efimov1973NPA} was predicted nearly 45 years ago 
in the context of nuclear physics. It is one of the most counterintuitive quantum phenomena present in a few particle system. 
In its purest form, the Efimov effect occurs for strongly interacting systems ($|a|\gg r_0$) and 
is associated with the appearance of an infinite number of weakly bound three-body states 
---today called {\em Efimov states}. In many ways, these states appear different from the common-sense 
expectation. 
For instance, a large number of Efimov states can exist even when the two-body 
interaction is not strong enough to support a 
%JPD
%\textcolor{red}
{{\em single} weakly bound $s$-wave bound state, a typical scenario found whenever $a<0$ and $|a|\gg r_0$.} 
And, if a weakly bound $s$-wave two-body state does exist ($a>0$), {\em weakening} the two-body 
interaction (for instance, by making the two-body potential well shallower) can only {\em increase} the number of
Efimov states. 
Perhaps more surprising, when a zero-energy $s$-wave two-body state exists
($|a|=\infty$), the number of Efimov states is rigorously {\em infinite}, even though the underlying two-body interactions are 
short-ranged and can only support a {\em finite} number of bound states.
These characteristics defy any simple and intuitive physical argument and, in fact, were what led to many attempts to disprove Efimov's 
original predictions.
However, these attempts mainly resulted in different ways to prove Efimov's predictions.

%JPD
%\textcolor{red}
{The key finding in Efimov's work was that, in the large $|a|$ limit, an attractive $1/R^2$ effective interaction
emerges in the range $r_0\ll R \ll |a|$, with a supercritical coefficient. As a result, this interaction causes the number of three-body states to 
increase to infinity as $a\rightarrow\pm\infty$.} (Here, $R$ describes the overall size of the three-body system.)
The specific $1/R^2$ form of the attractive effective interaction also naturally leads to one of the key features of the 
Efimov effect: three-body observables associated with the Efimov effect have geometric-scaling properties 
characterized by the {\em geometric scaling factor} $e^{\pi/s_0}$, with $s_0$ being intimately 
related to the strength of the attractive $1/R^2$ effective interaction. The $1/R^2$ form of the interaction results in log-periodic properties for
any low-energy three-body observable associated with the Efimov effect. 
A prime example is the geometric scaling of energies, $E_{\rm 3b}$, of Efimov states at $|a|=\infty$: 
\begin{align}
&E_{\rm 3b}^{(i+1)}=E_{\rm 3b}^{(i)}/(e^{\pi/s_0})^2,\label{EnergyN}
\end{align}
with $i=0$, 1, 2, ..., $\infty$. 
The {\em Efimov coefficient}, $s_0$, is a universal number depending only on very general properties of the system
such as its total angular momentum, number of resonant pairs of interactions, and the mass ratios 
between the particles \cite{efimov1973NPA}.
For instance, for three-identical bosons in their lowest total angular momentum state $s_0\approx1.0062378$ 
and the corresponding geometric scaling $e^{\pi/s_0}\approx22.694384$.
In addition to expressing the geometric scaling properties for the energies, Eq.~(\ref{EnergyN}) also 
%reveals one
%JPD
%\textcolor{red}
{manifests an}
important, and general, characteristic of the Efimov effect: three-body observables are determined in terms of a single 
{\em three-body parameter}, which depends on the short distance behavior of the three-body interactions \cite{braaten2006PRep}. 
In the case of Eq.~(\ref{EnergyN}), knowing the energy of the Efimov ground state ($i=0$) allows for the determination 
of the energy of all other states by simply rescaling this known value. 
%JPD
%\textcolor{red}
{Without the three-body parameter, one can only find the ratios but not where the ladder of values sits.}

In ultracold quantum gases, the main signatures for the Efimov effect arise from low-energy scattering observables 
that are measurable through atomic and molecular losses. 
Here too, observables have log-periodic properties.
In scattering processes, however, characteristic Efimov features originate from interference and resonant effects 
caused by the formation of Efimov states as $a$ is changed by the multiplicative factor $e^{\pi/s_0}$
\cite{esry1999PRL,nielsen1999PRLb,bedaque2000PRL,braaten2001PRL}.
The first experimental observation of the Efimov effect was made by measuring atomic losses in an ultracold 
gas of $^{133}$Cs atoms \cite{kraemer2006NT}; this observation triggered many other experimental explorations. 
Currently, signatures of Efimov states have been observed in ultracold gases of $^{133}$Cs \cite{kraemer2006NT,knoop2009NTP,
ferlaino2009PRL,knoop2010PRL,berninger2011PRL,ferlaino2011FBS,zenesini2012NJP,zenesini2013NJP,zenesini2014PRA,huang2014PRL,
huang2015PRA}, $^{39}$K \cite{zaccanti2009NTP,roy2013PRL}, $^{7}$Li \cite{pollack2009Sci,gross2009PRL,
gross2010PRL,gross2011CRP,machtey2012PRLb,dyke2013PRA}, $^{6}$Li \cite{ottenstein2008PRL,huckans2009PRL,
williams2009PRL,lompe2010PRL,lompe2010Sci,nakajima2010PRL,nakajima2011PRL,huang2014PRA}, $^{85}$Rb 
\cite{wild2012PRL,klauss2017PRL}, $^{4}$He \cite{knoop2012PRA,kunitski2015Sci}, and heteronuclear gas mixtures of $^{41}$K-$^{87}$Rb 
\cite{barontini2009PRL}, $^{40}$K-$^{87}$Rb \cite{bloom2013PRL}, $^{6}$Li-$^{133}$Cs \cite{pires2014PRL,tung2014PRL,
ulmanis2016PRA,ulmanis2016PRL,johansen2017NTP}, and $^{7}$Li-$^{87}$Rb \cite{maier2015PRL}. (Note that some debate 
\cite{hu2014PRA,wacker2016PRL}
still exists on the observations in Ref.~\cite{barontini2009PRL}.)
This frenetic pace of activity has deepened and expanded our knowledge of this exotic phenomena and provided 
the field of few-body physics with a broad range of outstanding problems and possibilities.

From a more general perspective, the physics obtained from Efimov's original work, or simply {\em Efimov physics}, 
has an even broader impact than the Efimov effect alone \cite{dincao2005PRL,dincao2006PRAb,dincao2008PRL}. 
One immediate result is that the Efimov physics allows for a classification scheme in which all three-body systems with 
strong $s$-wave interactions fall into one of two categories:  {\em three-body attractive systems}, where
the attractive $1/R^2$ effective interaction occurs (that is characteristic of the Efimov effect) and {\em three-body repulsive systems},
in which, instead, a repulsive $1/R^2$ effective interaction emerges.
Systems within the same class have similar scattering and bound state properties.
For attractive systems, scattering observables display unique signatures due to the formation of Efimov states and are 
generally associated to strong atomic and molecular losses causing great instability and short lifetimes in ultracold gases. 
In contrast, although no Efimov states occur in repulsive three-body systems, scattering observables 
also display a signature characteristic that traces back to Efimov physics: the suppression of inelastic transitions and
the consequent reduction of atomic and molecular losses.
For such systems, the repulsive $1/R^2$ effective interaction prevents atoms from approaching each other at short 
distances, which, in turn, leads to a suppression of inelastic transitions and various universal properties for 
the system \cite{petrov2004PRLb}. 
Therefore, the suppression of loss rates depends intimately on the strength of the Efimov repulsive interaction
and is an extremely beneficial effect for ultracold gases that allows for long lifetimes and stability.
This effect has been observed in two-component Fermi gases of $^{40}$K~\cite{greiner2003NT} and
$^{6}$Li~\cite{jochim2003Sci,cubizolles2003PRL,jochim2003PRL,zwierlein2003PRL,strecker2003PRL,bartenstein2004PRLb,
regal2004PRL,bourdel2004PRL}, as well as in heteronuclear Fermi-Fermi and Bose-Fermi mixtures of 
$^{40}$K-$^{87}$Rb~\cite{zirbel2008PRL}, $^{40}$K-$^{6}$Li~\cite{spiegelhalder2009PRL}, $^{6}$Li-$^{174}$Yb
~\cite{khramov2012PRA}, $^{23}$Na-$^{40}$K~\cite{wu2012PRL} and $^6$Li-$^7$Li \cite{laurent2017PRL}. 
Note that the non-existence of Efimov states does not mean that other universal states
can occur in repulsive systems. For a subclass of such systems, the interaction beyond the range in which Efimov 
potentials are repulsive ($r_0\ll R \ll a$) can be attractive and support bound states. These states, today 
known as Kartavtsev-Malykh (KM) states \cite{kartavtsev2007JPB}, can be extremely long-lived, have a very
different nature than Efimov states, and have an important impact on the scattering properties of the system 
\cite{kartavtsev2007JPB,endo2011FBS,endo2012PRA,kartavtsev2014YF}. 

In this tutorial, we provide an analysis of strongly interacting three-body systems from a simple and 
conceptually clear perspective based on the adiabatic hyperspherical representation. 
It is {\em not} our goal to review the theoretical approaches used to explore few-body problems,
because more comprehensive reviews can be found in Refs.~\cite{ferlaino2011FBS,nielsen2001PRep,jensen2004RMP,
braaten2006PRep,braaten2007AP,hammer2010ARNPS,rittenhouse2011JPB,blume2012RPP,wang2013Adv,mitroy2013RMP,
wang2015ARCAM,naidon2017RPP,greene2017RMP}. 
Rather, our goal is to focus on the broadness of Efimov physics and provide an intuitive picture of 
its impact on three-body scattering processes and, consequently, ultracold quantum gases.
This tutorial is organized as follows. 
In Section \ref{HypWay} we provide a brief discussion of the adiabatic
hyperspherical representation and set the stage for the analyses in the next sections. 
In Section \ref{ThreeBodySec}, by assuming contact interactions between
atoms, we analyze the solutions of the three-body problem. Revisiting Efimov's original work 
\cite{efimov1970PLB,efimov1971SJNP,efimov1972JETPL,efimov1973NPA}, 
we 
%JPD
%establish the classification scheme provided by Efimov physics and discuss the energy spectrum
%for both Efimov and KM states.
%\textcolor{red}
{provide a classification scheme based on Efimov physics and discuss the energy spectrum
for both Efimov and KM states.}
In Section \ref{Collisions}, using a pathway analysis for scattering processes
\cite{dincao2005PRL,dincao2006PRAb,dincao2008PRL,dincao2009PRL,colussi2014PRL,colussi2016JPB} 
combined with a simple WKB formulation \cite{berry1966Proc.Phys.Soc.}, we derive both the energy and scattering 
length dependence of three-body scattering observables for all three-body systems with strong $s$-wave 
interactions. This pathway analysis provides a clear and intuitive physical picture of how the main signatures of Efimov physics appear in 
scattering processes.
In Section \ref{EfimovPhysics}, we analyze the impact of Efimov physics in ultracold quantum gases, including finite temperature
effects. In Section \ref{EfimovUniversality}, we briefly review the universal relations for Efimov features 
\cite{braaten2006PRep,gogolin2008PRL,helfrich2010PRA,helfrich2011JPB} and discuss recent progress in understanding the 
origin of the universality of the three-body parameter and its four-body counterpart. 
In Section \ref{Summary}, we summarize our analysis.

% !TEX root = ./TutorialJPB.tex

%%%%%%%%%%%%%%%%%%%%%%%%%%%%%%%%%%%%%%%%%%%%%%%%%%%%
\section{The hyperspherical way} \label{HypWay}

The key idea behind the adiabatic hyperspherical representation \cite{macek1968JPB,fano1981PRA,fano1976PT} 
is to explore the quasi-separability of the corresponding hyperradial and hyperangular motions for systems with three or
more particles.
In this representation the {\em hyperradius}, $R$, determines the overall size of the system, while all other degrees 
of freedom are represented by a set of {\em hyperangles}, $\Omega$, describing the system's internal motion
%JPD
%\textcolor{red}
{and overall rotations}.
In many instances, this representation allows for the description of few-body systems 
in a similar way to a two-particle problem, but now in terms of the collective coordinate, the 
hyperradius $R$, thus providing a simple and conceptually clear physical picture for both bound and scattering processes. 
Although different choices of hyperspherical coordinates exist 
\cite{lin1995PRep,delves1959NP,delves1960NP,smith1960PR,whitten1968JMP,johnson1980JCP,lepetit1990CPL,kendrick1999JCP,
suno2002PRA,kuppermann1997JPCA,aquilanti1997JCSFT,lepetit2006JCP,rittenhouse2011JPB,avery1989}, 
the three-body Schr\"odinger equation can aways be written in a quasi-separable form 
%JPD
%(atomic units will be used unless otherwise noted) 
as:
\begin{eqnarray}
\left[-\frac{\hbar^2}{2\mu}\frac{\partial^2}{\partial R^2}
+H_{\rm ad}(R,\Omega)\right]\psi(R,\Omega)=E\psi(R,\Omega),\label{schreq} 
\end{eqnarray} 
\noindent
which is obtained after separating the center-of-mass motion and rescaling the total wave 
function for the relative motion according to: $\Psi=\psi/R^{5/2}$. 
In Eq.~(\ref{schreq}), $\mu$ is the three-body reduced mass, 
$\mu^2={m_{1}m_{2}m_{3}/M}$ with $M=m_{1}+m_{2}+m_{3}$ being the total mass
and $m_i$ the mass of the $i^{\rm th}$ particle and $E$ is the total (relative) 
energy of the system. The adiabatic Hamiltonian, $H_{\rm ad}$, is given by
\begin{eqnarray}
H_{\rm ad}(R,\Omega)=\frac{\Lambda^2(\Omega)+15/4}{2\mu R^2}\hbar^2
+V(R,\Omega),\label{Had}
\end{eqnarray}
and contains the hyperangular part of the kinetic energy as expressed through the grand-angular momentum operator, 
$\Lambda^2(\Omega)$, and all the interparticle interactions via $V(R,\Omega)$.
Although not necessary, it is typically assumed that the interactions are given in terms of a pairwise sum of the form
\begin{equation}
V(R,\Omega)=v(r_{12})+v(r_{23})+v(r_{31}), \label{Int}
\end{equation} 
where $r_{ij}=|\vec{r}_i-\vec{r}_j|$ is the distance between particles $i$ and $j$. 
In most cases, nonadditive three-body interactions can be easily introduced in Eq.~(\ref{Int}) with 
effectively no cost to the calculations \cite{dincao2009JPB}.

The quasi-separability of the hyperradial and hyperangular parts of the Schr\"odinger 
equation (\ref{schreq}) is exploited by imposing an adiabatic expansion of the total 
wave function as
\begin{equation}
\psi(R,\Omega)=\sum_{\nu}F_{\nu}(R)\Phi_{\nu}(R;\Omega),
\label{chfun}
\end{equation}
\noindent 
where $F_{\nu}(R)$ and $\Phi_{\nu}(R;\Omega)$ are the hyperradial wave functions 
and channel functions, respectively, with $\nu$ representing
all quantum numbers necessary to specify each channel. The channel functions $\Phi_{\nu}(R;\Omega)$
form a complete set of orthonormal functions at each value of $R$ and 
are eigenfunctions of $H_{\rm ad}$, 
\begin{equation}
H_{\rm ad}(R,\Omega)\Phi_{\nu}(R;\Omega)
=U_{\nu}(R)\Phi_{\nu}(R;\Omega).\label{poteq}
\end{equation}
\noindent
The eigenvalues $U_{\nu}(R)$ are the three-body hyperspherical potentials from which 
one can define three-body {\em effective potentials} for the hyperradial motion [see Eq.~(\ref{EffPot})
below]. 

By substituting the total wave function [Eq.~(\ref{chfun})] into the three-body Schr\"odinger equation [Eq.~(\ref{schreq})] 
and projecting out the channel functions $\Phi_{\nu'}$, we obtain the hyperradial Schr\"odinger equation,  
\begin{align}
&\left[-\frac{\hbar^2}{2\mu}\frac{d^2}{dR^2}+W_{\nu}(R)-E\right]F_{\nu}(R) \nonumber\\
&~~~~~~-\frac{\hbar^2}{2\mu}\sum_{\nu'\ne\nu}W_{\nu\nu'}(R)F_{\nu'}(R)
=0,\label{radeq}
\end{align}
\noindent
which describes the three-body hyperadial motion under the influence of the
{\em effective potentials}, $W_\nu(R)$ and  {\em nonadiabatic couplings}, $W_{\nu\nu'}(R)$, 
defined, respectively, as
\begin{align}
W_{\nu}(R)&=U_{\nu}(R)-\frac{\hbar^2}{2\mu}Q_{\nu\nu}(R), \label{EffPot}\\
W_{\nu\nu'}(R)&=2P_{\nu\nu'}(R)\frac{d}{dR}+Q_{\nu\nu'}(R), \label{Wcoup}
\end{align}
where the first and second derivative couplings are determined from
the hyperradial dependence of the channel functions,
\begin{align}
P_{\nu\nu'}(R)=\Big\langle\Phi_{\nu}\Big|\frac{\partial \Phi_{\nu'}}{\partial R}\Big\rangle~\mbox{and}~
Q_{\nu\nu'}(R)=\Big\langle\Phi_{\nu}\Big|\frac{\partial^2 \Phi_{\nu'}}{\partial R^2}\Big\rangle.\label{puvquv}
\end{align} 
The brackets denote integration over the hyperangular coordinates $\Omega$ only. 
%JPD
%Although in general the nonadiabatic couplings are directly responsible for the mixing of 
%channels and inelastic transitions in scattering processes, including $Q_{\nu\nu}(R)$ in the 
%definition of the effective potential $W_{\nu}(R)$ is crucial for obtaining potentials with the correct
%behavior at large distances \cite{fedorov2001JPA,nielsen2001PRep,wang2013Adv}. 
%\textcolor{red}
{Although nonadiabatic couplings are generally associated to  
channel mixing and inelastic transitions in scattering processes, including $Q_{\nu\nu}(R)$ in the 
definition of the effective potential $W_{\nu}(R)$ [see Eq.~(\ref{EffPot})] is crucial for obtaining potentials with 
the correct behavior at large distances \cite{fedorov2001JPA,nielsen2001PRep,wang2013Adv}.}
As it stands, Eq.~(\ref{radeq}) is exact. However, in practice, the sum 
over channels must be truncated, and the number of channels retained increased until one 
achieves the desired accuracy. 

\begin{figure}[htbp]
\begin{center}
\includegraphics[width=3.4in]{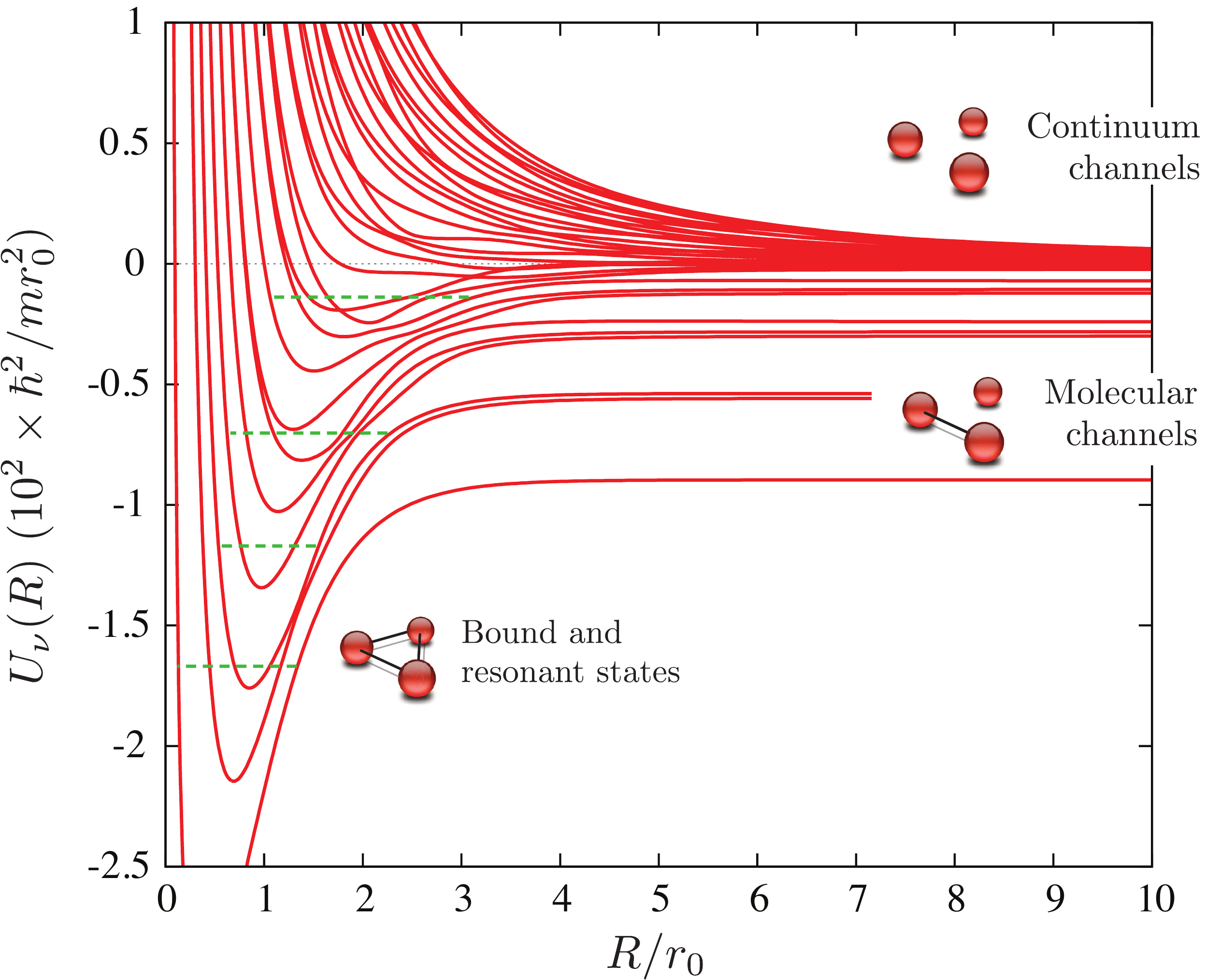}
\caption{Three-body hyperspherical potentials for a system of three identical bosons
interacting via the two-body interaction model: $v(r)=-D\mbox{sech}^2(r/r_0)$ \cite{dincao2005PRA}. For this calculation
we adjust $D$ to support a total of 12 diatomic bound states.
At distances $R\gg r_0$, the hyperspherical potentials can be classified into molecular and continuum 
channels describing atom-molecule collisions and collisions between three free atoms.}\label{3BPotentials}     
\end{center}
\end{figure}

Solving the hyperangular part of the problem, i.e., the adiabatic equation (\ref{poteq}),
is the {\em core}, and often the hardest part, of the adiabatic hyperspherical representation.
Nevertheless, once this task is performed, the effective potentials and nonadiabatic couplings, 
Eqs.~(\ref{EffPot})-(\ref{puvquv}), determine all bound and scattering properties of the system 
through the solutions of the hyperradial Schr\"odinger equation (\ref{radeq}) with proper boundary conditions, 
while still offering a simple and conceptually clear physical picture---which is no different than the usual radial two-body problem. 
Figure~\ref{3BPotentials} displays a typical set of hyperspherical potentials for a system of three identical bosons \cite{dincao2005PRA}
interacting via a simple two-body interaction: $v(r)=-D\mbox{sech}^2(r/r_0)$, with $D$ determining
the potential depth and $r_0$ the characteristic range of the interaction. In Fig.~\ref{3BPotentials}
we adjust $D$ to support a total of 12 diatomic bound states.

From the structure of the three-body potentials, various properties of the system become evident, thus providing
a clear physical picture for the system's bound and resonant spectra as well as scattering processes.
In the asymptotic regime, i.e., when $R$ is much larger than any
other length scale in the problem, the three-body hyperspherical potentials can be uniquely identified with particular 
physical processes. Three-body potentials that converge asymptotically to the energy of diatomic 
molecular states are associated with molecular channels and represent atom-molecule scattering states. On the other hand, 
potentials converging to the zero-energy three-body threshold (black-dotted horizontal line in Fig.~\ref{3BPotentials}) 
are associated with three-body continuum channels and 
describe collisions of three free atoms. Molecular and three-body continuum potentials have their asymptotic expansion 
given by, respectively,
\begin{equation}  
W_{\nu}(R)
\simeq-{E}_{b}+\frac{l(l+1)}{2\mu R^2}\hbar^2,\label{bcI}
\end{equation}  
and
\begin{equation}  
W_{\nu}(R)
\simeq\frac{\lambda(\lambda+4)+15/4}{2\mu R^2}\hbar^2,
\label{chI}
\end{equation}  
where $E_b$ is the binding energy of the diatomic molecular state, $l$ is the relative angular momentum between the 
molecule and the atom, and $\lambda$ is a positive integer number labeling the eigenstates of the 
hyperangular kinetic energy operator $\Lambda^2(\Omega)$ in Eq.~(\ref{Had}). 
As we will show in Section~\ref{Collisions},
the asymptotic form of the three-body potentials [Eqs.~(\ref{bcI}) and (\ref{chI})] determines the energy
dependence of both low-energy elastic and inelastic scattering processes. 
%JPD
%which, in turn, can be viewed as proceeding through a number of different pathways from 
%the relevant initial and final collision channels.
In Fig.~\ref{3BPotentials}, bound and resonant states (schematically represented by the green-dashed lines) 
are defined for states with energies below and above the lowest molecular channel, respectively. 
This characterization of states is consistent with the fact that true bound states do not decay, while resonant states will have a finite lifetime due 
to transitions to energetically open molecular channels. Although not shown in Fig.~\ref{3BPotentials}, the importance of 
nonadiabatic couplings can be, in part, deduced from the structure of avoided-crossings. As usual, broad avoided-crossings
imply in strong inelastic transitions while narrow ones indicate weak transitions. Although narrow avoided-crossings represent
weak transitions, they are numerically challenging to describe. Nevertheless, simple changes of the adiabatic representation
can fully address such problems \cite{wang2011PRA}.

In the next section, we apply the adiabatic hyperspherical representation for the three-body problem 
with two-body regularized Fermi pseudopotentials \cite{fermi1934NC,huang1957PR}, which allow us to account for 
$s$-wave interactions only and completely neglect finite-range effects. 
This interaction model, combined with the Faddeev representation for the total wave function 
\cite{faddeev1961JETP,efimov1973NPA,fedorov1993PRL} in Delves' hyperspherical coordinates 
\cite{delves1959NP,delves1960NP,smith1960PR}, allows for the determination of semianalytical solutions for 
the problem of three strongly interacting atoms and represents an ideal model for studying universal properties 
related to the Efimov physics.

% !TEX root = ./TutorialJPB.tex

%%%%%%%%%%%%%%%%%%%%%%%%%%%%%%%%%%%%%%%%%%%%%%%%%%%%
\section{Strongly interacting three-body systems} \label{ThreeBodySec}

In this section, we provide an analysis of the solutions of the three-body problem with $s$-wave 
zero-range interactions, implemented in terms of specific boundary conditions. The zero-range model is the basis 
of several other approaches to few-body systems and we refer to such works throughout this section. 
Evidently, zero-range model interactions can only capture the correct three-body physics at distances
larger than the typical range of interatomic interactions. 
Nevertheless, this constrain is enough to allow for the exploration of many universal properties of the system. 
In fact, within the adiabatic hyperspherical representation, one can still incorporate modifications to these 
universal results that originate from the finite-range aspects of more realistic interactions based on simple 
physical arguments. 
Here, we revisit Efimov's original approach and provide some modifications to it in order to impose 
the proper symmetrization of the total wave function. We also analyze the conditions for the occurrence 
of the Efimov effect, establish a universal classification scheme for {\em all} three-body systems with 
strong $s$-wave interactions, and discuss the structure of effective potentials and nonadiabatic couplings 
as well as the origin of universal three-body states.

%%%%%%%%%%%%%%%%%%%%%%%%%%%%%%%%%%%%%%%%%%%%%%%%%%%%
%\subsection{Three-body problem and contact interaction}
\subsection{Contact interactions}
\label{ZRP}

In his original seminal work~\cite{efimov1970PLB,efimov1971SJNP,efimov1972JETPL,efimov1973NPA}, 
Efimov demonstrated that three-body effective interactions are strongly modified once 
the short-range two-body interactions become resonant, i.e., when the two-body $s$-wave scattering length, $a$,
becomes much larger than the characteristic range of the two-body interactions $r_{0}$, $|a|\gg r_{0}$.
Efimov's work revealed that in the resonant regime, the scattering length becomes an important length 
scale in the problem, setting the range within which three-body interactions assume an unusual form
for systems interacting via short-range potentials.
When three particles are within distances smaller than, or comparable to, $|a|$ but still larger than $r_0$, 
the three-body interactions assume a long-range form that is proportional to $1/R^2$ and have a universal character, 
regardless of the details that cause the two-body interactions to be resonant.
In this case, the three-body effective potentials $W_\nu$ are found to be conveniently parameterized by  
\begin{equation}
W_{\nu}(R)=\frac{\xi^2_{\nu}-{1}/{4}}{2\mu R^2}\hbar^2~~~~\mbox{($r_{0}\ll R \ll |a|$)},
\label{efpot}
\end{equation}
where $\xi$ is a universal quantity that determines the strength as well as the attractive and repulsive character of 
three-body interactions; $\xi$ depends only on very general properties of the system, such as the 
total angular momentum and parity, $J^{\pi}$, the number of resonant pairs of interactions, and the mass ratios 
of the particles \cite{efimov1973NPA}. 
Efimov's central motivation was to analyze the cases in which the effective interaction is attractive, which is 
obtained when $\xi=i s_0$ is a purely imaginary number and inevitably leading to Efimov effect 
\cite{efimov1970PLB,efimov1971SJNP,efimov1972JETPL,efimov1973NPA}.
Nevertheless, the solutions leading to repulsive interactions, i.e., when $\xi$ is a real number, were not fully 
explored and, as we will show throughout this manuscript, they have an important impact in ultracold 
quantum gases.

Therefore, it is clear that the key quantity that controls Efimov physics is the coefficient $\xi$ in Eq.~(\ref{efpot}). 
To determine $\xi$, we follow closely Efimov's original work~\cite{efimov1973NPA} and present 
results in terms of the system's permutation symmetry, mass ratio, and total angular momentum. 
One convenient way to perform this analysis is to write the total wave function in terms of the Faddeev components
\cite{faddeev1961JETP,efimov1973NPA,fedorov1993PRL} in the following form:
\begin{align}
\psi(R,&\Omega)=c_{1}\psi^{(1)}(\vec{\rho}_{23},\vec{\rho}_{1}) \nonumber\\
&+c_{2}\psi^{(2)}(\vec{\rho}_{31},\vec{\rho}_{2})+
c_{3}\psi^{(3)}(\vec{\rho}_{12},\vec{\rho}_{3}).
\label{twf}
\end{align}
Each Faddeev component, $\psi^{(i)}$, favorably describes a particular configuration of the three-body system in terms of its 
mass-scaled Jacobi vectors $\vec{\rho}_{ik}$ and $\vec{\rho}_{l}$ \cite{delves1959NP,delves1960NP,smith1960PR}, 
defined in terms of the individual particle's positions $\vec{r}_i$ and masses $m_i$ as
\begin{align}
\vec{\rho}_{ij}&={d_{k}^{-1}}\left(\vec{r}_{i}-\vec{r}_{j}\right), \nonumber\\
\vec{\rho}_{k}&=d_{k}\left[\vec{r}_{k}-
\left(\frac{m_{i}\vec{r}_{i}+m_{j}\vec{r}_{j}}{m_{i}+m_{j}}\right)\right],
\end{align}
where $d^2_{k}=({m_{k}}/{\mu})(1-m_{k}/M)$. Indices $\{i,j,k\}$ are cyclic permutations of $\{1,2,3\}$.
Note that the different sets of Jacobi coordinates $\{\vec{\rho}_{ij},\vec{\rho}_{k}\}$  in the Faddeev decomposition of the total 
wave function (\ref{twf}) can be related to 
each other by means of a ``kinematic rotation''~\cite{efimov1973NPA,nielsen2001PRep}
\begin{align}
\vec{\rho}_{ij}&=-\vec{\rho}_{ki}\cos\gamma_{jk}+\vec{\rho}_{j}\sin\gamma_{jk},\nonumber\\
\vec{\rho}_{k}&=-\vec{\rho}_{ki}\sin\gamma_{jk}-\vec{\rho}_{j}\cos\gamma_{jk},\label{KRot}
\end{align}
where $\tan\gamma_{ij}=(m_{k}M/m_{i}m_{j})^{1/2}\sigma_{ijk}$, with $\sigma_{ijk}$ being the sign
of the permutation $\{i,j,k\}$. 
The Faddeev components are, in general, determined from the solutions of a set of integro-differential Faddeev
equations \cite{fedorov1993PRL,nielsen2001PRep}. However, as we will see next, the use of zero-range interactions
greatly simplifies the problem.

One of the main advantages of using the zero-range interaction model is that this model
allows for semianalytical solutions for the three-body problem~\cite{efimov1973NPA}. 
The form of the $s$-wave zero-range potential that we use \cite{fermi1934NC,huang1957PR} is 
\begin{eqnarray}
v(r_{ij})=\frac{4\pi a_{ij}\hbar^2}{2\mu_{ij}}\delta^{(3)}(\vec{r}_{ij})\frac{\partial}{\partial r_{ij}}r_{ij},\label{zrpot}
\end{eqnarray}
where $\mu_{ij}=m_im_j/(m_i+m_j)$ is the two-body reduced mass, $a_{ij}$ the
scattering length, and $\delta$ the usual three-dimensional Dirac-$\delta$ function. 
One important characteristic of this potential model is that it can support only a single bound state 
for $a_{ij}>0$ that has a binding energy equal to $\hbar^2/2\mu_{ij}a_{ij}^2$. 
More importantly, it can be shown that the effect of this potential model is equivalent to imposing
the Bethe-Peierls boundary condition on the total three-body wave function \cite{efimov1973NPA}, 
\begin{equation}
\frac{1}{(r_{ij}\psi)}\frac{\partial}{\partial r_{ij}}(r_{ij}\psi)=
-\frac{1}{a_{ij}}, ~~~\mbox{at $r_{ij}=0$.}
\label{bcZRPTotal}
\end{equation}
This boundary condition, along with the corresponding boundary conditions for the pairs $jk$ and $ki$,
fully represents all the interactions in the problem. Consequently, the determination of the Faddeev components becomes equivalent
to solving the problem of three free particles, representing a tremendous simplification.

Within the hyperspherical framework, although the hyperradius, $R=({\rho}^2_{ij}+{\rho}^2_{k})^{1/2}$, 
is independent of the choice of Jacobi coordinates, each set of Jacobi vectors defines a set of hyperspherical 
coordinates \cite{delves1959NP,delves1960NP,smith1960PR}, denoted collectively by 
$\Omega_k\equiv\{\alpha_k,\hat\rho_{ij},\hat\rho_k\}$, where
\begin{eqnarray}
\rho_{ij}&=&R\sin\alpha_k,\nonumber\\
\rho_{k}&=&R\cos\alpha_k, \label{HypCood}
\end{eqnarray}
define the hyperangle $\alpha_k$, and $\hat\rho_{ij}$ and $\hat\rho_k$ are the usual spherical angles associated with the 
Jacobi vectors $\vec\rho_{ij}$ and $\vec{\rho}_k$, respectively. [The relationship between different sets of hyperspherical
coordinates can also be obtained from the same kinematic relation between different Jacobi vectors (\ref{KRot}) 
(see also Ref.~\cite{nielsen2001PRep}).]
%JPD
%\textcolor{red}
{Here, using the adiabatic hyperspherical representation, the total wave function is still represented 
by the adiabatic expansion (\ref{chfun})
\begin{align}
\psi(R,\Omega)=\sum_{\nu}F_{\nu}(R)\Phi_\nu(R;\Omega), \label{TotalWF}
\end{align}
while the Faddeev decomposition in Eq.~(\ref{twf}) is now imposed to the channel functions, 
\begin{align}
\Phi(R;\Omega)&=c_{1}\phi^{(1)}(R;\Omega_1) \nonumber\\
&+c_{2}\phi^{(2)}(R;\Omega_2)+c_{3}\phi^{(3)}(R;\Omega_3).
\label{twfhyp}
\end{align}}
Each of the Faddeev components is now determined from the adiabatic 
equation (\ref{poteq}) for three noninteracting atoms,
\begin{align}
\left[\frac{\Lambda^2(\Omega_k)+15/4}{2\mu R^2}\hbar^2-U_{\nu}(R)\right]{\phi^{(k)}_{\nu}(R;\Omega_k)}=0.
\label{HypSchEq}
\end{align}
In Eq.~(\ref{HypSchEq}) the hyperangular kinetic energy operator, $\Lambda^2$, is given by 
\begin{align}
\Lambda^2(\Omega_k)=-\frac{\partial^2}{\partial\alpha_{k}^2}&-2(\cot\alpha_{k}-\tan\alpha_{k})\frac{\partial}{\partial\alpha_{k}}\nonumber\\
&+\frac{{L}_{ij}^2(\hat{\rho}_{ij})}{\sin^2\alpha_{k}}+\frac{{L}_{k}^2(\hat{\rho}_{k})}{\cos^2\alpha_{k}}, \label{LambdaK}
\end{align}
where ${L}_{ij}$ and ${L}_{k}$ are the angular momentum operators for the pair $ij$ and for the relative motion 
of the $k^{\text{th}}$ particle to the center of mass of the $ij$ pair, respectively. Note that the hyperradial 
dependence in Eq.~(\ref{HypSchEq}) originates exclusively from the boundary condition (\ref{bcZRPTotal}), 
which is now expressed as
\begin{equation}
\frac{1}{(\sin\alpha_k\Phi)}\frac{\partial}{\partial \alpha_{k}}(\sin\alpha_k\Phi)=
-\frac{d_{k}R}{a_{ij}},~~~\mbox{at $\alpha_{k}=0$.}
\label{bcZRP}
\end{equation}

In a more general problem, where atoms interact through some finite-range potential, each of 
the Faddeev components of the channel functions [Eq.~(\ref{twfhyp})] should be expanded in 
terms of the eigenstates of the total angular momentum $J$ and azimuthal projection $M_J$, 
which are given by the Clebsch-Gordan series
\begin{align}
{\cal Y}^{JM_J}_{l_{ij}l_{k}}(\hat{\rho}_{ij},\hat{\rho}_{k})=\sum_{m_{l_{ij}}m_{l_{k}}}&
\langle l_{ij}m_{l_{ij}}l_{k}m_{l_{k}}|JM_J\rangle \nonumber\\
&\times Y_{l_{ij}m_{l_{ij}}}(\hat{\rho}_{ij})Y_{l_{k}m_{l_{k}}}(\hat{\rho}_{k}),
\label{harm}
\end{align}
where $Y_{lm}$ is the spherical harmonic. As usual, the eigenstates of the total angular momentum [Eq.~(\ref{harm})]
satisfy the rules for angular momentum addition: $|l_{ij}-l_{k}|\le J \le l_{ij}+l_{k}$ and 
$M_J=m_{l_{ij}}+m_{l_{k}}$. 
Within the $s$-wave zero-range interaction model (\ref{zrpot}), however, 
only states with $l_{ij}=0$ and $l_{k}=J$ interact and, as a result, only ``parity-flavorable" angular momentum
states, i.e., $J^{\pi}$ states with $\pi=(-1)^J$ are relevant (see Ref.~\cite{suno2002PRA,gasaneo2002JPB}). 
%JPD
%\textcolor{red}
{Therefore, the $s$-wave interaction approximation} allows us to write the Faddeev components simply as
\begin{align}
\phi^{(k)}(R;\Omega_k)=
\frac{\varphi^{J}_{\xi}(\alpha_{k})}{\sin\alpha_{k}\cos\alpha_{k}}
Y_{00}(\hat{\rho}_{ij})Y_{JM_{J}}(\hat{\rho}_{k}),
\label{sol}
\end{align}
where $\varphi$ is the eigenfunction of the hyperangular kinetic energy operator $\Lambda^2$ in Eq.~(\ref{LambdaK}), 
given by \cite{efimov1973NPA}
\begin{align}
\varphi^{J}_{\xi}(\alpha_{k})={\cos^{J}\alpha_{k}}
\left[\frac{d}{d\alpha_{k}}\frac{1}{\cos\alpha_{k}}\right]^{J}
\sin\left[\xi\left(\frac{\pi}{2}-\alpha_{k}\right)\right],
\label{ang}
\end{align}
satisfying the boundary condition $\varphi(\pi/2)=0$. The corresponding eigenvalues are equal to 
$(\xi^2-4)$, allowing us to write [using Eqs.~(\ref{EffPot}) and (\ref{HypSchEq})] the
corresponding effective potential as
\begin{align}
W_\nu(R)=\frac{\xi^2_\nu(R)-1/4}{2\mu R^2}\hbar^2+\frac{\hbar^2}{2\mu}Q_{\nu\nu}(R),\label{WQ}
\end{align}
in terms of the, as yet unknown, parameter $\xi$. General expressions for the nonadiabatic couplings $P$ and $Q$ [Eq.~(\ref{puvquv})] 
can also be derived in terms of $\xi$, as shown in Refs.~\cite{kartavtsev2007JPB,rittenhouse2010PRA}. 
For the purpose of our present study, however, we won't need to determine the nonadiabatic couplings. We 
only need to know their general behavior, i.e., when and where they are important. For instance, for 
$R\ll|a|$, since the boundary condition (\ref{bcZRP}) becomes independent of $R$, and so are $\xi$ and 
the channel functions (\ref{twfhyp}), the nonadiabatic couplings vanish, allowing us to neglect 
$Q$ in Eq.~(\ref{WQ}). 

Substituting Eqs.~(\ref{sol}) and (\ref{twfhyp}) in the boundary condition (\ref{bcZRP}), 
one obtains a set of transcendental equations from which $\xi$ can be finally determined. 
From the kinematic relations~(\ref{KRot}), one can show that for $\alpha_k=0$, the other two hyperangles are 
simply given by $\alpha_{i}=\gamma_{ki}$, $\alpha_{j}=\gamma_{kj}$, while the spherical angles are related 
by $\hat\rho_i=\hat\rho_j=-\hat\rho_{k}$. Using these relations and a bit of algebra, one can rewrite the 
boundary condition (\ref{bcZRP}) as
\begin{align}
\dot\varphi^{J}_{\xi}(0)~c_{k}
&+\frac{(-1)^{J}\varphi^{J}_{\xi}(\gamma_{ki})}{\sin\gamma_{ki}\cos\gamma_{ki}}c_{i}\nonumber\\
&~~~+\frac{(-1)^{J}\varphi^{J}_{\xi}(\gamma_{kj})}{\sin\gamma_{kj}\cos\gamma_{kj}} c_{j}
=-d_{k}\frac{R}{a_{ij}}\varphi^{J}_{\xi}(0)~c_{k},
\label{bcNoSym}
\end{align}
where the derivative of $\varphi$ at $\alpha_{k}=0$ can be determined analytically for even values of $J$,
%\begin{align}
%\dot\varphi^{J}_{\xi}(0)=
%\begin{cases} 
%(-1)^{\frac{J}{2}+1}\xi{\displaystyle\prod_{j=2}^{J}}\left(\xi^2-j^2\right)\cos\left(\frac{\pi}{2}\xi\right),
%& \mbox{\hspace{-0.1in}($J$ even),} \\
%(-1)^{\frac{J+1}{2}}{\displaystyle\prod_{j=1}^{J}}\left(\xi^2-j^2\right)\sin\left(\frac{\pi}{2}\xi\right),
%& \mbox{\hspace{-0.1in}($J$ odd),} 
%\end{cases}
%\end{align}
%where $j$ runs over even (odd) values for $J$ even (odd).
\begin{align}
\dot\varphi^{J}_{\xi}(0)\underset{J\mbox{\scriptsize-even}}{=}(-1)^{\frac{J}{2}+1}\xi{\prod_{j=1}^{\scriptscriptstyle J/2}}\left[\xi^2-(2j)^2\right]\cos\left(\frac{\pi}{2}\xi\right),
\label{angdJeven}
\end{align}
and for odd values of $J$,
\begin{align}
\dot\varphi^{J}_{\xi}(0)\underset{J\mbox{\scriptsize-odd}}{=}(-1)^{\frac{J+1}{2}}{\prod_{j=1}^{\scriptscriptstyle(J+1)/2}}\left[\xi^2-(2j-1)^2\right]\sin\left(\frac{\pi}{2}\xi\right).
\label{angdJodd}
\end{align}
Equation (\ref{bcNoSym}) defines a homogeneous system of linear equations for the $c_{i}$ coefficients 
of Eq.~(\ref{twfhyp}), where imposing that the corresponding determinant vanishes results in a transcendental
equation determining all possible values of $\xi$. Nevertheless, since the channel functions (\ref{twfhyp}) 
have no permutation symmetry, this transcendental equation contains all solutions. 
In the next section, we explicitly define the form of the transcendental equations resulting from 
the boundary condition (\ref{bcZRP}) that accounts for the proper permutation symmetry. We then analyze 
the solutions $\xi$ and characterize their strength as well as the attractive and repulsive aspects
of the three-body potentials for all relevant three-body systems with $s$-wave interactions. 

%%%%%%%%%%%%%%%%%%%%%%%%%%%%%%%%%%%%%%%%%%%%%%%%%%%%
%\subsection{Boundary conditions for three-body systems} 
\subsection{Boundary conditions} 
\label{TransEq}

So far, our approach is completely general and does not take into account the specific permutational symmetry of 
the wave function. Doing so is a requirement for distinguishing the solutions for three-body systems with 
three identical bosons, $BBB$,  two  identical bosons, $BBX$, and two identical fermions, 
$FFX$. Hereafter, $B$ and $F$ denote bosonic and fermionic atoms, respectively, while $X$ (or in some cases, 
$Y$ or $Z$) denotes a distinguishable atom that can be either a boson or fermion with different mass or in a 
different spin state. Within the Faddeev approach, the proper permutation symmetry can be imposed directly in the total wave 
function [Eqs.~(\ref{TotalWF}) and (\ref{twfhyp})] in a simple and intuitive manner. 
Moreover, in the scenario relevant for ultracold quantum gases, atoms are found in a single spin state, which allows for 
the permutation symmetry be imposed only on the spacial part 
of the wave function. Also, within the adiabatic hyperspherical representation, since the hyperradius $R$ is invariant 
under any permutation, the symmetry properties of the total wave function (\ref{TotalWF}) are imposed in the 
channel functions (\ref{twfhyp}) only. 
It is important to emphasize that in general, for problems in which the internal spin structure or the 
multichannel nature of the interactions becomes important, the symmetrization of the total wave function
must be done more carefully \cite{,colussi2014PRL,colussi2016JPB,mehta2008PRA}.

As we will see next, various properties of three-body systems will depend
only on the mass ratio between the distinguishable particles rather than on the values of the
individual masses, 
%JPD 
%\textcolor{red}
{leading us to define $\delta_{XY}=m_{X}/m_{Y}$ as the mass ratio 
between the atomic species $X$ and $Y$.} Besides determining the proper symmetrized boundary conditions
for each system (expressed in terms of a transcendental equation), we will also analyze various properties
of Efimov physics in terms of the mass ratio and total angular momentum $J$.
In all cases, however, we will see that the boundary conditions admit at most a {\em single} purely
imaginary solution and an {\em infinite} number of real solutions.

\paragraph{Three identical bosons ($BBB$).} 
An advantage of the Faddeev representation for the total wave function is that the symmetrization 
procedure is simple. To impose the desired permutational symmetry, one needs to determine
the proper relation between $c_{1}$, $c_{2}$ and $c_{3}$ in the Faddeev decomposition of the channel function
(\ref{twfhyp}) or, equivalently, in the total wave function (\ref{twf}). 
For $BBB$ systems, requiring $c_{1}=c_{2}=c_{3}$ is enough to obtain a completely symmetric 
wave function. In this case, using the symmetrized solution 
[Eq.~(\ref{twfhyp}) with $c_{1}=c_{2}=c_{3}=c_{B}$] and proceeding as with the derivation of Eq.~(\ref{bcNoSym}), 
we find that, for $BBB$ systems, $\xi$ is determined by solving the transcendental equation
\begin{equation}
\dot\varphi^{J}_{\xi}(0)
+2\frac{(-1)^{J}\varphi^{J}_{\xi}(\gamma_{B})}{\sin\gamma_{B}\cos\gamma_{B}}
=-\frac{2^{1/2}}{3^{1/4}}\frac{R}{a_{BB}}\varphi^{J}_{\xi}(0),
\label{bc3ib}
\end{equation}
where $\gamma_{B}=\pi/3$, and $\varphi$ and $\dot\varphi$ are given by Eqs.~(\ref{ang}), 
(\ref{angdJeven}) and (\ref{angdJodd}), respectively. In the range of $R$ where $R/|a_{BB}|\ll1$, this 
boundary condition becomes $R$ independent, and $\xi$ becomes constant. 
As a result, since nonadiabatic couplings vanish in this case, the effective potentials in Eq.~(\ref{WQ}) are simply
proportional to $R^{-2}$ within the range $r_{0}\ll R\ll |a_{BB}|$. 
The actual repulsive or attractive aspect of the three-body interaction within 
this range, i.e., whether Eq.~(\ref{bc3ib}) admits a purely imaginary solution, can be determined 
by inspection. In fact, solving Eq.~(\ref{bc3ib}) for different values of $J$ allows us to conclude that
the Efimov effect can only occur for identical bosons in the $J^{\pi}=0^+$ state and where it is associated with the
imaginary solution $\xi\approx 1.00624i$ \cite{efimov1970PLB,efimov1971SJNP,efimov1972JETPL,
efimov1973NPA}, obtained by substituting $\xi=i s_0$ into Eq.~(\ref{bc3ib}) and solving it for $s_0$. 
For all other values of $J^{\pi}\ne0$, Eq.~(\ref{bc3ib}) only admits real solutions for $\xi$ and, consequently, 
the three-body potentials are repulsive. Because of the trigonometric nature of the terms in Eq.~(\ref{bc3ib}), 
there exist an infinite number of such real solutions.

\paragraph{Two identical bosons ($BBX$).}
For bosonic $BBX$ systems, the most general case involves two types of resonant interactions
with associated large values for the $a_{BB}$ and $a_{BX}$ scattering lengths. 
In ultracold quantum gases, this situation can occur when there exists an overlap between two 
Feshbach resonances~\cite{dincao2009PRL}. In this case, to determine the boundary condition with 
proper permutation symmetry, we use the symmetrized form of the wave function (\ref{twfhyp}) 
obtained by requiring that $c_{1}=c_{X}$ and $c_{2}=c_{3}=c_{B}$.  
The boundary condition (\ref{bcNoSym}) now leads to a system of equations for $c_X$ and $c_B$ given by
%%\begin{widetext}
%%\begin{align}
%%\dot\varphi^{J}_{\xi}(0)~c_{B}
%%+\frac{(-1)^{J}\varphi^{J}_{\xi}(\gamma_{B})}{\sin\gamma_{B}\cos\gamma_{B}}c_{X}
%%+\frac{(-1)^{J}\varphi^{J}_{\xi}(\gamma_{X})}{\sin\gamma_{X}\cos\gamma_{X}}c_{B}
%%&=-\frac{(\delta_{XB}+1)^{1/2}}{[\delta_{XB}(\delta_{XB}+2)]^{1/4}}\frac{R}{a_{BX}}\varphi^{J}_{\xi}(0)~c_{B},
%%\label{bc2ib1}
%%\\
%%\dot\varphi^{J}_{\xi}(0)~c_{X}
%%+ 2\frac{(-1)^{J}\varphi^{J}_{\xi}(\gamma_{B})}
%%{\sin\gamma_{B}\cos\gamma_{B}}c_{B}
%%&=-\frac{2^{1/2}\delta_{XB}^{1/4}}{(\delta_{XB}+2)^{1/4}}\frac{R}{a_{BB}}\varphi^{J}_{\xi}(0)~c_{X},
%%\label{bc2ib2}
%%\end{align}
%%\end{widetext}
%\begin{align}
%\dot\varphi^{J}_{\xi}(0)~c_{B}&
%+\frac{(-1)^{J}\varphi^{J}_{\xi}(\gamma_{B})}{\sin\gamma_{B}\cos\gamma_{B}}c_{X}
%+\frac{(-1)^{J}\varphi^{J}_{\xi}(\gamma_{X})}{\sin\gamma_{X}\cos\gamma_{X}}c_{B}
%\nonumber\\
%&~~~=-\frac{(\delta_{XB}+1)^{1/2}}{[\delta_{XB}(\delta_{XB}+2)]^{1/4}}\frac{R}{a_{BX}}\varphi^{J}_{\xi}(0)~c_{B},
%\label{bc2ib1}
%\\
%\dot\varphi^{J}_{\xi}(0)~c_{X}&
%+ 2\frac{(-1)^{J}\varphi^{J}_{\xi}(\gamma_{B})}
%{\sin\gamma_{B}\cos\gamma_{B}}c_{B}
%\nonumber\\
%&~~~=-\frac{2^{1/2}\delta_{XB}^{1/4}}{(\delta_{XB}+2)^{1/4}}\frac{R}{a_{BB}}\varphi^{J}_{\xi}(0)~c_{X},
%\label{bc2ib2}
%\end{align}
\begin{align}
\dot\varphi^{J}_{\xi}(0)~c_{B}&
+\frac{(-1)^{J}\varphi^{J}_{\xi}(\gamma_{B})}{\sin\gamma_{B}\cos\gamma_{B}}c_{X}
+\frac{(-1)^{J}\varphi^{J}_{\xi}(\gamma_{X})}{\sin\gamma_{X}\cos\gamma_{X}}c_{B}
\nonumber\\
%&~~~=-\frac{(\delta_{XB}+1)^{1/2}}{(\tan\gamma_X)^{1/2}}\frac{R}{a_{BX}}\varphi^{J}_{\xi}(0)~c_{B},
&~~~=-\left(\frac{\delta_{XB}+1}{\tan\gamma_X}\right)^{1/2}\frac{R}{a_{BX}}\varphi^{J}_{\xi}(0)~c_{B},
\label{bc2ib1}
\\
\dot\varphi^{J}_{\xi}(0)~c_{X}&
+ 2\frac{(-1)^{J}\varphi^{J}_{\xi}(\gamma_{B})}
{\sin\gamma_{B}\cos\gamma_{B}}c_{B}
\nonumber\\
%&~~~=-\frac{2^{1/2}}{(\tan\gamma_B)^{1/2}}\frac{R}{a_{BB}}\varphi^{J}_{\xi}(0)~c_{X},
&~~~=-\left(\frac{2}{\tan\gamma_B}\right)^{1/2}\frac{R}{a_{BB}}\varphi^{J}_{\xi}(0)~c_{X},
\label{bc2ib2}
\end{align}
where $\tan\gamma_{B}=[(\delta_{XB}+2)/\delta_{XB}]^{1/2}$, $\tan\gamma_{X}=[\delta_{XB}(\delta_{XB}+2)]^{1/2}$,
and $\delta_{XB}=m_{X}/m_{B}$. For this homogeneous linear system, requiring the determinant to vanish
determines the value of $\xi$. As one can see, Eqs.~(\ref{bc2ib1}) and (\ref{bc2ib2}) depend on the mass ratio 
$\delta_{XB}$ instead of the individual atomic masses, as do their solutions $\xi$. Nevertheless,
since Eqs.~(\ref{bc2ib1}) and (\ref{bc2ib2}) depend on more than one scattering length, the structure of the
three-body potentials can be, in general, more complicated \cite{dincao2009PRL}.

We can, however, simplify our analysis for $BBX$ systems if we assume that the interaction between 
identical particles is not resonant, and set $a_{BB}=0$. In this case, one obtains $c_{X}=0$ 
[from Eq.~(\ref{bc2ib2})] and the boundary condition in Eq. (\ref{bc2ib1}) reduces to
\begin{equation}
\dot\varphi^{J}_{\xi}(0)
+\frac{(-1)^{J}\varphi^{J}_{\xi}(\gamma_{X})}
{\sin\gamma_{X}\cos\gamma_{X}}
%=-\frac{(\delta_{XB}+1)^{1/2}}{[\delta_{XB}(\delta_{XB}+2)]^{1/4}}\frac{R}{a_{BX}}\varphi^{J}_{\xi}(0).
=-\left(\frac{\delta_{XB}+1}{\tan\gamma_X}\right)^{1/2}\frac{R}{a_{BX}}\varphi^{J}_{\xi}(0).
\label{bc2ibBX}
\end{equation}
Analysis of the solutions of this transcendental equation shows that the Efimov effect 
occurs for $J^{\pi}=0^+$ [i.e., Eq.~(\ref{bc2ibBX}) admits a purely imaginary solution] 
for all values of the mass ratio $\delta_{XB}$. For even values of $J^{\pi}\ne0$,
the Efimov effect can also occur, but only for values of the mass ratio below a critical value, $\delta_{c}$.
We analyze these cases in more detail, along with their physical interpretation in Section~\ref{EfimovConditions}.
On the other hand, for odd values of $J^{\pi}$, Eq.~(\ref{bc2ibBX}) only permits real solutions, and the corresponding 
three-body systems are repulsive. 
A similar analysis can also be performed if the interaction between the dissimilar particles is not resonant, i.e., $a_{BX}=0$. 
In this case, the only resonant pair is the one involving the identical bosons. This results in $c_{B}=0$ [from Eq.~(\ref{bc2ib1})], 
and Eq.~(\ref{bc2ib2}) reduces to
\begin{equation}
\dot\varphi^{J}_{\xi}(0)
%=-\frac{2^{1/2}\delta_{XB}^{1/4}}{(\delta_{XB}+2)^{1/4}}\frac{R}{a_{BB}}\varphi^{J}_{\xi}(0).
=-\left(\frac{2}{\tan\gamma_B}\right)^{1/2}\frac{R}{a_{BB}}\varphi^{J}_{\xi}(0).
\label{bc2ibBB}
\end{equation}
This equation does not allow for purely imaginary solutions for $\xi$ for any value of the mass 
ratio $\delta_{XB}$ and total angular momentum $J$. 
In fact, in the limit $R/|a_{BB}|\rightarrow0$, the solutions of Eq.~(\ref{bc2ibBB}) are quite simple, 
$\xi=J+1,J+3,J+5,...$, irrespective of the value of the mass ratios~\cite{dincao2008PRL}. 

\paragraph{Two identical fermions ($FFX$).} 
For fermionic $FFX$ systems, the antisymmetric form of the wave function (\ref{twfhyp}), or, equivalently,
the total wave function (\ref{twf}), is obtained by requiring $c_{2}=-c_{3}=c_{F}$. 
Evidently, since $s$-wave interactions are not allowed for identical fermions,
one has to set the coefficient $c_{1}$, corresponding to the pair $FF$, to zero in Eq.~(\ref{twfhyp}).
Therefore, for these systems the only possible scattering length is $a_{FX}$, and
the boundary condition [Eq. (\ref{bcNoSym})] is considerably simplified, 
\begin{equation}
\dot\varphi^{J}_{\xi}(0)
-\frac{(-1)^{J}\varphi^{J}_{\xi}(\gamma_{F})}
{\sin\gamma_F\cos\gamma_F}
%=-\frac{(\delta_{XF}+1)^{1/2}}{[\delta_{XF}(\delta_{XF}+2)]^{1/4}}\frac{R}{a_{FX}}\varphi^{J}_{\xi}(0),
=-\left(\frac{\delta_{XF}+1}{\tan\gamma_{F}}\right)^{1/2}\frac{R}{a_{FX}}\varphi^{J}_{\xi}(0),
\label{bc2if}
\end{equation}
where $\tan\gamma_{F}=[\delta_{XF}(\delta_{XF}+2)]^{1/2}$ and $\delta_{XF}=m_{X}/m_{F}$.
Here, again the transcendental equation depends on the mass ratio
$\delta_{XF}$ instead of the individual atomic masses, and so do their solutions $\xi$. 
The analysis of Eq.~(\ref{bc2if}) shows that for all even values of $J^{\pi}$, 
no purely imaginary solutions exist. Thus their resulting effective potentials are always repulsive within the 
range $r_{0}\ll R\ll |a_{FX}|$. For odd values of $J^{\pi}$, however, the Efimov effect can in fact occur for $FFX$ systems, 
but only for mass ratios below a certain critical value, $\delta_{c}$, in a way similar to $BBX$ systems with even values
of $J^\pi$. We discuss such cases in more detail in Section~\ref{EfimovConditions}.

\begin{figure}[htbp]
\includegraphics[width=3.3in]{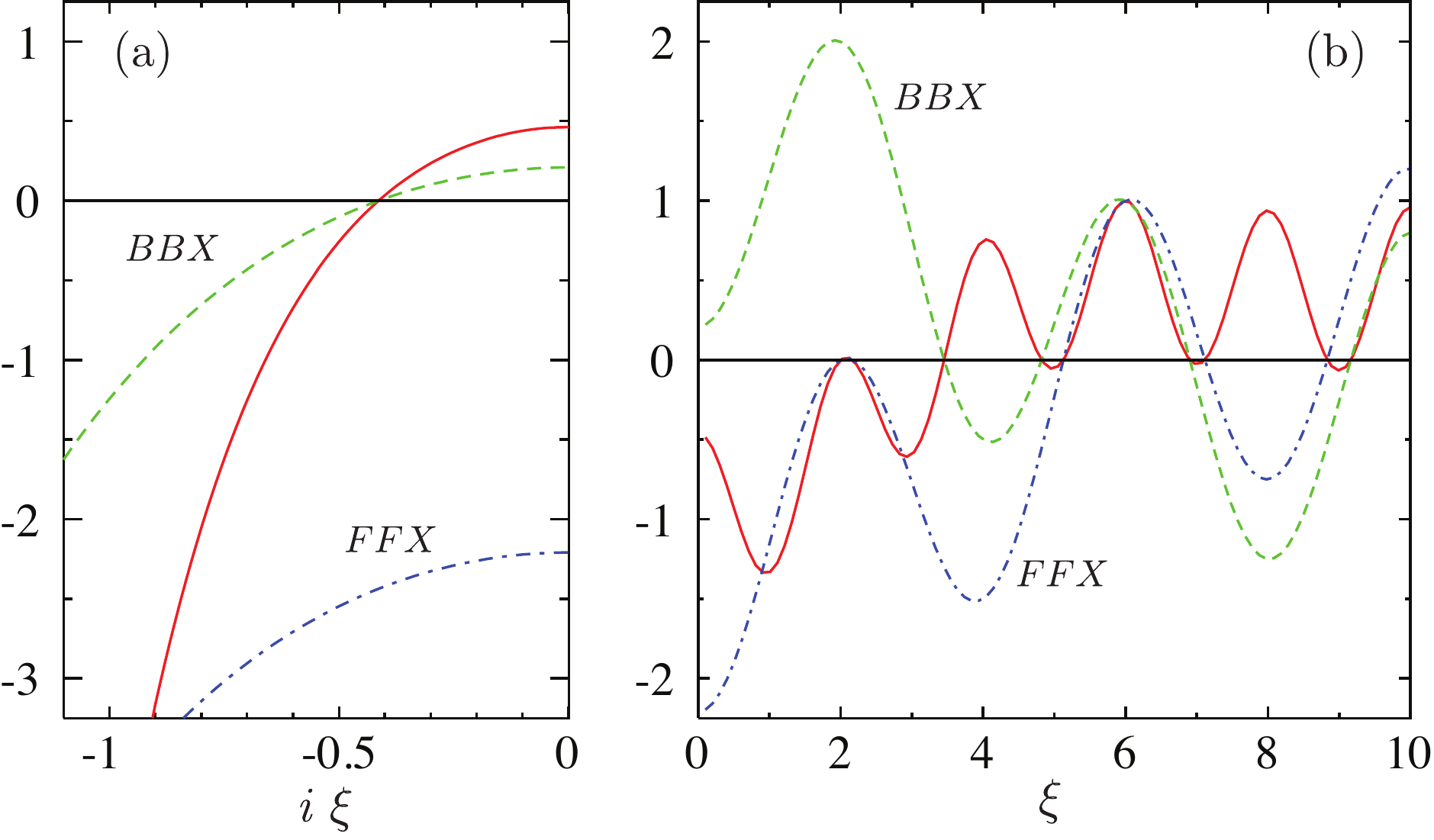}
\caption{Solutions of the boundary conditions for $BBX$ systems [Eq.~(\ref{bc2ibBX})] and $FFX$ systems 
[Eq.~(\ref{bc2if})] with $J^{\pi}=0^+$ and $\delta_{XB}=\delta_{XF}=1$, green-dashed and blue dot-dashed lines, 
respectively. For both complex (a) and real (b) solutions the boundary conditions for $BBX$ and $FFX$ systems
reproduce the results obtained from a nonsymmetrized wave function  [see Eq.~(\ref{bcNoSym})].}\label{Sols}     
\end{figure}

To illustrate the approach developed in this section, 
Figs.~\ref{Sols}(a) and \ref{Sols}(b) show a plot of the left-hand side of 
Eqs.~(\ref{bc2ibBX}) and (\ref{bc2if}) for $BBX$ and $FFX$ systems, respectively, 
for $J^{\pi}=0^+$ and $\delta_{XB}=\delta_{XF}=1$ (see green-dashed and blue dot-dashed lines in Fig.~\ref{Sols}).
The zeros in these plots represent the solutions of Eqs.~(\ref{bc2ibBX}) and (\ref{bc2if}) in the $R/|a|\rightarrow0$ limit. 
The red solid line in Fig.~\ref{Sols} corresponds to the determinant obtained from the nonsymmetrized boundary 
condition (\ref{bcNoSym}) and shows that our symmetrized boundary conditions (\ref{bc2ibBX}) and (\ref{bc2if}) correctly 
reproduce the full spectrum of solutions.

%%%%%%%%%%%%%%%%%%%%%%%%%%%%%%%%%%%%%%%%%%%%%%%%%%%%
\subsection{Conditions for the Efimov effect}\label{EfimovConditions}

In this section we discuss in more detail three-body systems in which the Efimov 
effect occurs. The analysis of solutions to the boundary conditions 
determined in the previous section gives us a precise idea of 
the conditions under which a particular three-body system will display
an attractive $1/R^2$ effective potential, which is characteristic of the Efimov effect.
Whenever a particular boundary condition allows for a purely imaginary solution for $\xi$
for $R/|a|\ll 1$, the Efimov coefficient $s_0=-i\xi$ determines the strength of the Efimov 
potential (\ref{efpot}), 
\begin{align}
W_{\nu}(R)=-\frac{s_0^2+1/4}{2\mu R^2}\hbar^2.
\end{align}
According to our discussion in the previous section, a necessary, but 
not sufficient, condition for the occurrence of the Efimov effect is that {\em at least} two of the pairs of 
interactions must be resonant. Efimov's original work~\cite{efimov1973NPA} also claims that 
the persistence of the Efimov effect is a result of a delicate balance 
between the repulsive and attractive contributions from the kinetic and potential energies, respectively. 
We discuss the cases in which the Efimov effect occurs in terms of this balance, 
giving further insight into the nature of this intriguing phenomena. 

\begin{figure}[htbp]
\begin{center}
\includegraphics[width=3.0in,angle=0,clip=true]{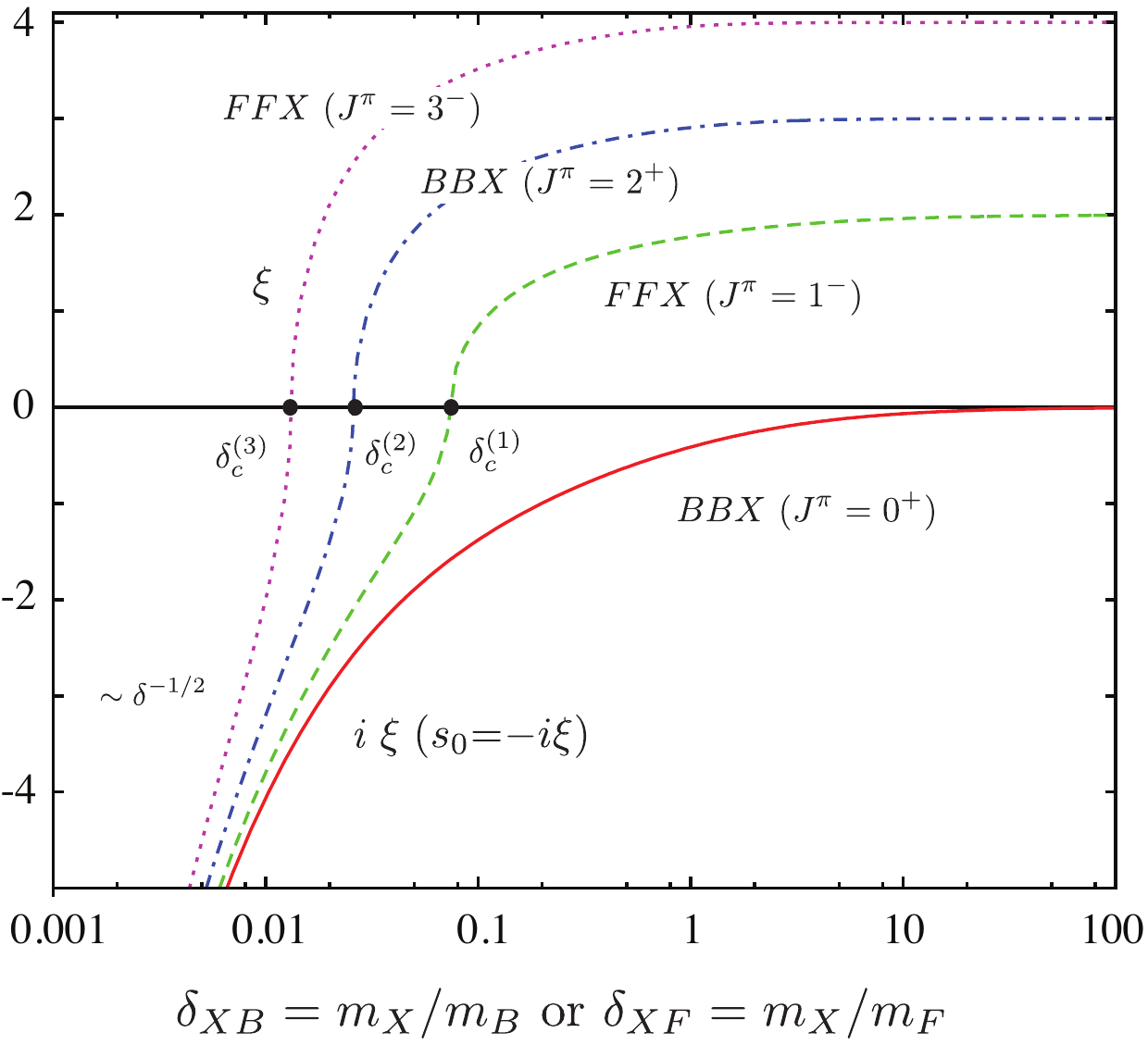}
\caption{Values of $\xi$ for $BBX$ and $FFX$ systems obtained from the solutions of Eqs.~(\ref{bc2ibBX})
and (\ref{bc2if}), respectively.  These results demonstrate that for $BBX$ systems the Efimov effect occurs for even partial waves, 
and the effect occurs for $FFX$ systems for odd partial waves and values of $\delta$ below a critical $\delta_{c}^{(J)}$
[$\delta_{c}^{(1)}\approx0.073492$, $\delta_{c}^{(2)}\approx0.025887$,$\delta_{c}^{(3)}\approx0.013159$].}\label{EfimovCond}     
\end{center}
\end{figure}

We begin with three identical bosons in the $J^{\pi}=0^+$ state. In this case, the Efimov effect is present 
and associated with the solution $\xi\approx 1.00624i$ of the boundary condition (\ref{bc3ib}) 
for $R/|a_{BB}|\ll1$. For the $J^{\pi}\ne0^+$ symmetry, 
Eq.~(\ref{bc3ib}) only allows real solutions, i.e., the system is described by repulsive $1/R^2$ effective potentials.
In contrast, for $BBX$ and $FFX$ heteronuclear systems, the Efimov effect occurs for even and odd values of $J^{\pi}>0$,
respectively, below a critical mass ratio, $\delta_{c}^{(J)}$. For $J^{\pi}=0^+$ $BBX$ systems, the Efimov effect occurs 
for {\em all} mass ratios \cite{efimov1973NPA}.
From Eqs. (\ref{bc2ibBX}) and (\ref{bc2if}), the critical value of the mass ratio
is found to be: $\delta_{c}^{(1)}\approx0.073492$, $\delta_{c}^{(2)}\approx0.025887$, and 
$\delta_{c}^{(3)}\approx0.013159$, for $J^\pi=1^{-},2^{+}$ and $3^{-}$, respectively.
Figure~\ref{EfimovCond} shows the values of $\xi$, obtained from Eqs.~(\ref{bc2ibBX}) and (\ref{bc2if}) 
in the regime in which $R/|a|\ll1$, as a function of the relevant mass ratios and indicates the 
corresponding values of $\delta_c$.
In Fig.~\ref{EfimovCond}, only the lowest solution for each value of $J^\pi$ is shown,
clearly displaying the transition from a repulsive (real $\xi$) to an attractive (imaginary $\xi$) 
interaction regime as the mass ratio decreases.
In fact, for small mass ratios, the corresponding values for the Efimov coefficient, $s_0=-i\xi$, 
diverge as $\delta^{-1/2}$ \cite{efimov1973NPA}. For large mass ratios, the attractive solution $
\xi=i s_0$ disappears or becomes vanishingly small.

\begin{figure}[htbp]
\begin{center}
\includegraphics[width=3.in,angle=0,clip=true]{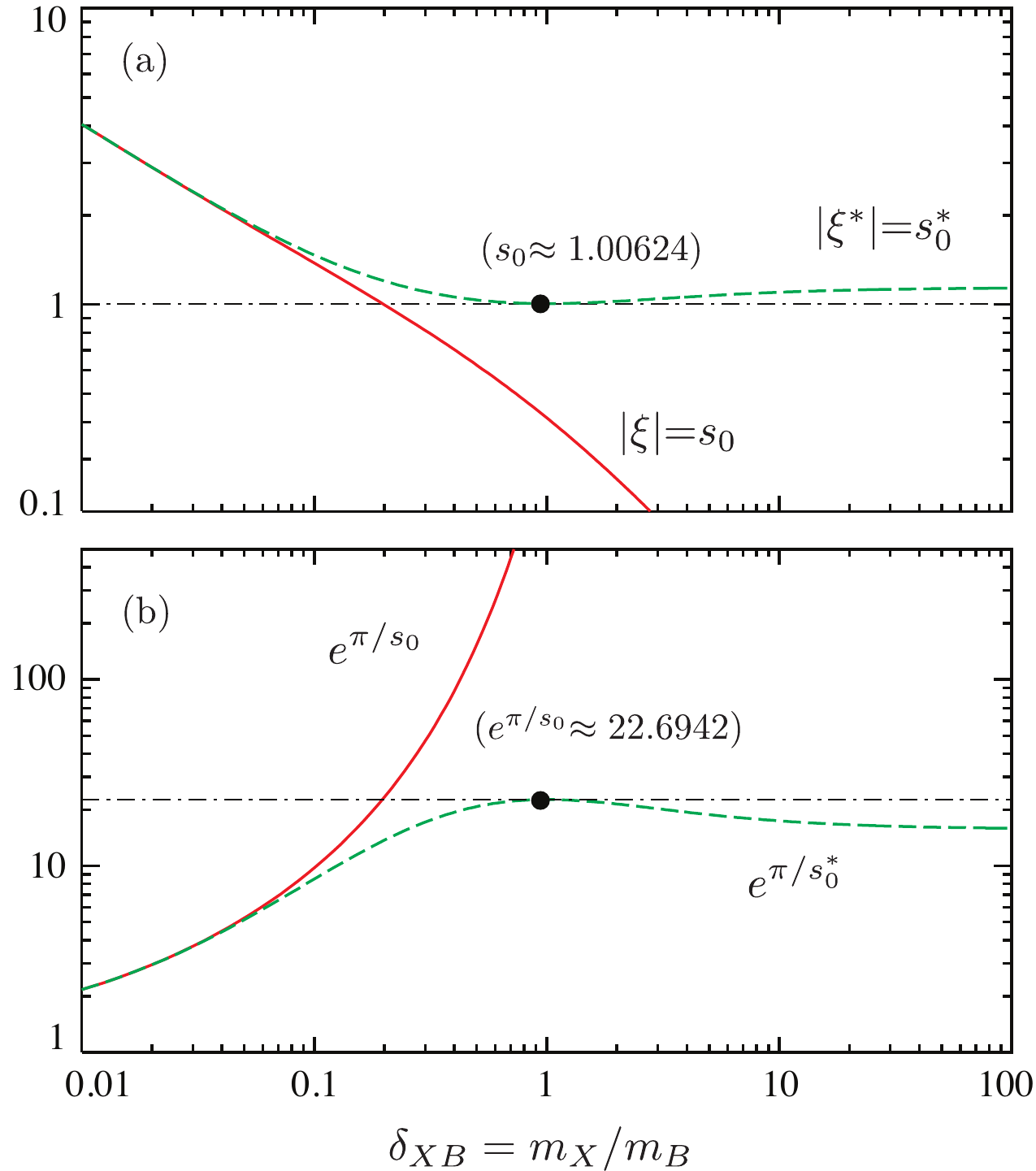}
\caption{(color online). 
Mass ratio dependence of the Efimov coefficients $s_{0}$ and $s_0^*$, (a), and geometric factors $e^{\pi/s_{0}}$ and $e^{\pi/s^*_{0}}$,
for $J^\pi=0^+$ $BBX$ systems with interspecies resonant interactions (solid red lines) and both interspecies and intraspecies resonant interactions
(green dashed lines), respectively.
For small mass ratios, $s_{0}$ and $s_0^*$ become identical, while for large mass ratios, $s_0$ vanishes and $s_0^*$ remains finite.}
\label{S0MassDep}
\end{center}
\end{figure}

The increase of the $s_0$ coefficient for $BBX$ and $FFX$ systems with small mass ratios 
implies that systems composed of two heavy and one light particles
possess a stronger attractive $1/R^2$ effective interaction and can be thought of as ``Efimov favorable" systems.
Systems with large values for $s_0$ have a denser energy spectrum and a smaller value for the 
geometric-scaling parameter $e^{\pi/s_0}$, thus resulting in a better scenario for experimental observations \cite{dincao2006PRA,dincao2006PRAb}. 
On the other hand, systems of two light and one heavy particles (large mass ratios)
have a weaker attraction and are thought of as ``Efimov unfavorable" systems.
For $J^\pi=0^+$ $BBX$ systems, where both  $a_{BB}$ and $a_{BX}$ scattering lengths are large, 
the scenario for the mass ratio dependence of $s_0=-i\xi$ is substantially changed \cite{dincao2009PRL}. 
In this case, setting $|a_{BB}|=|a_{BX}|=\infty$ in Eqs. (\ref{bc2ib1}) and (\ref{bc2ib2}),
we find that the values for the purely imaginary solution, $\xi^*=i s_0^*$, still diverge for small $\delta_{XB}$;
however, the values remain finite for large $\delta_{XB}$. Figure \ref{S0MassDep} shows the results for the Efimov constant $s_0^*$
and geometric scaling factor $e^{\pi/s_0^*}$ (green dashed lines). For comparison, Fig.~\ref{S0MassDep} also 
shows the corresponding results for $BBX$ systems with $a_{BB}=0$, and $|a_{BX}|=\infty$,  $s_0$ and
$e^{\pi/s_0}$ (solid red lines). 
We find that the smallest value for $s_0^*$ occurs at $\delta_{XB}=1$ and corresponds to the value obtained
for $BBB$ systems, $s_0\approx1.00624$ [see Fig.~\ref{S0MassDep}(a)]. As a result, the largest
value for $e^{\pi/s_0^*}$ also corresponds to $e^{\pi/s_0}\approx22.6942$ at $\delta_{XB}=1$  [see 
Fig.~\ref{S0MassDep}(b)]. 
Therefore, the effect of the resonant intraspecies interaction is to make the attractive $1/R^2$ effective interaction 
substantially stronger to the extent that even ``Efimov unfavorable" systems (two light and one heavy
particles) become more favorable. 
%For other values of $J^\pi\ne0$, our analysis does not show a substantial 
%difference between the values for $s_0^*$ and $s_0$.

\begin{figure}[htbp]
\begin{center}
\includegraphics[width=3.0in,angle=0,clip=true]{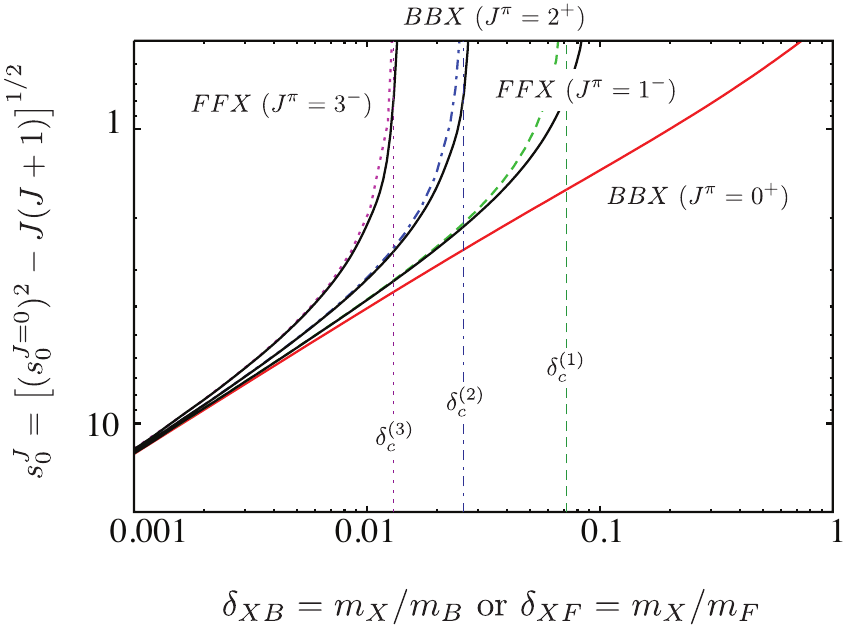}
\caption{Comparison between the exact values for $s_0$ from Eqs. (\ref{bc2ibBX}) and (\ref{bc2if}) and the ones obtained 
considering contributions from potential and kinetic energy (solid lines), as shown in
Eq.~(\ref{PotKinC}).}\label{EfimovCondBar}     
\end{center}
\end{figure}

From this analysis, it is clear that the occurrence of the Efimov effect covers a broad class of systems
and symmetries. Efimov, in his original work~\cite{efimov1973NPA}, interpreted the critical value of the mass ratio 
where the attractive $1/R^2$ effective potential for $J^{\pi}\ne 0$ emerges, $\delta_{c}$, as the 
point at which the attraction induced by the potential energy overcomes the kinetic energy centrifugal barrier. 
This interpretation implies that the effective interaction for $J^{\pi}\ne0$ can be written as, 
\begin{equation}
-\frac{(s_{0}^{J})^2+1/4}{2\mu R^2}\hbar^2\approx-\frac{(s_0 ^{J=0})^2+1/4}{2\mu R^2}\hbar^2+\frac{J(J+1)}{2\mu R^2}\hbar^2,
\label{PotKinC}
\end{equation}
where $s_0^J$ is the Efimov constant for the system in the $J^\pi$ state.
Equation (\ref{PotKinC}) decomposes the effective interaction in terms of the 
$J^{\pi}=0^+$ interaction and a kinetic energy term 
proportional to $J(J+1)$. This decomposition is, of course, only an approximation. 
Nevertheless, it allows us to determine $s_0^J$ as
\begin{equation}
s_0^{J}=\left[(s_0^{J=0})^2-J(J+1)\right]^{1/2},
\label{xiPotKinC}
\end{equation}
and compare it to the exact values obtained from the boundary conditions in Eqs. (\ref{bc2ibBX}) and (\ref{bc2if}). 
This comparison is shown in Fig.~\ref{EfimovCondBar}, where results from Eq.~(\ref{xiPotKinC}) are shown as 
solid lines, and the exact results for $J^\pi=1^-$, $2^+$ and $3^-$ are shown are shown as dashed, dot-dashed 
and dotted lines, respectively. This qualitative agreement clearly indicates that the kinetic energy is the key
factor that prohibits the Efimov effect for $J^\pi\ne0$ states with a mass ratio beyond $\delta_{c}$.

%%%%%%%%%%%%%%%%%%%%%%%%%%%%%%%%%%%%%%%%%%%%%%%%%%%%
\subsection{Universal classification scheme} \label{SecClass}

In Section \ref{EfimovConditions}, we showed that, as a result of the Efimov physics, 
%JPD
%the effective potentials 
%can only be attractive or repulsive $1/R^2$ effective potentials in the range $r_{0}\ll R\ll|a|$, with strengths 
%depending on the number of resonant pairs, permutation symmetry, angular momentum state, and the mass ratios.
%\textcolor{red}
{the three-body effective potentials are proportional to $1/R^2$ in the range $r_{0}\ll R\ll|a|$, 
with their strength, as well as their attractive or repulsive character, 
depending on the number of resonant pairs, permutation symmetry, angular momentum state, and mass ratios.}
This fact motivated us to propose in Ref.~\cite{dincao2005PRL} a classification scheme of the Efimov physics
where all three-body systems with resonant $s$-wave interactions can be sorted out in one of two categories:
{\em three-body attractive systems} (i.e., systems in which the Efimov effect occurs) and 
{\em three-body repulsive systems} (i.e., systems in which the Efimov effect is absent). 
Within this universal classification scheme, discussed further in Section.~\ref{Collisions}, 
three-body systems falling into the same category have qualitatively 
similar ultracold scattering properties~\cite{dincao2005PRL,dincao2006PRAb}. Here we detail our 
classification scheme by characterizing the effective three-body potentials and the important 
properties of nonadiabatic couplings between the relevant three-body channels.

\subsubsection{Effective potentials and couplings} \label{EffPotW}

%JPD
%Using 
%\textcolor{red}
{Taking account of}
the analysis and discussions presented in the previous sections, the
three-body effective potentials can be divided into three distinct regions in $R$, 
where different physics controls the specific form of the interactions: the {\em short-range region}, $R\lesssim r_0$,
the {\em Efimov region}, $r_0\ll R \ll |a|$, and the {\em asymptotic region}, $R\gg |a|$.
For distances smaller than the characteristic range of the interatomic interactions, $R\lesssim r_{0}$, 
all particles are close to each other, and the effective potentials depend on the details of the 
interatomic potentials. Note that the zero-range model explored in the previous sections does not account
for the proper behavior of the interatomic interactions at short distances.
For this reason, and the fact that  the potentials for $R\lesssim r_{0}$ are nonuniversal, i.e.,
they can vary from system to system, we will not specify the potentials in this region. 
However, the importance of nonuniversal physics will be made clear 
whenever appropriate.

In contrast, the effective potentials within the Efimov region, $r_{0}\ll R\ll |a|$, are universally determined. 
The fact that the effective potentials in this region can only be attractive or repulsive 
$1/R^2$ effective potentials is what establishes our classification scheme. Instead of parameterizing the 
potentials in terms of $\xi$, as shown in Sections \ref{ZRP} and \ref{TransEq}, it is more convenient to parameterize the 
potentials in a way that reflects the two different categories. For {\em attractive systems},
the potentials within the Efimov region are parameterized by the coefficients $s_{\nu}$:   
\begin{equation}
W_{\nu}(R)=-\frac{s^{2}_{0}+{1}/{4}}{2\mu R^2}\hbar^2,
\hspace{0.25cm}\mbox{and}\hspace{0.25cm}
W_{\nu}(R)=\frac{s^{2}_{\nu}-{1}/{4}}{2\mu R^2}\hbar^2,
\label{efbosonic} 
\end{equation}
where $s_0$ is the usual Efimov coefficient, and $s_\nu$ ($\nu>0$) are the coefficients 
associated with all other repulsive potentials. 
Similarly, for {\em repulsive systems} the potentials are now parameterized by the coefficients $p_{\nu}$: 
\begin{equation}
W_{\nu}(R)=\frac{p^{2}_{0}-{1}/{4}}{2\mu R^2}\hbar^2,
\hspace{0.25cm}\mbox{and}\hspace{0.25cm}
W_{\nu}(R)=\frac{p^{2}_{\nu}-{1}/{4}}{2\mu R^2}\hbar^2,
\label{effermionic} 
\end{equation}
where $p_0<p_\nu$, characterizing all repulsive potentials within this category.
We note that these potentials only describe the physics for channels
in which $s$-wave interactions are resonant. Other three-body channels
associated with higher partial-wave states are, to a very good approximation, insensitive to variation of 
the scattering length.

In the asymptotic region, $R\gg |a|$, the form of the effective potentials is also universal. 
The effective potentials in this region, however, can describe two types of asymptotic behavior.
They can be associated with molecular channels that describe atom-molecule scattering 
and with three-body continuum channels that describe collisions of three free atoms. 
The asymptotic form of the effective potentials for molecular and continuum channels can be determined 
analytically. Their leading-order terms are given, respectively, by~\cite{dincao2005PRA,nielsen2001PRep,fedorov2001JPA} 
\begin{equation}  
W_{\nu}(R)
%\underset{R\gg |a|}{\longrightarrow}
\simeq{E}_{vl'}+\frac{l(l+1)}{2\mu R^2}\hbar^2,\label{bc}
\end{equation}  
and
\begin{equation}  
W_{\nu}(R)
%\underset{R\gg |a|}{\longrightarrow}
\simeq\frac{\lambda(\lambda+4)+15/4}{2\mu R^2}\hbar^2.
\label{ch}
\end{equation}  
In Eqs.~(\ref{bc}) and (\ref{ch}), the diatomic molecular state energy ${E}_{vl'}$ is labeled by its ro-vibrational 
quantum numbers $v$ and $l'$; $l$ is the relative angular momentum between the molecule and the atom; 
and $\lambda=2n_{\lambda}+l+l'$, and $n_{\lambda}$ are positive integers labeling the eigenstates of the 
hyperangular kinetic energy operator~(\ref{LambdaK}), 
%JPD
%. They are 
%\textcolor{red}
{determined by the total angular 
momentum $J^\pi$ and permutation symmetry.}
Note, again, that here we are only interested in channels representing strong $s$-wave interactions.
In this case, since $l$ and $l'$ satisfy the triangular inequality, $|l-l'|\le J\le l+l'$, 
for an $s$-wave state, we have $l'=0$ and $l=J$, restricting the number of relevant molecular
and continuum channels. In fact, although, in general, many molecular channels (\ref{bc}) can exist, 
the only one relevant for strong $s$-wave interactions is the one associated with the molecular state 
for $a>0$, with energy $E_{v0}^*\propto-1/a^2$. Channels associated with non $s$-wave
interactions are largely insensitive to $a$ and can be simply described by the potentials in Eqs.~(\ref{bc})
and (\ref{ch}) for $R\gg r_0$. These channels correspond, respectively, to deeply bound 
molecular channels and continuum states associated with states where $l'\ne0$ \cite{nielsen2001PRep}.

The connection between the potentials in the Efimov and asymptotic regions
is made in a one-to-one correspondence according to their energy
and occur smoothly across $R=|a|$ \cite{dincao2005PRA}.
For instance, the lowest potential in the Efimov region smoothly connects 
to the lowest potential in the asymptotic region. 
The physics implied from these connections, however, depends on 
the sign of  $a$, because of the presence of the $s$-wave weakly bound diatomic molecule for
$a>0$. 
For instance, for three-body attractive systems, when $a>0$, the attractive $1/R^2$ effective potential in 
Eq.~(\ref{efbosonic}) occurs in the weakly bound molecular channel (\ref{bc}),
while all repulsive $1/R^2$ effective potentials in Eq.~(\ref{efbosonic}) connect to the continuum 
channels (\ref{ch}).
For $a<0$, however, the attractive $1/R^2$ effective potential occurs in the lowest 
three-body continuum channel in Eq.~(\ref{ch}) ---associated with the lowest-allowed value of $\lambda$. 
All other repulsive potentials in Eq.~(\ref{efbosonic}) connect to the other higher-lying continuum 
channels in Eq.~(\ref{ch}). A similar analysis applies for the repulsive systems.

\begin{figure*}[htbp]
\begin{center}
\includegraphics[width=6.8in,angle=0,clip=true]{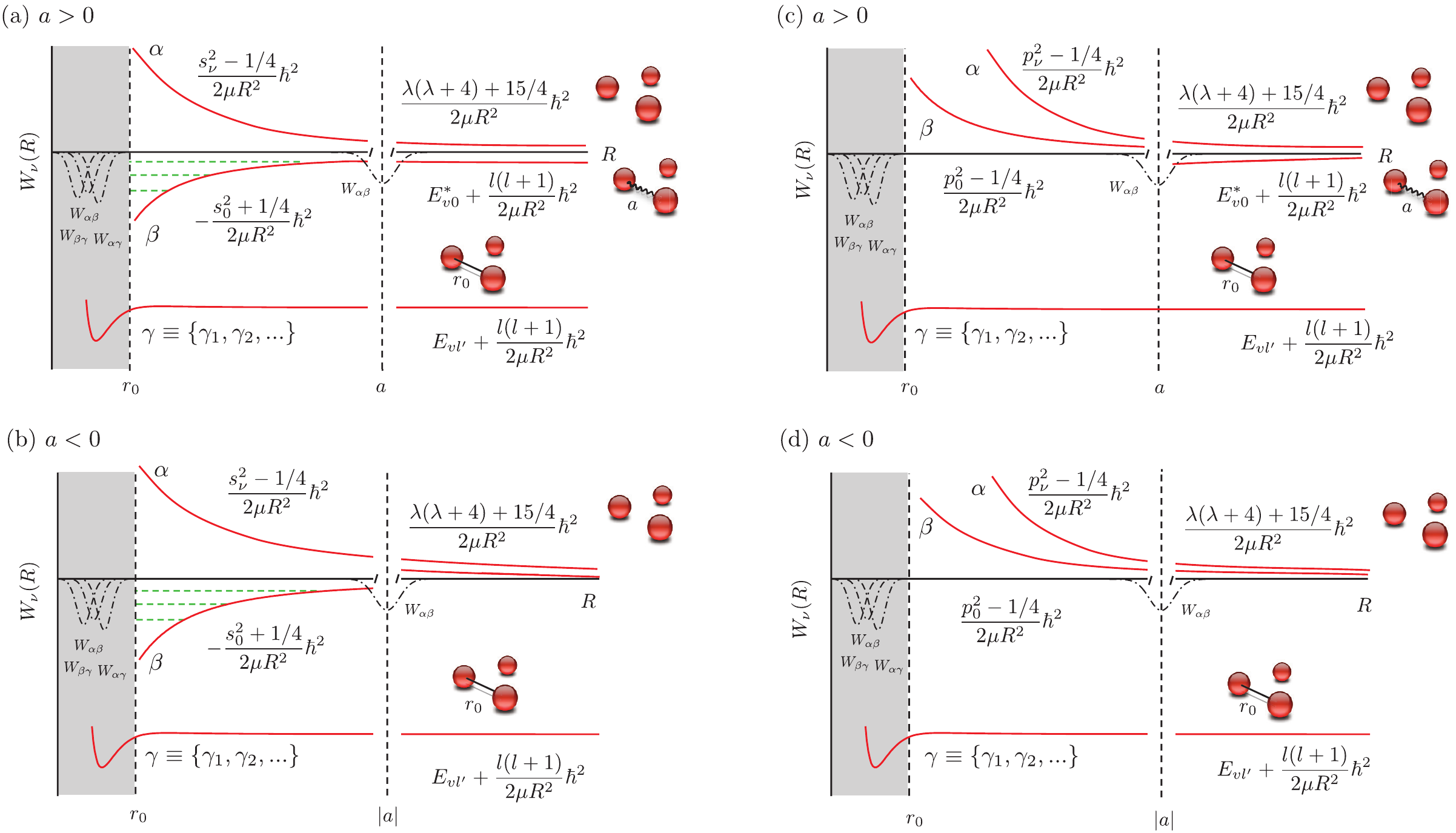}
\caption{Schematic representation of the effective potentials and couplings for
 attractive, (a) and (b), and repulsive, (c) and (d), three-body systems. 
For $a > 0$, (a) and (c), the channel ``$\alpha$'' denotes the lowest continuum channel and represents
collisions of three free atoms. ``$\beta$'' denotes the weakly bound molecular channel, representing atom-molecule collisions, 
and ``$\gamma\equiv\{\gamma_{1}$, $\gamma_{2},...\}$'' collectively denotes deeply bound (nonresonant) molecular channels. 
For $a < 0$ ((b) and (d)), however, since there are no weakly bound molecules, both ``$\alpha$'' and ``$\beta$'' channels describe
three free atoms. }
  \label{SchPot}   
\end{center}
\end{figure*}

Figure~\ref{SchPot} shows a schematic representation of the 
relevant low-lying potentials for both attractive [Figs.~\ref{SchPot}(a) and (b)] and repulsive 
[Figs.~\ref{SchPot}(c) and (d)] systems with $a>0$ and $a<0$. 
Note that the form of the potentials near $R=|a|$ is omitted in this schematic representation 
since 
%JPD
%they form 
%\textcolor{red}
{their actual form}
are different from Eqs.~(\ref{bc})-(\ref{effermionic}). The potentials in this region, however, are smooth functions
of $R$ and are also expected to be universal. We also omitted the form of the potentials in the $R\lesssim r_0$ (see shaded region in 
Fig.~\ref{SchPot}), because the potentials within this region are expected to be nonuniversal.
The connection between the effective potentials in the Efimov and asymptotic regions
can be derived by following the proper labels in the figure.
For $a>0$ [Figs.~\ref{SchPot}(a) and (c)], the channel $\alpha$ connects asymptotically
to the lowest continuum channel, while the channel $\beta$ connects asymptotically to the
$s$-wave weakly bound molecular channel.
For $a<0$ [Figs.~\ref{SchPot}(b) and (d)], since the weakly bound molecular channel is absent, 
the $\alpha$ and $\beta$ channels connect asymptotically to the second-lowest and lowest 
continuum channels, respectively. 
The channel $\gamma$ in Fig.~\ref{SchPot} represents collectively all possible (nonresonant) deeply
bound molecular channels, $\gamma\equiv\{\gamma_{1},\gamma_{2},...\}$, that are insensitive to variations 
of $a$. 
Figure~\ref{SchPot} also represents schematically the nonadiabatic couplings for the relevant channels, i.e.,
$W_{\alpha\beta}$, $W_{\alpha\gamma}$, and $W_{\beta\gamma}$, in the regions where inelastic transitions 
are most likely. At short distances, $R\lesssim r_0$, since all particles are close together, nonadiabatic
couplings are expected to be strong between all channels. 
Near $R=|a|$, however, since only the $\alpha$ and $\beta$ channels vary strongly with $a$, 
only the $W_{\alpha\beta}$ coupling is expected to be important. 
We have verified all these assumptions numerically \cite{dincao2005PRA}. 

Figure~\ref{SchPot} represent all relevant channels and couplings necessary to describe 
three-body scattering processes at ultracold energies, as will be further explored in Section \ref{Collisions}. 
The particular form of the potentials and couplings near $R=r_{0}$ and $R=|a|$ 
will be shown to be encapsulated in specific constants within our model.
Simply knowing the leading-order terms of the potentials and the general form of the 
nonadiabatic couplings is enough to determine the pathways for elastic and inelastic 
collisions as well as the resulting energy and scattering length dependence 
of the ultracold three-body scattering observables.

\subsubsection{Summary: Classification scheme}

Accordingly to the discussion in the previous section, the classification of strongly interacting ($s$-wave)
three-body systems relies on the understanding of Efimov physics, which determines whether a system 
falls into the category for attractive or repulsive systems.
There exist, however, only four types of systems that allow for $s$-wave interactions, namely: 
$BBB$, $BBX$, $FFX$ and $XYZ$ (i.e., three dissimilar particles). 
In the case of homonuclear systems, $BBX$, $FFX$ and $XYZ$
systems will be represented by $BBB'$, $FFF'$ and $BB'B''$ (or $FF'F''$), 
respectively, where the prime indicates a different internal spin state.
For homonuclear systems, Table~\ref{TabI} lists the two lowest Efimov coefficients, $s_{\nu}$ and $p_{\nu}$
controlling the strength of the $1/R^2$ effective potentials for attractive and repulsive systems [Eqs.~(\ref{efbosonic}) 
and (\ref{effermionic})], respectively. These results were obtained by solving 
Eqs.~(\ref{bc3ib})--(\ref{bc2if}). Table~\ref{TabI} also lists the quantum numbers $l$ and $\lambda$ 
characterizing the system's asymptotic behavior [see Eqs.~(\ref{bc}) and (\ref{ch})].
The only homonuclear systems that fall into the category of attractive systems are 
$BBB$ and $BBB'$ ($a_{BB'}$) in the $J^\pi=0^+$ state. 
The table also displays the solutions for $XYZ$ ($a_{XY}$) systems, since those have the same 
values for $p_\nu$ as $BBB'$($a_{BB}$) systems, regardless of the mass ratios. 

\begin{table}[htbp]
\caption{Coefficients and asymptotic quantum numbers characterizing the effective potentials for 
homonuclear three-body systems. In the range $r_{0}\ll R\ll |a|$, the potentials are characterized by the 
Efimov coefficients, $s_{\nu}$ and $p_{\nu}$, [see Eqs.~(\ref{efbosonic}) and (\ref{effermionic})]. For $R\gg|a|$ the 
$s$-wave weakly bound molecular channel is characterized by the relative atom-molecule angular momentum 
$l(=J)$, and three-body continuum channels are determined by the values of $\lambda$ [see Eqs.~(\ref{bc}) and (\ref{ch})]. Note that
the quantum numbers $l$ and $\lambda$ do not depend on the atomic mass ratios.}   
\label{TabI}
\begin{ruledtabular}
\begin{tabular}{ccllcl} 
 & & 
\multicolumn{2}{c}{\hspace{0.0in}$r_{0}\ll R \ll |a|$} & 
\multicolumn{2}{l}{\hspace{0.0in}$R \gg |a|$}  \\ 
 System &$J^{\pi}$ & $s_{0}$ & $s_{1}$ & $l$ & $\lambda$  \\ \hline 
$BBB$($a_{BB}$) 
      & 0$^+$ & 1.0062378 & 4.4652946  & 0 & 0, 4, ... \\
      & 1$^-$  & 2.8637994($p_0$)& 6.4622044($p_{1}$) & 1 & 3, 5, ...  \\
      & 2$^+$ & 2.8233419($p_0$)& 5.5082494($p_{1}$) & 2 & 2, 4, ... \\ [0.05in]
$BBB'$($a_{BB'}$) 
       & 0$^+$ & 0.4136973 & 3.4509891 & 0 & 0, 2, ... \\
      & 1$^-$  & 2.2787413($p_0$)& 3.6413035($p_{1}$) & 1 & 1, 3, ... \\ 
       & 2$^+$ & 2.9073507($p_0$)& 5.2232519($p_{1}$) & 2 & 2, 4, ...  \\ [0.025in] \hline\hline 
 System & $J^{\pi}$ & $p_0$ & $p_1$ & $l$ & $\lambda$\\ \hline 
$FFF'$($a_{FF'}$) 
        & 0$^+$  & 2.1662220 & 5.1273521 & 0 & 2, 4, ... \\
        & 1$^-$  & 1.7727243 & 4.3582493 & 1 & 1, 3, ... \\
        & 2$^+$  & 3.1049769 & 4.7954054 & 2 & 2, 4, ... \\ [0.025in]
$BBB'$($a_{BB}$)
        & 0$^+$ & 1.0000000 & 3.0000000 & 0 & 0, 2, ...   \\
$[XYZ(a_{XY})]$        & 1$^-$ & 2.0000000 & 4.0000000 & 1 & 1, 3, ... \\ 
        & 2$^+$ & 3.0000000 & 5.0000000 & 2 & 2, 4, ...  \\ 
\end{tabular}
\end{ruledtabular}
\end{table}

\begin{figure*}[htbp]
\begin{center}
\includegraphics[width=6.8in,angle=0,clip=true]{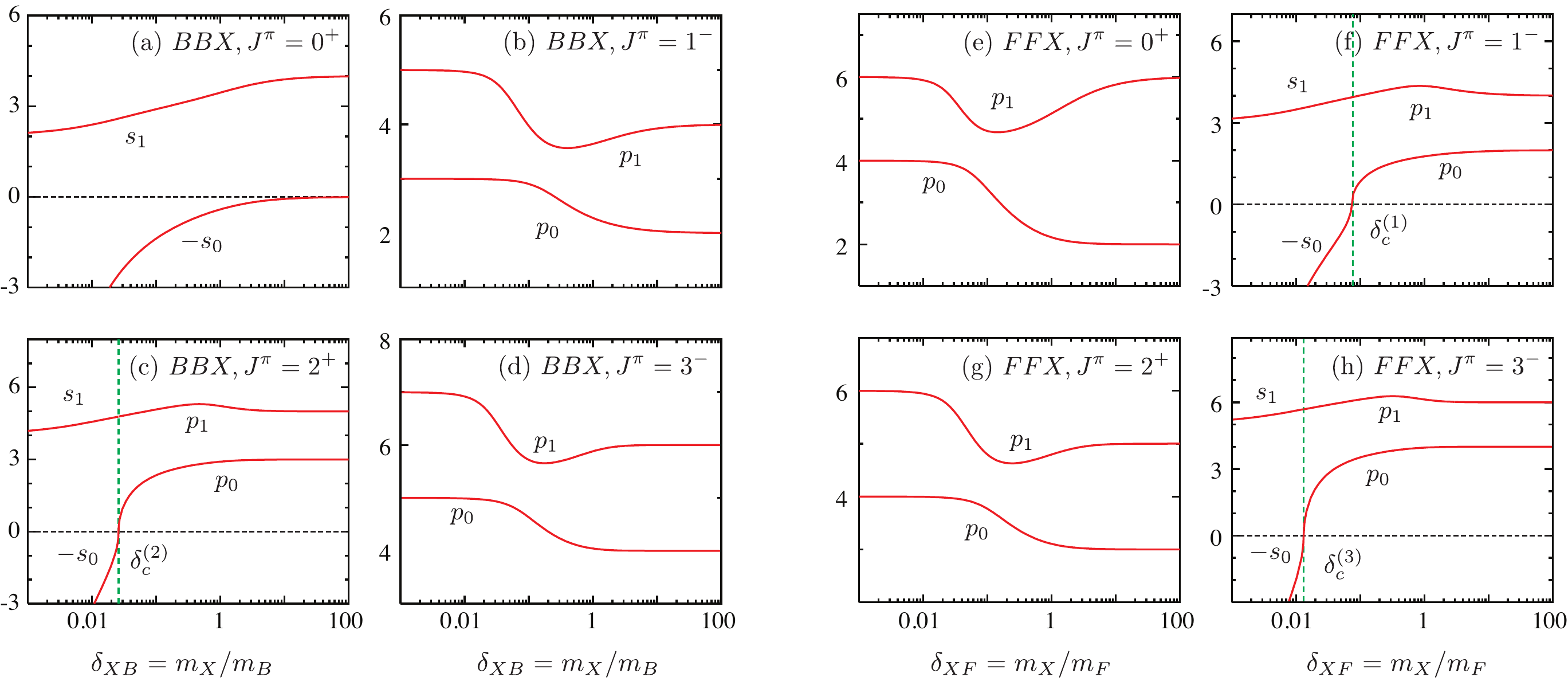}
\end{center}
\caption{Coefficients $s_{\nu}$ and $p_{\nu}$ for the effective potentials Eqs.~(\ref{efbosonic}) and  (\ref{effermionic}) for $BBX$($a_{BX}$), 
(a)--(d), and $FFX$($a_{FX}$), (e)--(h), systems as a function of the mass ratio $\delta$. 
The figure shows that $BBX$ systems in $J^\pi>0$(even) states display an attractive interaction for $\delta<\delta_c$, while in
$FFX$ systems, the attractive interaction appears for $\delta<\delta_c$ and $J^\pi$(odd).
 Here, $\delta_{c}^{(1)}\approx0.073492$, $\delta_{c}^{(2)}\approx0.025887$ and $\delta_{c}^{(3)}\approx0.013159$}.  
\label{CoeffsBBXFFX}
\end{figure*}

For heteronuclear systems, the classification of three-body systems into attractive and repulsive systems,
as well as the strength of their interactions, depends on the value of the mass ratios. In fact, three-body effective 
potentials can change their nature from repulsive to attractive for mass ratios smaller than the critical value, 
$\delta_{c}$. 
This change in nature is shown in Fig.~\ref{CoeffsBBXFFX} for the mass ratio dependence of the coefficients 
$s_\nu$ and $p_\nu$ for $BBX$ and $FFX$ systems, which can be obtained by solving Eqs.~(\ref{bc2ibBX}) and (\ref{bc2if}), 
respectively. In this figure, while $J^\pi=0^+$ $BBX$ systems always fall into the category of attractive systems, $J^\pi$-even 
states are only attractive systems for $\delta<\delta_c$. A similar result applies to $FFX$ systems, 
where the system is attractive only for $J^\pi$-odd states with $\delta<\delta_c$ (see Section~\ref{EfimovConditions}).
For all other cases, the systems fall into the category of repulsive interactions.
Therefore, to establish the classification of heteronuclear systems, one needs to account for their proper mass
ratio dependence. This accounting is done in Table~\ref{Class}, which summarizes the classification of three-body systems 
with $s$-wave resonant interactions. 
The only restriction made while classifying the systems listed in Table~\ref{Class} is that only one 
type of interaction is resonant, i.e., interspecies and intraspecies interactions are not simultaneously 
resonant. When the system has strong interspecies and intraspecies interactions, e.g., when different 
Feshbach resonances overlap~\cite{dincao2009PRL}, our classification scheme does not 
necessarily apply, since now the system can have mixed characteristics of both categories 
(see Ref.~\cite{dincao2009PRL}). 
 
The present classification scheme allows general results to be obtained for the relevant scattering observables 
in three-body systems, as it will more become evident in Section \ref{Collisions}. 
In fact, we find that systems that fall into the same category share similar scattering properties
because of their shared potential and coupling structure. For instance in attractive systems,
scattering observables display a nontrivial dependence on the scattering length that is intimately 
related to the Efimov effect. For repulsive systems, scattering observables have, in general, 
a simpler scattering length dependence characterized by the suppression of inelastic processes. 
 In both cases, however, the behavior of three-body scattering observables is a direct consequence of Efimov physics.
 
\begin{table}[htbp]
\caption{Summary of the classification of three-body systems with $s$-wave
  resonant interactions in terms of Efimov physics. Attractive and repulsive systems
  are indicated by categories Attract. and Repuls., respectively. We also have indicated the relevant
  scattering length $a$ specifying which pair interaction is resonant.} 
\label{Class}
\begin{ruledtabular}
\begin{tabular}{ccc}
 System   & Category & Total Angular Momentum, $J^{\pi}$ \\ \hline 
$BBB$($a_{BB}$) & Attract.:&  $0^+$\\ 
                 & Repuls.:& $J^\pi>0$ \\ [0.05in]
 $BBX$($a_{BX}$) & Attract.:  &$0^+,~J^\pi\mbox{(even)}>0~[\delta<\delta_{c}^{(J)}]$ \\
                 & Repuls.: &$J^\pi\mbox{(even)}>0~[\delta>\delta_{c}^{(J)}],J^\pi\mbox{(odd)}$ \\ [0.05in]
$FFX$($a_{FX}$) & Attract.:  &$J^\pi\mbox{(odd)}~[\delta<\delta_{c}^{(J)}]$ \\
                 & Repuls.: &$J^\pi\mbox{(odd)}~[\delta>\delta_{c}^{(J)}],~J^\pi\mbox{(even)}$ \\ [0.05in]
$BBX$($a_{BB}$) & Repuls.:&  All $J^\pi$ \\ 
$XYZ(a_{XY})$ & Repuls.:&   All $J^\pi$\\                  
\end{tabular}
\end{ruledtabular} 
\end{table}

%%%%%%%%%%%%%%%%%%%%%%%%%%%%%%%%%%%%%%%%%%%%%%%%%%%%
\subsection{Efimov and other universal states} \label{Spectrum}

Before we move towards analyzing the relevant scattering processes in ultracold gases,
which are described in detail in the next section, it is instructive to build up an intuition of some of their
properties derived from the energy spectrum for strongly interacting three-body systems.
Understanding the energy spectrum gives an intuitive idea of when important resonant effects
appear in ultracold scattering observables.

\begin{figure*}[htbp]
\begin{center}
\includegraphics[width=6.2in,angle=0,clip=true]{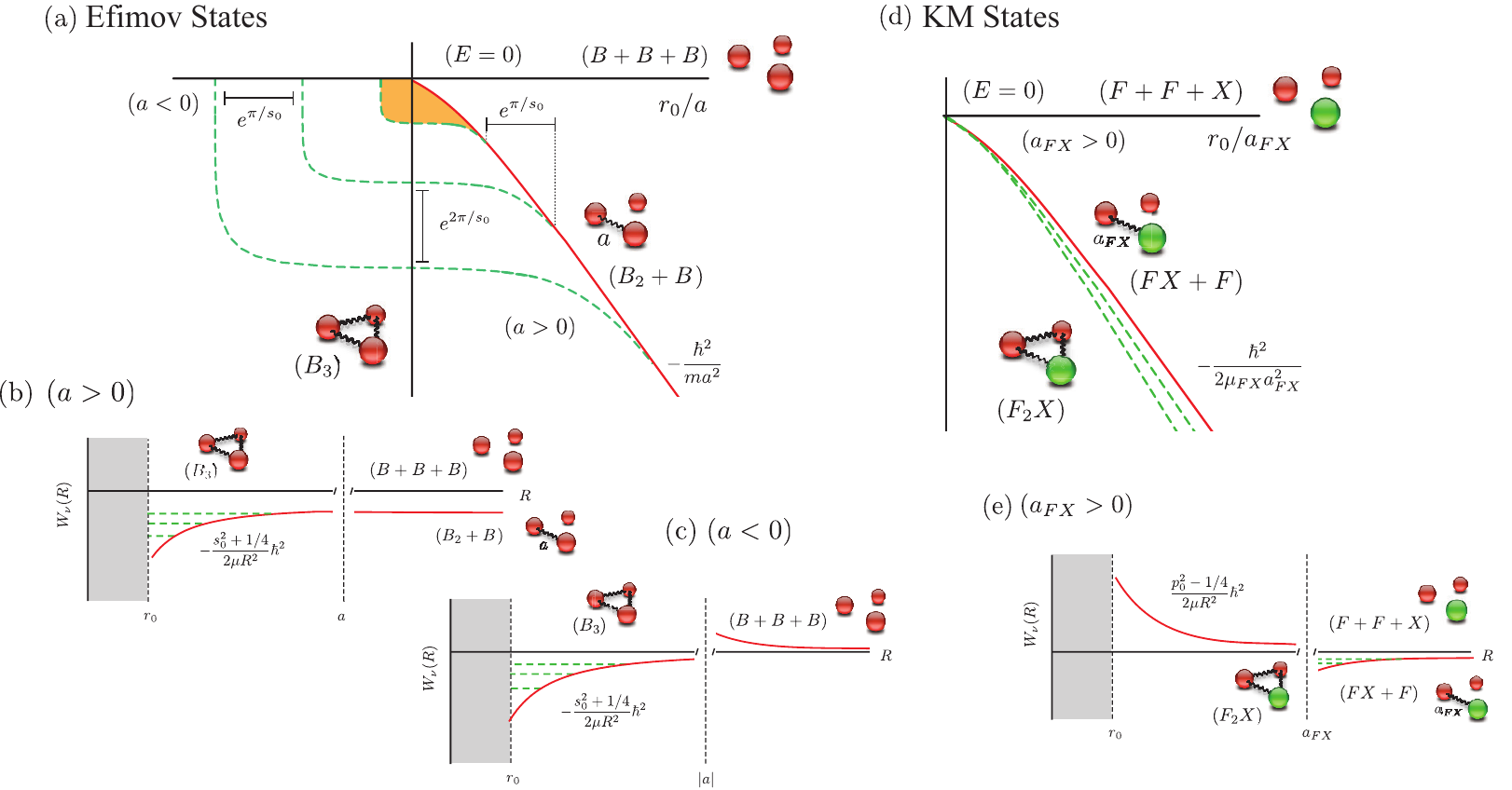}
\caption{(a) Schematic representation of the energy spectrum of Efimov states (green dashed lines). 
For $a<0$ Efimov states are bound below the three-body continuum ($E=0$), while for $a>0$
they are bound below the atom-molecule threshold (solid red line). 
(b) and (c) show a schematic representation for effective potentials for $a<0$ and $a>0$, respectively.
(d) Schematic representation of the KM states' energy spectrum (green dashed lines) for a $FFX$ system. 
KM states exist only for $a_{FX}>0$ and are bound below the atom-molecule threshold (solid red line). 
(e) shows a schematic representation of effective potentials in this class of three-body systems.}     
\label{EfimovStates}
\end{center}
\end{figure*}

For systems in which the Efimov effect occurs (see Table~\ref{Class}), it is important to realize that the attractive $1/R^2$ effective interaction 
only holds for $r_{0}\ll R\ll |a|$. In that sense, an infinity of Efimov states is only obtained 
when $|a|=\infty$. For large, but finite, values of $a$ only a {\em finite} number of states 
can exist.
In Fig.~\ref{EfimovStates}(a), we show schematically the energies of the lowest three Efimov states 
(green dashed lines) ---the shaded area in the figure indicates the region of the
spectrum in which other high-lying states would appear.  Note that although we indicate that
such a spectrum corresponds to a $BBB$ system, the same general form of the spectrum is also valid for {\em any} 
attractive system.
As expected, Fig.~\ref{EfimovStates}(a) shows that as $|a|$ increases,
the number of Efimov states increases. In fact, an estimate of the number of Efimov states can be obtained from \cite{efimov1973NPA}
\begin{equation}
N\approx \frac{s_{0}}{\pi}\ln(|a|/r_{0}).
\label{NumStates}
\end{equation}
This result implies a fundamental property of the Efimov effect:
whenever $a$ is increased by the geometric factor $e^{\pi/s_{0}}$, a new Efimov state is formed. 
This property is also shown in Fig.~\ref{EfimovStates}(a). 
For $a>0$, Efimov states emerge from the atom-molecule threshold ($B_{2}+B$) in Fig.~\ref{EfimovStates}(a), corresponding
to the weakly bound molecular channel, as illustrated in Fig.~\ref{EfimovStates}(b). 
For $a<0$, however, Efimov states emerge from the three-body breakup threshold ($B+B+B$), which now corresponds to 
the lowest three-body continuum, as schematically represented in Fig.~\ref{EfimovStates}(c). As we can see in both cases, 
as $|a|$ increases, so does the range in which the attractive $1/R^2$ effective potential holds ($r_0\ll R\ll |a|$), thus causing the increase on the number
of Efimov states the system can support.
In terms of the scattering properties, it is evident that the formation of Efimov states 
as $a$ increases should lead to characteristic Efimov features, such as interference 
minima and/or resonance peaks, in the low-energy three-body scattering observables. 
Moreover, Efimov features in scattering observables should be expected to 
log-periodic, since a new state is formed any time $a\rightarrow a\times e^{\pi/s_0}$.
From Figs.~\ref{EfimovStates}(b) and (c), it is clear that resonant features associated
with the formation of an Efimov state should lead to an enhancement of atom-molecule collisions ($a>0$)
and collisions involving three-free atoms ($a<0$). In Section~\ref{Collisions} we will see that this is, in fact, 
the case. We will also see that other less obvious interference effects can occur in scattering 
observables that are also related to Efimov states.
%\begin{figure}[htbp]
%\begin{center}
%\includegraphics[width=3.0in,angle=0,clip=true]{FigSpectrumEfimov.pdf}
%\caption{(a) Schematic representation of the energy spectrum of Efimov states (green dashed lines). 
%For $a<0$ Efimov states are bound below the three-body continuum ($E=0$), while for $a>0$
%they are bound below the atom-molecule threshold (solid red line). 
%Plots (b) and (c) show a schematic representation for effective potentials for $a<0$ and $a>0$, respectively.}     
%\label{EfimovStates}
%\end{center}
%\end{figure}

For three-body systems in which Efimov physics leads to a repulsive interaction within the 
range $r_0\ll R\ll |a|$ (Table~\ref{Class}), the expectation is that no three-body bound states 
would be allowed. However, for some cases, other equally universal three-body states do exist. 
A new class of states was discovered by Kartavtsev and Malykh \cite{kartavtsev2007JPB} for $FFX$ 
systems in the $J^\pi=1^-$ state and with mass ratios $\delta_{XF}>\delta_c\approx0.073492$. 
For a given value of $a_{BX}>0$, they found that up to two three-body states can exist for 
$\delta_c\le\delta_{XF}\le0.12236$. 
Recent studies \cite{endo2011FBS,endo2012PRA,kartavtsev2014YF} have 
shown that such Kartartsev-Malykh (KM) states can also occur for $BBX$ ($J^\pi\ne0$-even) systems
and other $FFX$ ($J^\pi$-odd) systems, showing that KM states represent a novel class of three-body 
states with interesting universal properties. 
We illustrate the typical energy spectrum of KM states (green dashed lines) in a $FFX$ system
in Fig.~\ref{EfimovStates}(d). One important feature of such KM states is that their energies are 
proportional to the diatomic molecular state, $-1/2\mu_{FX}a_{FX}^2$ (where $\mu_{FX}$ is the two-body reduced mass). 
In fact, in the adiabatic hyperspherical representation, KM states occur in the weakly bound molecular channel
and lie in a potential well near $R=a$, as represented schematically in Fig.~\ref{EfimovStates}(e).
As a result, atoms in KM states can not all be enclosed within distances comparable 
with $r_0$, in strong contrast to Efimov states, whose atoms can access distances well within the region 
$r_0\ll R\ll |a|$. 
In terms of the scattering properties for systems displaying KM states,
since such states exist for all values of $a>0$, their signatures can only be determined via 
their mass dependence. In fact, signatures exist for both collisions between three free 
atoms \cite{petrov2003PRA,kartavtsev2007JPB} and atom-molecule collisions 
\cite{endo2011FBS,endo2012PRA}. Depending on how close the KM state is to the atom-molecule 
threshold, enhancements and interference effects can occur. As we see next, 
such effects can be incorporated in terms of a universal, mass-ratio-dependent, phase.

%\begin{figure}[htbp]
%\begin{center}
%\includegraphics[width=2.8in,angle=0,clip=true]{FigSpectrumOthers.pdf}
%\caption{(a) Schematic representation of the KM states energy spectrum (green dashed lines) for a $FFX$ system. 
%KM states exist only for $a_{FX}>0$ and are bound below the atom-molecule threshold (solid red line). 
%(b) shows a schematic representation for effective potentials in this class of three-body systems.}
%\label{OtherStates}
%\end{center}
%\end{figure}

.pdf% !TEX root = ./TutorialJPB.tex

%%%%%%%%%%%%%%%%%%%%%%%%%%%%%%%%%%%%%%%%%%%%%%%%%%%%
%\section{Efimov physics and ultracold collisions} 
%\section{Pathways for ultracold three-body collisions} 
\section{Pathways for ultracold collisions} 
\label{Collisions}

In this section we provide a detailed analysis of a framework for obtaining 
ultracold three-body rates in a simple and unifying picture. 
Our method \cite{dincao2005PRL,dincao2006PRAb,dincao2008PRL,dincao2009PRL,colussi2014PRL,colussi2016JPB}
relies primarily on recognizing the dominant pathways in which a particular collision process can occur.
This formulation not only makes evident the main physical mechanisms controlling the 
collision processes, but also demonstrates the pervasive influence of Efimov physics in determining
the scattering length dependence of the collision rates.
In our approach, the limiting steps controlling the ultracold collision rates are tunneling, transmission, 
and reflection through three-body effective potentials.
A simple WKB  analysis~\cite{berry1966Proc.Phys.Soc.} is sufficient to 
determine both the energy and scattering length dependence of the rates and to elucidate the origin of universal 
behavior.

%%%%%%%%%%%%%%%%%%%%%%%%%%%%%%%%%%%%%%%%%%%%%%%%%%%%
\subsection{The general idea}

We analyze both elastic and inelastic scattering observables using the classification
scheme described in Section \ref{SecClass} and derive general results for all
relevant three-body systems with resonant $s$-wave interactions.
As we will see, Efimov physics is associated with striking differences between collision processes 
for attractive and repulsive systems. 
In attractive systems (see Table~\ref{Class}), the formation of Efimov states 
as the scattering length increases leads to characteristic interference minima and resonance 
peaks in the three-body scattering observables. Such Efimov features appear 
whenever the scattering length is changed by the geometric factor $e^{\pi/s_{0}}$. 
For repulsive systems (see Table~\ref{Class}), Efimov physics also impacts the 
scattering length dependence of the three-body rates and in various cases, is responsible
for the suppression of the inelastic collision rates and other universal properties.

Within our model, scattering processes at ultracold energies proceed mainly via 
tunneling in the initial collision channel, following by transitions, transmission and/or reflection 
depending on the structure of the effective potentials, $W_\nu$, and nonadiabatic couplings, 
$W_{\nu\nu'}$ [see Eqs.~(\ref{EffPot}) and (\ref{Wcoup})]. This simple physical picture can 
easily be applied to the schematic representation of effective potentials and couplings, as described 
in Fig.~\ref{SchPot}. This figure shows that, since couplings can peak in more than one place, and effective potentials 
can change in different regions, the probability for a particular scattering process depends on contributions 
from multiple pathways.
At ultracold energies, however, only a few pathways will dominate
the scattering event, where the dominance can simply determined by the WKB tunneling probability 
\begin{equation}
P^{(\nu)}_{x\rightarrow y}=\exp\left[-2\Big|\int_{y}^{x}\sqrt{\frac{2\mu}{\hbar^2}\left(W_{\nu}(R)+\frac{{1}/{4}}{2\mu R^2}\hbar^2\right)}dR\:\Big|\right],
\label{TransProb}
\end{equation}
calculated along the different parts of the potentials. 
[In Eq.~(\ref{TransProb}), $x$ and $y$ are the usual inner and outer 
classical turning points, and $(1/4)\hbar^2/(2\mu R^2)$ corresponds to the semiclassical Langer correction \cite{berry1966Proc.Phys.Soc.}.]
Since the tunneling probability in Eq.~(\ref{TransProb}) depends on the effective potentials, the relative importance of each 
pathway, as well as possible interference between them, will strongly depend on the nature of the three-body interactions.
Moreover, since in all cases the couplings peak near $R=|a|$ and/or $R=r_0$, the evaluation of the tunneling probabilities 
through a particular pathway will cover regions that span both kinds of potentials characterizing the two categories of 
three-body systems [Eqs.~(\ref{efbosonic})--(\ref{ch})].

One of the advantages of using a pathway analysis for scattering processes is that it allows
for the identification of the main physical mechanisms that control scattering events.
Moreover, a pathway analysis also lets us determine whether a particular collision 
process has a universal character or not. 
For instance, collision processes whose pathways only involve transitions and/or reflections occurring near 
$R=|a|$ ($|a|\gg r_0$) are expected to be universal. 
On the other hand, processes where pathways are affected by the physics 
near $R=r_0$ will be expected to have nonuniversal contributions that depend on the details of the interatomic 
interactions. 
Note that our piecewise representation of the three-body effective potentials in Fig.~\ref{SchPot} 
makes it simple to obtain analytical results for tunneling probabilities (\ref{TransProb}) 
and other important phases (responsible for interference and resonant effects) 
involved in the determination of scattering observables.
However, while determining the probability for scattering through a particular pathway, 
the actual nonpiecewise character of the effective potentials will be encapsulated into 
specific constants that often have a simple physical interpretation. 
Such constants, whenever having a universal character (originating from the physics near $R=|a|$), will be represented 
by capital letters ($A$, $B$, $C$, etc.). Constants depending on the short-range form of the three-body interactions
(originating from the physics near $R=r_0$) will be represented by capital letters with an index 
$\eta$ ($A_{\eta}$, $B_{\eta}$, $C_{\eta}$, etc.) and will be referred as nonuniversal.
Further studies, however, are needed to fully determine whether such constants have any residual universal character. 
Although our present simple pathway analysis is not capable of determining the values of 
such constants, it offers a unifying picture in which the importance of the Efimov
physics in determining the scattering length dependence of scattering observables becomes clear and intuitive. 

%%%%%%%%%%%%%%%%%%%%%%%%%%%%%%%%%%%%%%%%%%%%%%%%%%%%
\subsection{Three-body inelastic collisions}
\label{InelasticCol}

Since the early years after the experimental realization of ultracold atomic and molecular gases, 
three-body inelastic processes have been largely recognized as the main mechanism
controlling the lifetime and stability of such systems \cite{anderson1995Sci,davis1995PRL,bradley1997PRL,burt1997PRL,
inouye1998NT,courteille1998PRL,stenger1999PRL,roberts2000PRL,marte2002PRL,weber2003Sci,weber2003PRL}.
Therefore, the general picture for three-body inelastic processes, and their relation to Efimov physics, provided in this
section, are necessary for determining the factors that control the stability of
a variety of strongly interacting atomic and molecular gases. For pure atomic gases,
the main inelastic loss process is {\em three-body recombination}.
Here, three free atoms collide to form a diatomic molecule and a free atom ($X+Y+Z\rightarrow XY+Z$),
releasing kinetic energy that corresponds to the molecular binding energy, which is sufficient to eject both atoms and molecules from
typical traps \cite{esry1999PRL,nielsen1999PRLb,bedaque2000PRL,braaten2001PRL,moerdijk1996PRA}.
Note that under some conditions, weakly bound molecules formed through recombination
can still remain trapped  \cite{jochim2003Sci,cubizolles2003PRL,jochim2003PRL,
zwierlein2003PRL} because of their extremely small binding energy. For that reason, we treat
recombination into weakly and deeply bound molecules separately.
For atom-molecule gas mixtures, {\em atom-molecule relaxation} plays a fundamental
role in determining the lifetime of the mixture. In this case, collision of atoms with weakly bound diatomic 
molecules can stimulate inelastic transitions to more deeply bound molecular states ($XY^*+Z\rightarrow XY+Z$),
again releasing enough kinetic energy to cause losses.  Atom-molecule relaxation is defined here as 
the process in which an atom collides with a molecule leading to a transition to {\em any} lower-lying diatomic 
molecular state. However, if the gas temperature is high enough 
or the molecular binding energy too small, {\em collision-induced dissociation} can also lead
to a loss of weakly bound molecules ($XY^*+Z\rightarrow X+Y+Z$). This process is the time reversal process of recombination,
but the final products can still remain trapped. Here we determine the scattering
length dependence of each of these inelastic processes in light of the classification scheme established
in Section \ref{SecClass}, making clear the relevance of Efimov physics. 

\subsubsection{Pathways for inelastic collisions}
\label{PathwayInelastic}

In general, for a typical three-body collision process there exist many possible pathways along which the collision
can proceed. This large number of pathways is associated with both the infinite number of continuum channels and 
the usually large number of diatomic molecular channels. At ultracold collision energies, however, 
it is possible to identify a small number of pathways from which the general energy and scattering length 
dependence of three-body scattering rates can be derived.
While the dominant pathways can be determined by simple energy arguments, based on the Wigner threshold 
laws \cite{esry2002PRA}, some of these pathways, which involve various subsequent inelastic transitions and/or additional tunneling effects,
will represent higher-order corrections to the scattering rates and will be neglected in this analysis. 
In fact, the effective potentials and couplings in Fig.~\ref{SchPot} display the only channels necessary to determine the dominant 
pathways for the processes we are interested in, and we follow closely the schematic representation in Fig.~\ref{SchPot} throughout 
this section.
Moreover, since the structure of nonadiabatic couplings is the same
for both attractive and repulsive systems (both peak near $R=r_0$ and/or $|a|$), 
the relevant pathways for our analysis will be the same regardless of the classification of the three-body 
system. The specific form of the potentials, however, is crucial in determining 
 the energy and scattering length dependence of the scattering observables.
Here, we simply describe the dominant pathways for each of the inelastic scattering 
processes relevant for ultracold quantum gases. The explicit form 
of the transition probabilities, and their resulting dependence on energy and scattering length
are left to the Sections \ref{K3Section}, \ref{D3Section}, and \ref{VRSection} .

\begin{figure*}[htbp]
\includegraphics[width=6.8in,angle=0,clip=true]{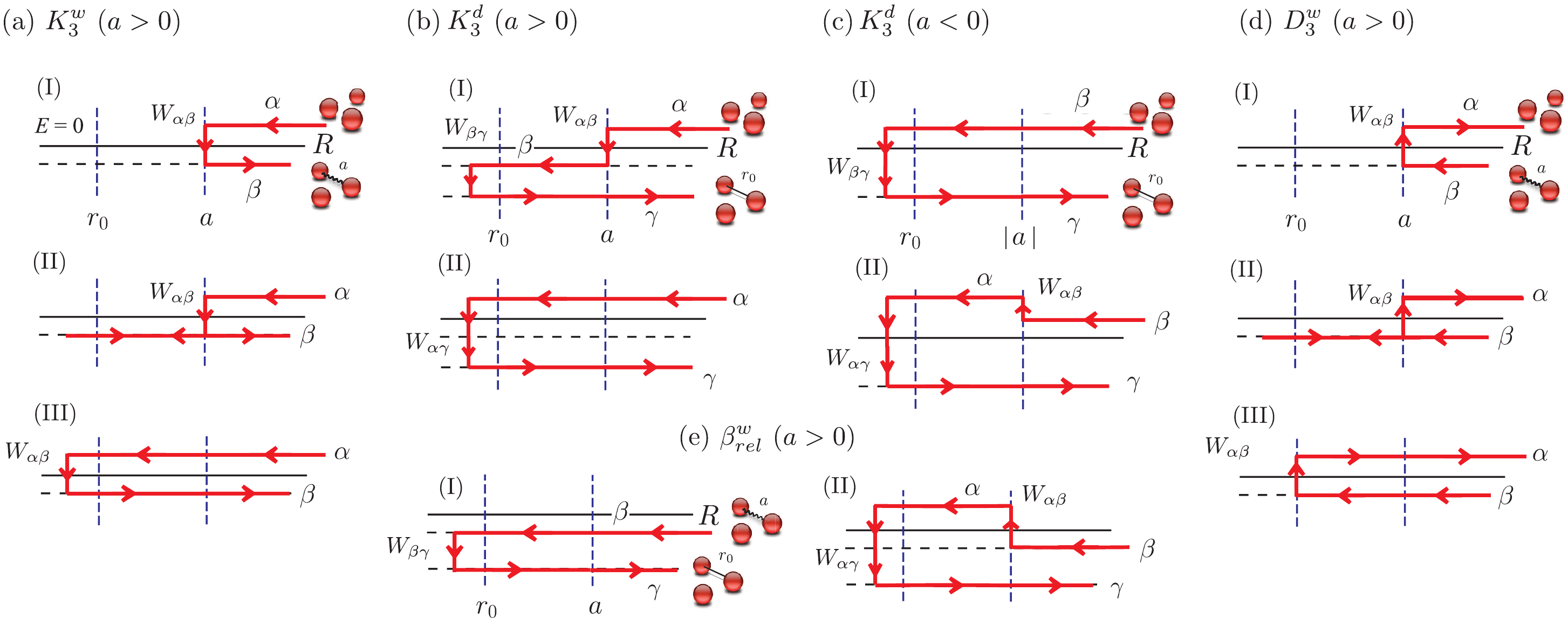}
\caption{Pathways for recombination, (a)-(c), dissociation, (d), and relaxation, (e).
The relevant nonadiabatic couplings, $W_{\nu\nu'}$, driving inelastic transitions along the pathways are 
indicated in the figure in the region in which they are expected to be enhanced (see Fig.~\ref{SchPot}).
For $a>0$, the channels $\alpha$, $\beta$, and $\gamma$
correspond to the lowest continuum, weakly, and deeply bound
molecular channels, respectively, and are represented in Fig.~\ref{SchPot}(a) and \ref{SchPot}(c).
For $a<0$, however, the channels $\alpha$ and $\beta$ correspond to the first-excited and lowest continuum channels 
represented in Figs.~\ref{SchPot}(b) and \ref{SchPot}(d).} 
\label{AllPaths}
\end{figure*}

\paragraph{Three-body recombination.} 
For three-body recombination, we separately describe the pathways for recombination processes 
leading to the formation of weakly and deeply bound molecules.
These two processes have a distinct dependence
on the scattering length because of fundamental differences in the collision pathways. 
For $a>0$, the initial channel for recombination in both cases is represented by the channel $\alpha$
in Figs.~\ref{SchPot}(a) and (c).
From the structure of the nonadiabatic couplings, the most dominant pathways for recombination into weakly 
bound molecules, i.e., pathways that start in channel $\alpha$ and finish in channel $\beta$, can be represented 
diagrammatically as in Fig.~\ref{AllPaths}(a). 
The pathways (I) and (II) both
describe three free-incoming atoms in the channel $\alpha$ undergoing an inelastic 
transition to channel $\beta$ near $R=a$, $W_{\alpha\beta}$.  While in pathway (I) molecules are formed just
after this transition, in pathway (II) atoms can still reach short distances in channel $\beta$ before forming a molecular 
state. In pathway (III), atoms approach each other at short distances ($R=r_0$) and undergo an inelastic transition 
to channel $\beta$, forming a weakly bound molecular state. 
For recombination into deeply bound molecules,
the two dominant pathways are shown in Fig.~\ref{AllPaths}(b). Here, pathway (I) represents atoms approaching each other, 
making an inelastic transition to channel $\beta$ near $R=a$, and following an inelastic transition to the final state
in channel $\gamma$. In pathway (II), the inelastic transition to channel 
$\gamma$ only occurs at short distances.
For $a<0$, no weakly bound molecular state exists, and only recombination into deeply bound molecules can happen 
[see Figs.~\ref{SchPot}(b) and (d)].
In that case, the initial channel for recombination is the channel $\beta$,
and the relevant pathways are shown in Fig.~\ref{AllPaths}(c). While in pathway (I) inelastic transitions
occur only at short distances, in pathway (II) an inelastic transition to an excited three-body continuum (channel $\alpha$)
occurs near $R=|a|$ before transitioning to the final channel $\gamma$.

\paragraph{Collision-induced dissociation.} At ultracold temperatures the only dissociation process
that can happen involves weakly bound molecules formed when $a>0$. In this case, dissociation is the time-reverse process
of three-body recombination into weakly bound molecules. Therefore, the pathways for dissociation are
the reverse of the pathways for recombination; they are shown in Fig.~\ref{AllPaths}(d). 
%JPD
%As shown later, 
%this will imply in a different energy and scattering length dependence than found for recombination.
%\textcolor{red}
{As shown later, however, the pathways for dissociation will imply a different energy and 
scattering length dependence than found for recombination, mostly due to the differences in their initial collision
states.}

 \paragraph{Atom-molecule relaxation.} The pathways for inelastic collisions between an atom and the weakly bound 
 molecular state for $a>0$ are shown in Fig.~\ref{AllPaths}(e). In pathway (I) atoms in channel $\beta$ closely approach each other and experience
 an inelastic transition to the final deeply bound molecular channel $\gamma$. In pathway (II), an additional inelastic transition
to the three-body continuum (channel $\alpha$) can occur near $R=a$ before the transition to the final channel $\gamma$ takes place.

 \paragraph{Phases and amplitudes.}
Within our pathway analysis we determine the transition probabilities assuming
that each pathway contributes with a probability amplitude given by $A_j=e^{i\phi_j}|A_j|$, where 
\begin{align}
\phi_{j}=\int\sqrt{-\frac{2\mu}{\hbar^2}\left(W(R)+\frac{1/4}{2\mu R^2}\right)}dR-\frac{\pi}{2}
\label{PhaseWKB}
\end{align}
is the WKB phase accumulated along the pathway \cite{berry1966Proc.Phys.Soc.}, calculated
whenever the system is in a classically allowed region, and $|A_j|^2$ is the corresponding probability for such
a pathway, determined simply from the tunneling probability in Eq.~(\ref{TransProb}).
The total transition probability, as usual, is determined from the coherent sum of the amplitudes
for each pathway, i.e.,
\begin{align}
|T_{fi}|^2&=|\sum_{j}A_j^{fi}|^2\nonumber\\
&=\sum_{j,k}|A_j^{fi}||A_k^{fi}|\cos(\phi_{j}-\phi_{k}),\label{GenTP}
\end{align}
where $j$ and $k$ (=I, II, III, ...) label the collision pathways, starting from the initial state $i$ to the final state $f$. 
Complications occur, however, when three-body
resonances appear in the initial collision channel. For such cases, we 
extend our pathway analysis to incorporate the proper resonant effects. 
 
\subsubsection{Three-body recombination} \label{K3Section}

The general difficulty of accurately describing three-body recombination, 
$X+Y+Z\rightarrow XY+Z$, is related to the infinite number of 
incoming continuum channels and the many possible final molecular states, 
which can make numerical calculations extremely challenging \cite{wang2013Adv,wang2011PRA,wang2011PRAc,wolf2017ARX}.
Within our pathway analysis, however, since we are mostly interested in the
energy and scattering length dependence of the rates, a number of simplifications are possible.
Formally, the recombination rate is given in terms of the $T$-matrix elements as \cite{suno2002PRA,dincao2004PRL}
\begin{equation}
K_{3} = n!\sum_{fi}  32\pi^2\hbar \frac{(2J+1)}{\mu  k^4}{|T_{fi}|^2},
\label{K3rate}
\end{equation}
where $n$ is the number of identical particles, $i$ and $f$ label all possible initial and final states, respectively,
and $k^2=2\mu E/\hbar^2$ is the incoming wave vector. Within our model, we interpret $|T_{fi}|^2$ as simple 
transition probabilities (\ref{GenTP}). Because of the Wigner threshold laws \cite{esry2002PRA,wang2011PRA},
the energy and scattering length dependence for recombination determined solely
from the properties of the lowest few continuum states, and the effects of all deeply bound states
can be encapsulated in a single parameter, greatly simplifying the problem. 

As a first step, the transition probabilities $|T_{fi}|^2$ in Eq.~(\ref{K3rate}) are expressed in terms of the 
contributions from the relevant recombination pathways shown in Fig.~\ref{AllPaths}.
For $a>0$, recombination into weakly bound states is described by 
the pathways (I), (II), and (III) in Fig.~\ref{AllPaths}(a), and the probability for recombination is
determined by
\begin{align}
|T_{\beta\alpha}|^2&=|A_{\rm I}^{\beta\alpha}|^2+|A_{\rm II}^{\beta\alpha}|^2+|A_{\rm III}^{\beta\alpha}|^2+2|A_{\rm I}^{\beta\alpha}||A_{\rm II}^{\beta\alpha}|\nonumber\\
&\times\cos(\phi_{\rm I}-\phi_{\rm II})+2|A_{\rm I}^{\beta\alpha}||A_{\rm III}^{\beta\alpha}|\cos(\phi_{\rm I}-\phi_{\rm III})\nonumber\\
&+2|A_{\rm II}^{\beta\alpha}||A_{\rm III}^{\beta\alpha}|\cos(\phi_{\rm II}-\phi_{\rm III}),
\label{K3aposweak}
\end{align}
where the corresponding amplitudes for recombination can be derived from the pathways in Fig.~\ref{AllPaths}(a) as,
\begin{align}
&|A_{\rm I}^{\beta\alpha}|^2=P^{(\alpha)}_{r_{c}\rightarrow a}P^{(\beta)}_{a\rightarrow\infty},
\nonumber\\
&|A_{\rm II}^{\beta\alpha}|^2=P^{(\alpha)}_{r_{c}\rightarrow a}P^{(\beta)}_{a\rightarrow r_{0}}P^{(\beta)}_{r_{0}\rightarrow a}P^{(\beta)}_{a\rightarrow \infty},
\nonumber\\
&|A_{\rm III}^{\beta\alpha}|^2=P^{(\alpha)}_{r_{c}\rightarrow a}P^{(\alpha)}_{a\rightarrow r_{0}}P^{(\beta)}_{r_{0}\rightarrow a}P^{(\beta)}_{a\rightarrow\infty}.
\label{AK3aposweak}
\end{align}
In Eq.~(\ref{AK3aposweak}), $r_{c}=(\lambda+2)/k$ is the outer classical turning point for recombination, 
determined from the asymptotic form of the effective potentials in Eq.~(\ref{ch}), including the 
Langer correction \cite{berry1966Proc.Phys.Soc.}. (Evidently, at ultracold energies one can assume that $r_{c}\gg |a|$.) 
Note that the notation used for the probabilities [see Eqs.~(\ref{AK3aposweak})] translate contributions from each pathway 
into a simple and precise form. 
For instance, in the transition probability in Eq.~(\ref{AK3aposweak}) corresponding to the pathway (I)
in Fig.~\ref{AllPaths}(a), $|A_{\rm I}|^2$, $P^{(\alpha)}_{r_{c}\rightarrow a}$ represents the probability of 
reaching $R=a$ from $R=r_{c}$ through the channel $\alpha$, while $P^{(\beta)}_{a\rightarrow \infty}$
indicates that the atom-molecule compound, described by channel $\beta$, was formed and represents
the exit channel contribution. 
Note that since there is no tunneling in the final (exit) channel between $R=a$ (or $R=r_0$) and $R=\infty$,
the corresponding probabilities will be simply constant.
The derivation of the pathways contributions to the transition probability
for recombination into deeply bound states is done in a similar way. In this case,
the relevant transition probability is determined from contributions of pathways (I) and (II) in  Fig.~\ref{AllPaths}(b) and is
given by
\begin{align}
|T_{\gamma\alpha}|^2=|A_{\rm I}^{\gamma\alpha}|^2+|A_{\rm II}^{\gamma\alpha}|^2
+2|A_{\rm I}^{\gamma\alpha}||A_{\rm II}^{\gamma\alpha}|\cos(\phi_{\rm I}-\phi_{\rm II}),
\label{K3aposdeep}
\end{align}
where, now, we have
\begin{align}
&|A_{\rm I}^{\gamma\alpha}|^2=P^{(\alpha)}_{r_{c}\rightarrow a}P^{(\beta)}_{a\rightarrow r_{0}}P^{(\gamma)}_{r_{0}\rightarrow \infty},
\nonumber\\
&|A_{\rm II}^{\gamma\alpha}|^2=P^{(\alpha)}_{r_{c}\rightarrow a}P^{(\alpha)}_{a\rightarrow r_{0}}P^{(\gamma)}_{r_{0}\rightarrow \infty},
\label{AK3aposdeep}
\end{align}
with $\gamma$ generically representing the set of deeply bound molecular channels (see Fig.~\ref{SchPot}). 
In principle, Eq.~(\ref{K3aposdeep}) should include all possible deeply bound channels and associated
values of $\gamma$. However, since all terms would then share the same probabilities from 
channels $\alpha$ and $\beta$, this sum can be encapsulated in a single term, namely, $P^{(\gamma)}_{r_{0}\rightarrow \infty}$.
Equations (\ref{AK3aposweak}) and (\ref{AK3aposdeep}) fully determine the transition probability for recombination processes
for $a>0$.

For $a<0$, only recombination into deeply bound molecules can occur, and the relevant pathways are shown in 
Fig.~\ref{AllPaths}(c). Although such recombination processes occur only at short distances ($R\approx r_{0}$),
they can proceed via a direct transition, as described by pathway (I), or via an indirect transition, described by the pathway
(II). These two pathways contribute to recombination into deeply bound molecules ($a<0$), resulting in the following expression 
for the recombination probability,
\begin{align}
|T_{\gamma\beta}|^2=|A_{\rm I}^{\gamma\beta}|^2+|A_{\rm II}^{\gamma\beta}|^2
+2|A_{\rm I}^{\gamma\beta}||A_{\rm II}^{\gamma\beta}|\cos(\phi_{\rm I}-\phi_{\rm II}),
\label{K3anegdeep}
\end{align}
where
\begin{align}
&|A_{\rm I}^{\gamma\beta}|^2=P^{(\beta)}_{r_{c}\rightarrow |a|}P^{(\beta)}_{|a|\rightarrow r_{0}}P^{(\gamma)}_{r_{0}\rightarrow \infty},
\nonumber\\
&|A_{\rm II}^{\gamma\beta}|^2=P^{(\beta)}_{r_{c}\rightarrow |a|}P^{(\alpha)}_{|a|\rightarrow r_{0}}P^{(\gamma)}_{r_{0}\rightarrow \infty}.
\label{AK3anegdeep}
\end{align}

The above formulas for the transition probabilities are quite general and were obtained by considering only 
the position where the coupling peaks. The actual shape of the three-body effective potentials reveals the 
energy and scattering length dependence for recombination. Within our model, however, 
the energy dependence of the transition probabilities is solely determined by the
tunneling probability in the initial channel, from $r_{c}=(\lambda+2)/k$ to $|a|$, and is present 
in all pathways [see Eqs.~(\ref{K3aposweak})-(\ref{AK3anegdeep})].
For recombination, since the potential $W_\nu$ in the initial channel is given by Eq.~(\ref{ch}), this tunneling
probability can be determined from Eq.~(\ref{TransProb}) as
\begin{equation}
P^{(\alpha)}_{r_{c}\rightarrow a} \:{\propto}\:
(ka)^{2\lambda+4} ~~\mbox{and}~~
P^{(\beta)}_{r_{c}\rightarrow |a|} \:{\propto}\: (k|a|)^{2\lambda+4},
\label{TPbeta} 
\end{equation}   
for $a>0$ and $a<0$, respectively. Therefore, for a given three-body system, 
the lowest value of $\lambda$, determined from the permutation 
symmetry (see Table~\ref{TabI}), dictates the leading-order energy behavior of 
recombination, in complete analogy to the Wigner threshold law for recombination \cite{esry2002PRA}. 
Note that Eqs.~(\ref{TPbeta}) already introduce 
some scattering length dependence for recombination. 
The complete scattering length dependence, however, can still be strongly modified by the tunneling probability 
in the Efimov region, $r_{0}\ll R\ll|a|$, thus depending on the category the system falls under. 
In the next two sections, recombination for each category is discussed, along with the determination
of contributions from Efimov physics.    

\paragraph{Recombination for attractive systems.}
Three-body recombination for systems that fall into the attractive class [see Figs.~\ref{SchPot}(a) and (c) and Table~\ref{Class}] 
is expected to be strongly affected by the Efimov effect.  For those cases, the effect of the attractive potential in the
region $r_{0}\ll R \ll |a|$ [Eq.~(\ref{efbosonic})] is mostly carried out through the WKB phases in Eq.~(\ref{PhaseWKB}).
However, since no tunneling occurs in this region, the corresponding probabilities within this region will be independent of energy and
scattering length. This phase accumulation will lead to the usual interference and resonant phenomena 
related to the Efimov effect.

For recombination into $s$-wave weakly bound molecules ($a>0$), the pathways (I), (II), and (III) in Fig.~\ref{AllPaths}(a)
lead to amplitudes (\ref{AK3aposweak}) given by
\begin{align}
&|A_{\rm I}^{\beta\alpha}|^2=A^2(ka)^{2\lambda+4},
\nonumber\\
&|A_{\rm II}^{\beta\alpha}|^2=A^2e^{-4\eta}(ka)^{2\lambda+4},
\nonumber\\
&|A_{\rm III}^{\beta\alpha}|^2=B^2e^{-4\eta}\left(\frac{r_{0}}{a}\right)^{2s_{1}}(ka)^{2\lambda+4},
\label{ABosonicK3aposweak}
\end{align}
and corresponding phase differences $\phi_{\rm I}-\phi_{\rm II}=-2[s_0\ln(a/r_0)+\Phi]$, $\phi_{\rm I}-\phi_{\rm III}=-[s_0\ln(a/r_0)+\Phi]$,
and $\phi_{\rm II}-\phi_{\rm III}=s_0\ln(a/r_0)+\Phi$, as determined from Eq.~(\ref{PhaseWKB}).
Note that in the definition of the phases from pathways (II) and (III), an unknown three-body phase $\Phi$ has been introduced. 
This phase encapsulates phase accumulated in the region near $R=r_{0}$ where the form of the potentials in Eq.~(\ref{efbosonic}) 
is not expected to hold.
Note also that the expressions for $A_{\rm II}$ and $A_{\rm III}$ account for the possibility of loss of amplitude in these
pathways due to decay into deeply bound molecular channels, whose probability is given by $(1-e^{-4\eta})$, where $\eta$ is the 
{\em inelasticity parameter} \cite{braaten2006PRep} ($\eta\ll1$ meaning low decay probability
and $\eta\gg1$ high decay probability).
The above considerations lead to the following expression for recombination into weakly bound 
molecules (\ref{K3aposweak}):
\begin{align}
K^{w}_{3}(a>0)=\frac{\hbar }{\mu} \bigg[\frac{4A_w^2}{e^{2\eta}}\left(\sin^2\left[s_{0}\ln({a}/{a_{+})}\right]+\sinh^2\eta\right)
\nonumber\\
+\frac{2A_wB_\eta}{e^{2\eta}}\frac{\sin\left[s_{0}\ln({a}/{a_{+})}\right]}{(1+e^{-2\eta})^{-1}}\left(\frac{r_{0}}{a}\right)^{s_{1}}
\nonumber\\
+\frac{B_{\eta}^2}{e^{4\eta}}\left(\frac{r_{0}}{a}\right)^{2s_{1}}\bigg]{k^{2\lambda}a^{2\lambda+4}},
\label{BosonicK3aposweak} 
\end{align}
where $A_w$ and $B_\eta$ relate to the coefficients $A$ and $B$ in Eq.~(\ref{ABosonicK3aposweak}), respectively, 
according to Eq.~(\ref{K3rate}). Here we introduce the so-called {\em three-body parameter} $a_+$, related to the three-body phase
$\Phi$ by $s_0\ln(a/a_+)=s_0\ln(a/r_0)+\Phi+\pi/2$, thus also encapsulating the short-range behavior of the three-body interactions. 
In Eq.~(\ref{BosonicK3aposweak}), the first term is the dominant contribution for recombination. It
contains the interference between pathways (I) and (II), leading to St\"uckelberg oscillations. 
Such St\"uckelberg interference is a signature of the Efimov effect and leads
to a log-periodic interference pattern in $K_3$, with minima obtained for values of $a/a_+$ equal to integer powers 
of the geometric scaling factor $e^{\pi/s_0}$. 
Equation (\ref{BosonicK3aposweak}) reproduces the well-known expression obtained for $J^\pi=0^{+}$ ($\lambda=0$) 
recombination of three identical bosons~\cite{nielsen1999PRL,esry1999PRL,bedaque2000PRL} by neglecting
the contribution from pathway (III) or, equivalently, setting $B_{\eta}=0$. 
Equation (\ref{BosonicK3aposweak}), however, generalizes recombination into weakly bound molecules to all 
attractive three-body systems and all values of $J^\pi$. 
Within our framework, the coefficient $A_{w}$ in Eq.~(\ref{BosonicK3aposweak}), which originates from pathways (I) and (II) 
in Fig.~\ref{AllPaths}(a), can be expected 
to be universal due to the fact that it accounts for inelastic transitions near $R=a$. In contrast, $B_{\eta}$ is expected to be 
nonuniversal because it is related to inelastic transitions near $R=r_{0}$.
The universal nature of the coefficient $A_w$ was noticed in the first studies on recombination in $BBB$ systems
\cite{nielsen1999PRL,esry1999PRL,bedaque2000PRL,dincao2004PRL,braaten2006PRep}, with numerical results consistent with the 
exact value $A_{w}^{2}\approx67.1177(\sqrt{3}/2)$ of Refs.~\cite{PetrovK3,macek2006PRA,gogolin2008PRL}. Results for $A_w$ for heteronuclear 
$BBX(J^{\pi}=0^+)$ systems are also known, but expressed in terms of the relevant mass ratio~\cite{helfrich2010PRA}. 

For three-body recombination into deeply bound states ($a>0$), the collision event is mainly controlled by the pathways (I) and (II), 
shown in Fig.~\ref{AllPaths}(b). For such pathways, the corresponding amplitudes [see Eq.~(\ref{AK3aposdeep})] are given by
\begin{align}
&|A_{\rm I}^{\gamma\alpha}|^2=A^2(1-e^{-4\eta})(ka)^{2\lambda+4},
\nonumber\\
&|A_{\rm II}^{\gamma\alpha}|^2=B^2(1-e^{-4\eta})\left(\frac{r_0}{a}\right)^{2s_1}(ka)^{2\lambda+4},
\label{ABosonicK3aposdeep}
\end{align}
and the phase difference is given by $\phi_{\rm I}-\phi_{\rm II}=s_0\ln(a/r_0)+\Phi$.
As a result, the probability of recombination is determined from Eq.~(\ref{K3aposdeep}), 
leading to
\begin{align}
K^{d}_{3}&(a>0)=\frac{\hbar }{\mu}
\bigg[A_{w}^2(1-e^{-4\eta})+B_{\eta}^2(1-e^{-4\eta})\left(\frac{r_{0}}{a}\right)^{2s_{1}}
\nonumber\\
&~~~+2A_wB_\eta\frac{\sin\left[s_{0}\ln({a}/{a_{+})}\right]}{(1-e^{-4\eta})^{-1}}\left(\frac{r_{0}}{a}\right)^{s_{1}}
\bigg]k^{2\lambda}a^{2\lambda+4},
\label{BosonicK3aposdeep}
\end{align}
where the first term is the dominant contribution from pathway (I) in Fig.~\ref{AllPaths}(b).
The second and third terms are corrections due to contributions from pathway (II), with $B_{\eta}$ being a nonuniversal constant 
that depends on the details of the interactions. Equation~(\ref{BosonicK3aposdeep}) is also consistent with previous results 
for three identical bosons with $J^{\pi}=0^+$~\cite{braaten2006PRep} and generalizes recombination in into deeply bound molecules
for all attractive three-body systems.

\begin{figure}[htbp]
\begin{center}
\includegraphics[width=3.4in,angle=0,clip=true]{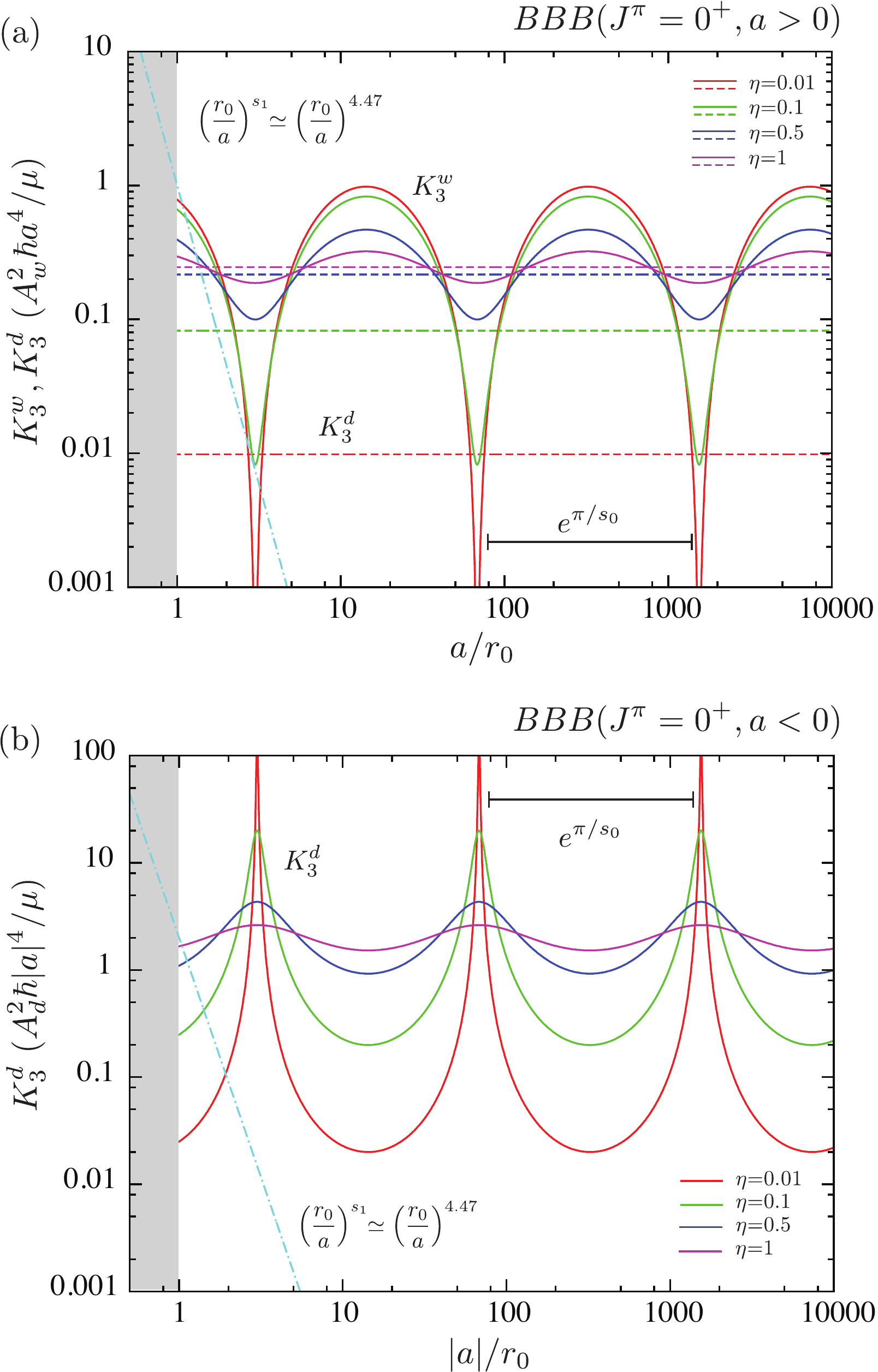}
\caption{Three-body recombination for $BBB$ ($J^\pi=0^+$, $\lambda=0$) systems
characterizing the log-periodic structure associated with the Efimov effect. (a) For $a>0$, the plot
shows recombination into weakly and deeply bound molecules, $K_3^w$ and $K_3^d$, respectively,
for different values of $\eta$. (b) For $a<0$, the plot shows a log-periodic resonant structure for
recombination into deeply bound molecules, $K_3^d$, for different values of $\eta$.}\label{K3Att}
\end{center}
\end{figure}

A similar analysis can determine recombination rates into deeply bound molecules for $a<0$. 
In this case, the two relevant pathways are shown in Fig.~\ref{AllPaths}(c), which leads to the recombination 
probability given in Eq.~(\ref{K3anegdeep}). 
A crucial difference between recombination for $a>0$ and $a<0$ is that for $a<0$, the attractive $1/R^2$ effective potential [Eq.~(\ref{efbosonic})] 
appears in the initial collision channel $\beta$, instead of the final channel. 
%JPD
%This 
%\textcolor{red}
{Having the attractive $1/R^2$ effective potential in the initial collision channel}
implies that, as $a$ increases and new Efimov states appear at the collision threshold, resonant 
effects should strongly enhance recombination for $a<0$. The appearance of an Efimov state in the entrance channel requires 
us to include many pathways similar to pathway (I) in Fig.~\ref{AllPaths}(c). These pathways account for the multiple
rattles within the range $r_0\ll R\ll |a|$, which is characteristic of a collision process near a resonant state. Therefore, 
Eq.~(\ref{K3anegdeep}) alone is clearly insufficient to capture the resonant physics. To account
for the proper resonant effects, we include all important pathways by defining
\begin{align}
|T_{\gamma\beta}|^2=&\sum_{j,k}|A_{\rm I,j}^{\gamma\beta}||A_{{\rm I},k}^{\gamma\beta}|\cos(\phi_{{\rm I},j}-\phi_{{\rm I},k})
\nonumber\\
&+|A_{{\rm II}}^{\gamma\beta}|^2+2|A_{{\rm II}}^{\gamma\beta}|\sum_{j}|A_{\rm I,j}^{\gamma\beta}|\cos(\phi_{{\rm I},j}-\phi_{{\rm II}}).
\label{K3anegdeepRes}
\end{align}
After some consideration, the amplitudes for the pathways described above can be determined,
as in Eq.~(\ref{AK3anegdeep}), to be
\begin{align}
&|A_{{\rm I},j}^{\gamma\beta}|^2=A^2(1-e^{-4\eta})(e^{-4\eta})^{j-1}(k|a|)^{2\lambda+4}, 
\nonumber\\
&|A_{{\rm II}}^{\gamma\beta}|^2=B^2(1-e^{-4\eta})\left(\frac{r_0}{|a|}\right)^{2s_1}(k|a|)^{2\lambda+4}, 
\label{AK3anegdeepRes}
\end{align}
where $j=1,2,...,\infty$ and corresponding WKB phases in Eq.~(\ref{K3anegdeepRes}) are given by
$\phi_{{\rm I},j}-\phi_{\rm I}=(2j-1)[s_0\ln(|a|/r_0)+\Phi]$ and $\phi_{{\rm I},j}-\phi_{{\rm I},k}=2(j-k)[s_0\ln(|a|/r_0)+\Phi]$.
Substituting Eq.~(\ref{AK3anegdeepRes}) into Eq.~(\ref{K3anegdeepRes}), one can perform the summations analytically.
Our present analysis, based on Eq.~(\ref{K3anegdeepRes}) and valid for all values 
of $J^\pi$, yields the same expression for recombination found in Ref.~\cite{braaten2006PRep} for $
J^\pi=0^+$ ($\lambda=0$) with a few additional modifications,
\begin{align} 
K^{d}_{3}&(a<0)=\frac{\hbar }{\mu}\bigg[\frac{A_{d}^2}{2}\frac{\sinh2\eta}{\sin^2\left[s_0\ln(|{a}/{a_-}|)\right]+\sinh^2\eta}
\nonumber\\
&+\frac{A_{d}B_{\eta}}{(1-e^{-2\eta})^{-1}}\frac{\sinh2\eta\cos\left[s_0\ln(|{a}/{a_-}|)\right]}{\sin^2\left[s_0\ln(|{a}/{a_-}|)\right]+\sinh^2\eta}\left(\frac{r_{0}}{|a|}\right)^{s_{1}}
\nonumber\\
&~~~~~~+B_{\eta}^2\left(\frac{r_{0}}{|a|}\right)^{2s_{1}}\bigg]k^{2\lambda}|a|^{2\lambda+4},
\label{BosonicK3anegdeep}
\end{align}
where $s_0\ln(|a/a_-|)=s_0\ln(|a|/r_0)+\Phi$.
The first and third terms in Eq.~(\ref{BosonicK3anegdeep}) correspond to contributions for recombination from pathways (I) and (II) 
in Eq.~(\ref{AK3anegdeepRes}), respectively, while the second term accounts for the interference between them. 
As expected, when $a$ increases by the geometric factor, $e^{\pi/s_0}$, the recombination rate (\ref{BosonicK3anegdeep})
displays resonant enhancements associated with the creation of Efimov states at the collision threshold.  
The width of the resonance in $K_{3}^{d}$ is controlled by the inelastic parameter $\eta$ \cite{braaten2006PRep}.
In Eq.~(\ref{BosonicK3anegdeep}), the {\em three-body parameter} $a_-$ is introduced;
it encapsulates the short-range behavior of the three-body interaction. As shown in Ref.~\cite{braaten2006PRep}, there
exists a universal relation between $a_-$ and $a_+$ in Eq.~(\ref{BosonicK3aposweak}). Such universal relations, along with the
universality of the three-body parameters themselves, are discussed in Sec.~\ref{Universality}.
From our analysis of the relevant pathways for recombination, the coefficient $B_{\eta}$ in Eq.~(\ref{BosonicK3anegdeep}) defines 
a nonuniversal contribution, originated from pathway (II) in Fig.~\ref{AllPaths}(c). 
The universality of recombination for $a<0$ is reflected in the constant $A_{d}$, 
which is related to the tunneling probability at $R\approx |a|$. Its numerical value, for $BBB$ ($J^\pi=0^+$) systems, 
was calculated in Ref.~\cite{braaten2006PRep} for identical bosons and found to be $A_{d}^2\approx4590(4\sqrt{3})$. 

Figure \ref{K3Att} shows the scattering length dependence of the recombination rates 
for $BBB$ ($J^\pi=0^+$, $\lambda=0$) systems, illustrating the log-periodic structure associated 
with the Efimov effect. For $a>0$, Fig.~\ref{K3Att}(a) shows recombination into weakly and deeply bound molecules via 
Eqs.~(\ref{BosonicK3aposweak}) and (\ref{BosonicK3aposdeep}), respectively, for different values of $\eta$. 
As one can see, the corresponding log-periodic minima occurring in recombination into weakly bound molecules (solid lines) 
reach a maximum contrast in the absence of deeply bound molecular states ($\eta=0$), while they tend to disappear in the limit 
of strong decay ($\eta\gg1$). In that case, the recombination rates for weakly and deeply bound states (dashed lines)
are approximately the same. For $a<0$, Fig.~\ref{K3Att}(b) shows the log-periodic resonant structure for
recombination into deeply bound molecules, Eq.~(\ref{BosonicK3anegdeep}). Again, the resonances reach
a maximum contrast in the regime of small decay ($\eta\ll1$), and they tend to disappear in the limit 
of strong decay ($\eta\gg1$). Note that in Fig.~\ref{K3Att}, we plot the leading-order term of 
 Eqs.~(\ref{BosonicK3aposweak}), (\ref{BosonicK3aposdeep}), and (\ref{BosonicK3anegdeep}), while plotting
 the scaling behavior of the next-to-leading-order term separately. Although this term, $(r_0/a)^{s_1}\simeq(r_0/a)^{4.47}$, 
 is strongly suppressed for large $a$,  for $BBX$ ($J^\pi=0^+$) systems with small mass ratios, where $s_1\simeq2$ 
 [see Fig.~\ref{CoeffsBBXFFX}(a)], 
 such terms can be more important. Note also that the shaded region in Fig.~\ref{K3Att} indicates the region in which finite-range
 corrections to the formulas (\ref{BosonicK3aposweak}), (\ref{BosonicK3aposdeep}), and (\ref{BosonicK3anegdeep}) are expected 
 to be important.

\paragraph{Recombination for repulsive systems.}
In contrast to recombination for three-body attractive systems, where the Efimov effect leads to 
interference and resonant effects, recombination for repulsive systems [see Figs.~\ref{SchPot}(b) and (d), and Table~\ref{Class}] 
is strongly influenced by tunneling effects, which are also controlled by Efimov physics. Specifically, these 
effects are controlled by the repulsive $1/R^2$ effective potentials (\ref{effermionic}) in the range $r_{0}\ll R \ll |a|$.
For $a>0$, recombination into weakly bound molecules can be determined
from the transition probabilities in Eq.~(\ref{K3aposweak}), which originates from the pathways in Fig.~\ref{AllPaths}(a). 
Here, however, since the contributions from pathway (III) are more strongly suppressed, we only consider contributions from 
pathways (I) and (II), whose amplitudes are determined from Eq.~(\ref{AK3aposweak}) to be
\begin{align}
&|A_{{\rm I}}^{\beta\alpha}|^2=A^2(ka)^{2\lambda+4}, 
\nonumber\\
&|A_{{\rm II}}^{\beta\alpha}|^2=B^2(e^{-4\eta})\left(\frac{r_0}{a}\right)^{4p_0}(ka)^{2\lambda+4}, 
\end{align}
and $\phi_{\rm I}-\phi_{\rm II}=0$. In that case, recombination into weakly bound
states is simply
\begin{align}
K^{w}_{3}(a>0)=&\frac{\hbar }{\mu}
\bigg[A_{w}^2+B_\eta^2e^{-4\eta}\left(\frac{r_{0}}{a}\right)^{4p_{0}}
\nonumber\\
&~~~+2A_{w}B_{\eta}e^{-2\eta}\left(\frac{r_{0}}{a}\right)^{2p_{0}}\bigg]k^{2\lambda}a^{2\lambda+4}.
\label{FermionicK3aposweak}
\end{align}
In Eq.~(\ref{FermionicK3aposweak}), $A_{w}$ is expected to be universal, since it corresponds to 
a pathway in which inelastic transitions occur only near $R=a$, while the coefficient $B_{\eta}$ will depend 
on the details of the interactions. This result for recombination reproduces the 
$k^2a^6$~\cite{petrov2003PRA} and $k^4a^8$~\cite{dincao2004PRL} scaling obtained for recombination of $FFX$ ($J^{\pi}=1^-$, $\lambda=1$) 
and $BBB$ systems ($J^{\pi}=2^+$, $\lambda=2$), respectively (see Table~\ref{TabI}).

Note that according to the discussion in Section~\ref{Spectrum}, for $BBX$ and $FFX$ systems,
Kartartsev-Malykh (KM) states \cite{kartavtsev2007JPB,endo2011FBS,endo2012PRA,kartavtsev2014YF} can occur
over a certain range of mass ratios, leading to important effects in scattering observables. 
KM states occur in the atom-molecule channel $\beta$ in Fig.~\ref{SchPot}(c), 
and interference effects are possible for recombination into weakly bound molecular states
(see Refs.~\cite{petrov2003PRA,kartavtsev2007JPB}). In this case, the phase controlling the interference 
phenomena, $\Phi_\delta$, is determined by the potentials near $R=a$ and, therefore, is universal---depending 
only on the mass ratio $\delta$. 
Our pathway analysis for this case follows in complete analogy to the derivation of Eq.~(\ref{BosonicK3aposweak}).
Here, however, we neglect contributions from pathway (III) in Fig.~\ref{AllPaths}(a), 
and replace $s_0\ln(a/a_+)\rightarrow\Phi_\delta+\pi/2$, and $e^{-4\eta}\rightarrow e^{-4\eta_0}=1-(1-e^{-4\eta})(r_0/a)^{2p_0}$,
incorporating the fact that, now, the probability for decay to deeply bound states is suppressed by tunneling through
the potentials in the range $r_0\ll R\ll a$. With these considerations, and expanding the results for small 
$\eta_0\approx(1/4)(1-e^{-4\eta})(r_0/a)^{2p_0}$, one finally can find that recombination into weakly 
bound states is given by
\begin{align}
K^{w}_{3\delta}&(a>0)\approx\frac{\hbar }{\mu} \bigg[
4A_w^2
\left[1-\frac{(1-e^{-4\eta})}{2}\left(\frac{r_0}{a}\right)^{2p_0}\right]
\cos^2\Phi_{\delta}
\nonumber\\
&+A_w^2\frac{(1-e^{-4\eta})}{2}\left(\frac{r_{0}}{a}\right)^{4p_{0}}\bigg]{k^{2\lambda}a^{2\lambda+4}}.
\label{FermionicK3aposweakKM} 
\end{align}
For $\eta=0$, our result is consistent with Refs.~\cite{petrov2003PRA,kartavtsev2007JPB}, 
where recombination is predicted to vanish for certain values of $\delta$ or, as in our case, for $\Phi_{\delta}=(2n+1)\pi/2$, with
$n=0$, 1, 2, etc. Equation (\ref{FermionicK3aposweakKM}), however, shows that in the presence of inelastic decay ($\eta\ne0$) ,
recombination into weakly bound molecules is not fully suppressed. In fact, analysis shows that contributions from the
neglected pathway (III) in Fig.~\ref{AllPaths}(a) would also lead to a non-vanishing recombination, with its minimum value proportional 
to $(r_0/a)^{2p_1+2p_0}k^{2\lambda}a^{2\lambda+2}$, even in the absence of decay ($\eta=0$).

Now, turning to recombination into deeply bound molecules ($a>0$), its energy and scattering length dependence 
are determined from Eq.~(\ref{K3aposdeep}), with amplitudes [Eq.~(\ref{AK3aposdeep})] given by
\begin{align}
&|A_{{\rm I}}^{\gamma\alpha}|^2=A^2(1-e^{-4\eta})\left(\frac{r_0}{a}\right)^{2p_0}(ka)^{2\lambda+4}, 
\nonumber\\
&|A_{{\rm II}}^{\gamma\alpha}|^2=B^2(1-e^{-4\eta})\left(\frac{r_0}{a}\right)^{2p_1}(ka)^{2\lambda+4}, 
\label{FermionicK3aposdeepA}
\end{align}
and the corresponding phase difference $\phi_{\rm I}-\phi_{\rm II}=0$. 
Substituting Eq.~(\ref{FermionicK3aposdeepA}) in Eq.~(\ref{K3aposdeep}), we find 
recombination into deeply bound molecules to be given by
\begin{align}
K^{d}_{3}&(a>0)=\frac{\hbar }{\mu}(1-e^{-4\eta})
\bigg[A_{\eta}^2\left(\frac{r_0}{a}\right)^{2p_0}
\nonumber\\
&+2A_{\eta}B_{\eta}\left(\frac{r_{0}}{a}\right)^{p_{0}+p_1}
+B_\eta^2\left(\frac{r_{0}}{a}\right)^{2p_{1}}
\bigg]k^{2\lambda}a^{2\lambda+4},
\label{FermionicK3aposdeep}
\end{align}
where amplitudes $A_{\eta}$ and $B_{\eta}$ depend on details of the 
interatomic interactions. Effects associated with the KM states can be obtained by 
following the derivation of Eq.~(\ref{BosonicK3aposdeep}), with the proper replacements:
$s_1\rightarrow p_1$, $s_0\ln(a/a_+)\rightarrow\Phi_\delta+\pi/2$, and 
$e^{-4\eta}\rightarrow1-(1-e^{-4\eta})(r_0/a)^{2p_0}$. Doing so, we obtain
\begin{align}
K^{d}_{3\delta}&(a>0)=\frac{\hbar }{\mu}(1-e^{-4\eta})
\bigg[A_{w}^2\left(\frac{r_0}{a}\right)^{2p_0}+B_{\eta}^2\left(\frac{r_{0}}{a}\right)^{2p_{1}}
\nonumber\\
&~~~+2A_wB_\eta\cos\Phi_{\delta}\left(\frac{r_{0}}{a}\right)^{p_0+p_{1}}
\bigg]k^{2\lambda}a^{2\lambda+4}.
\label{FermionicK3aposdeepKM}
\end{align}
Comparing the scattering length dependence for recombination into weakly
and deeply bound states [Eqs.~(\ref{FermionicK3aposweak}), (\ref{FermionicK3aposdeep}), and (\ref{FermionicK3aposdeepKM}),
respectively], we see that recombination in deeply bound states has a much weaker dependence on 
$a$, and, as a result, most of the recombination processes for $a>0$ should lead to weakly bound molecules. 
For instance, for $FFF'$ ($J^{\pi}=1^-$) systems, recombination into weakly bound molecules 
scales as $k^2a^6$, while recombination in deeply bound molecules 
scales approximately as $k^2a^{2.455}$.
This dominance of recombination into weakly bound states is in contrast with 
attractive systems where recombination into weakly and deeply bound states has the same $a$ scaling 
[see Eqs.~(\ref{BosonicK3aposweak}) and (\ref{BosonicK3aposdeep})].

This same suppression for recombination into deeply bound states is also observed for $a<0$. 
In this case, recombination is determined from Eq.~(\ref{K3anegdeep}), with amplitudes [Eq.~(\ref{AK3anegdeep})]
given by
\begin{align}
&|A_{{\rm I}}^{\gamma\beta}|^2=A^2(1-e^{-4\eta})\left(\frac{r_0}{|a|}\right)^{2p_0}(k|a|)^{2\lambda+4}, 
\nonumber\\
&|A_{{\rm II}}^{\gamma\beta}|^2=B^2(1-e^{-4\eta})\left(\frac{r_0}{|a|}\right)^{2p_1}(k|a|)^{2\lambda+4}, 
\end{align}
and $\phi_{\rm I}-\phi_{\rm II}=0$. In this case, recombination into deeply bound
molecules is given by
\begin{align}
K^{d}_{3}&(a<0)=\frac{\hbar }{\mu}(1-e^{-4\eta})
\bigg[A_{\eta}^2\left(\frac{r_0}{|a|}\right)^{2p_0}+B_\eta^2\left(\frac{r_{0}}{|a|}\right)^{2p_{1}}
\nonumber\\
&~~~+2A_{\eta}B_{\eta}\left(\frac{r_{0}}{|a|}\right)^{p_{0}+p_1}\bigg]k^{2\lambda}|a|^{2\lambda+4},
\label{FermionicK3anegdeep}
\end{align}
where both amplitudes $A_{\eta}$ and $B_{\eta}$ depend on details of the 
interatomic interactions.

\begin{figure}[htbp]
\begin{center}
\includegraphics[width=3.4in,angle=0,clip=true]{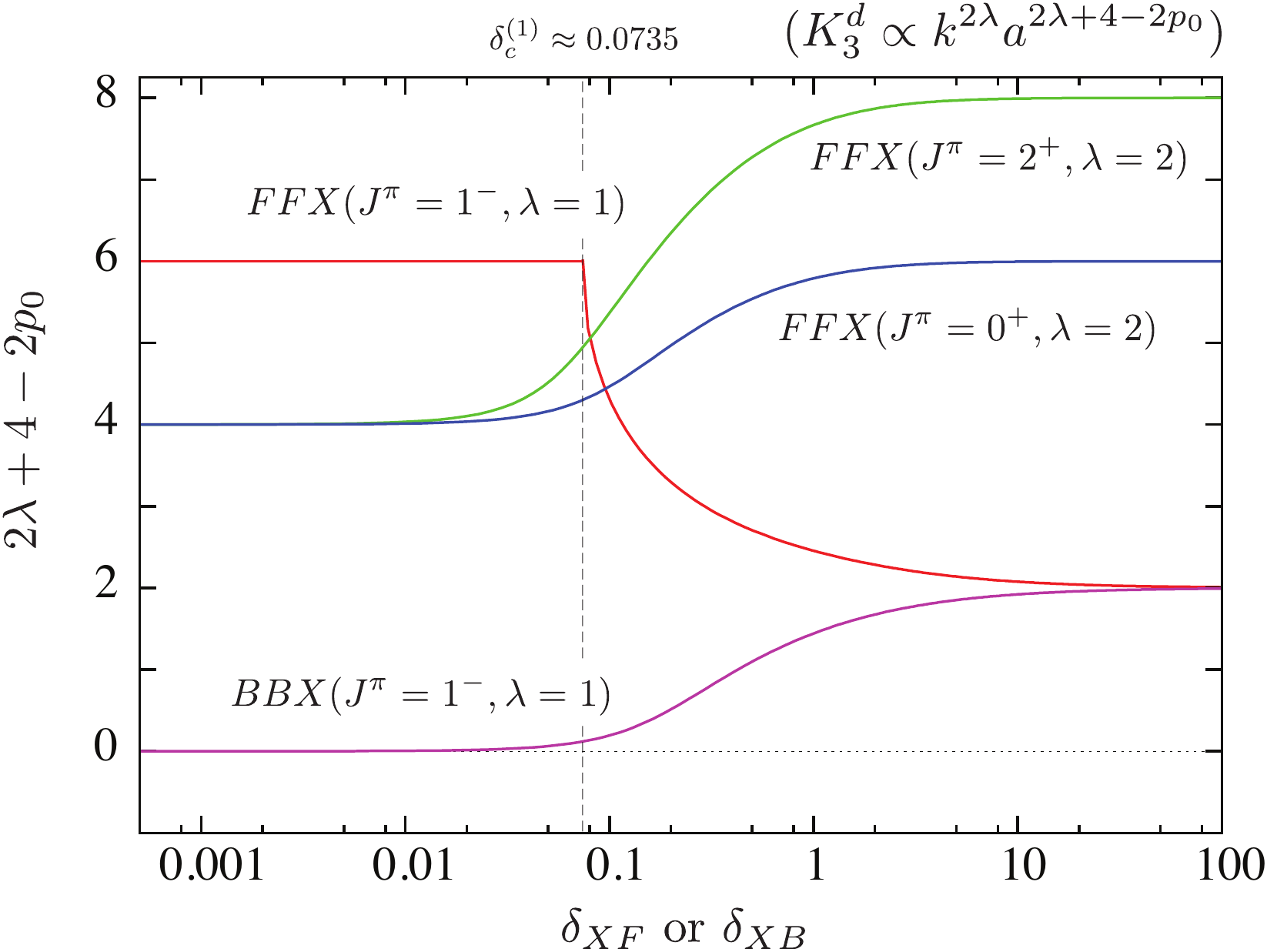}
\caption{Coefficient $2\lambda+4-2p_0$ determining the scattering length dependence for 
recombination into bound molecules [Eqs.~(\ref{FermionicK3aposdeep}) and (\ref{FermionicK3anegdeep})] for different
values of the relevant mass ratios. For $FFX$ ($J^\pi=1^-$) systems, the suppression occurs more strongly for large mass ratios. 
For the other cases shown in the figure, the suppression is more pronounced for small mass ratios.}
\label{K3Rep}
\end{center}
\end{figure}

The results in Eqs.~(\ref{FermionicK3aposweak})-(\ref{FermionicK3anegdeep}) thus predict an asymmetry in 
recombination into weakly bound molecules ($K_3^w\propto k^{2\lambda}a^{2\lambda-4}$) and deeply bound molecules
($K_3^d\propto k^{2\lambda}r_0^{2p_0}|a|^{2\lambda+4-2p_0}$), which is a direct consequence of the Efimov physics
manifested through its dependence on the Efimov coefficient $p_0$ \cite{dincao2005PRL}.
This asymmetry is illustrated in Fig.~\ref{K3Rep}, which shows the coefficient $2\lambda+4-2p_0$ (controlling the scattering length dependence) for $FFX$ and
$BBX$ repulsive systems as a function of the relevant mass ratios. 
For large mass ratios $FFX$ ($J^\pi=1^-$), recombination into deeply bound states tends to be more
strongly suppressed while it becomes proportional to $ka^6$ for $\delta_{XF}\le\delta_c$, i.e.,
when the system changes its characteristic from repulsive to attractive, as shown in Fig.~\ref{K3Rep}. For
these cases, the suppression of recombination is more pronounced for small mass ratios.

\subsubsection{Collision-induced dissociation} \label{D3Section}

In this section, the energy and scattering length dependence of 
collision-induced dissociation processes, $XY+Z\rightarrow X+Y+Z$, is determined. 
In ultracold quantum gases, collision-induced dissociation 
is allowed only if the collision energy of the incoming collision partners ($XY+Z$) is greater or equal to the corresponding
molecular binding energy. In that case, the three-body continuum channels ($X+Y+Z$) are energetically 
open and dissociation can proceed. 
As a result, at ultracold temperatures only dissociation of weakly bound molecules 
are relevant to experiments.
The dissociation rate of weakly bound molecules is given by \cite{suno2002PRA}
\begin{equation}
D_{3} = \sum_{f}4\pi\hbar\frac{(2J+1)}{\mu_{ad} k_{ad}}{|T_{fi}|^2}, 
\label{D3rate}
\end{equation}
where $k^2_{ad}=2\mu_{ad}(E-E_{vl'})/\hbar^2$ is the wave vector in the initial atom-molecule collision channel, and $\mu_{ad}$ is the 
atom-molecule reduced mass. For dissociation, however, the important wave 
vector for the threshold law is $k^2=2\mu E/\hbar^2$, which is associated with the final channels and originates from the 
energy dependence of the dissociation probability, $|T_{fi}|^2$.
In the ultracold regime ($ka\ll1$), $k_{ad}\propto 1/a$, and the formula (\ref{D3rate}) 
already contain a linear dependence in $a$. 

Extending the analysis performed for recombination in Section~\ref{K3Section}, the dissociation of weakly bound molecules 
can be determined from pathways (I), (II), and (III) in Fig.~\ref{AllPaths}(d). 
As a result, the probability for dissociation is given by
\begin{align}
|T_{\alpha\beta}|^2&=|A_{\rm I}^{\alpha\beta}|^2+|A_{\rm II}^{\alpha\beta}|^2+|A_{\rm III}^{\alpha\beta}|^2+2|A_{\rm I}^{\alpha\beta}||A_{\rm II}^{\alpha\beta}|\nonumber\\
&\times\cos(\phi_{\rm I}-\phi_{\rm II})+2|A_{\rm I}^{\alpha\beta}||A_{\rm III}^{\alpha\beta}|\cos(\phi_{\rm I}-\phi_{\rm III})\nonumber\\
&+2|A_{\rm II}^{\alpha\beta}||A_{\rm III}^{\alpha\beta}|\cos(\phi_{\rm II}-\phi_{\rm III}),
\label{D3aposweak}
\end{align}
where the corresponding amplitudes for dissociation can be derived from the pathways in Fig.~\ref{AllPaths}(d) to be:
\begin{align}
&|A_{\rm I}^{\alpha\beta}|^2=P^{(\beta)}_{\infty\rightarrow a}P^{(\alpha)}_{a\rightarrow r_{c}},
\nonumber\\
&|A_{\rm II}^{\alpha\beta}|^2=P^{(\beta)}_{\infty\rightarrow a}P^{(\beta)}_{a\rightarrow r_{0}}P^{(\beta)}_{r_{0}\rightarrow a}P^{(\alpha)}_{a\rightarrow r_{c}},
\nonumber\\
&|A_{\rm III}^{\alpha\beta}|^2=P^{(\beta)}_{\infty\rightarrow a}P^{(\beta)}_{a\rightarrow r_{0}}P^{(\alpha)}_{r_{0}\rightarrow a}P^{(\alpha)}_{a\rightarrow r_{c}}.
\label{AD3aposweak}
\end{align}
Note that for dissociation, there is no tunneling in the initial channel for $R>a$.  As a result,
the term $P^{(\beta)}_{\infty\rightarrow a}$ above is constant.
All pathways for dissociation include tunneling processes in the final channel in the
region $a\ll R\ll r_{c}$, where $r_{c}=(\lambda+2)/k$ is the outer classical turning point 
in the final collision channel $\alpha$ of Figs.~\ref{SchPot}(a) and (c). In this case, the tunneling probability calculated 
using Eq.~(\ref{TransProb}) is the same irrespective of the category into which a system falls into, i.e.,
\begin{equation}
P^{(\alpha)}_{a\rightarrow r_{c}} \:{\propto}\:
(ka)^{2\lambda+4},
\label{TPbetad} 
\end{equation}   
thus determining the threshold law for collision-induced dissociation~\cite{esry2002PRA} and some of 
the scattering length dependence for dissociation.
The scaling with $a$, however, can be strongly modified because of Efimov physics, providing the 
main signature of such effects in dissociation. 

In fact, Eq.~(\ref{AD3aposweak}) fully determines the scattering length dependence for dissociation of weakly bound
molecules ($a>0$). For attractive systems, the dissociation rate is determined
from Eq.~(\ref{D3aposweak}) with amplitudes
\begin{align}
&|A_{\rm I}^{\alpha\beta}|^2=A^2(ka)^{2\lambda+4},
\nonumber\\
&|A_{\rm II}^{\alpha\beta}|^2=A^2e^{-4\eta}(ka)^{2\lambda+4},
\nonumber\\
&|A_{\rm III}^{\alpha\beta}|^2=B^2e^{-4\eta}\left(\frac{r_{0}}{a}\right)^{2s_{1}}(ka)^{2\lambda+4},
\label{ABosonicD3aposweak}
\end{align}
and corresponding phases $\phi_{\rm I}-\phi_{\rm II}=-2[s_0\ln(a/r_0)+\Phi]$, $\phi_{\rm I}-\phi_{\rm III}=-[s_0\ln(a/r_0)+\Phi]$, and 
$\phi_{\rm II}-\phi_{\rm III}=s_0\ln(a/r_0)+\Phi$, determined from Eq.~(\ref{PhaseWKB}).
These considerations lead to the following expression for dissociation of weakly bound 
molecules:
\begin{align}
D^{w}_{3}(a>0)&=\frac{\hbar }{\mu_{ad}} \bigg[\frac{4A_w^2}{e^{2\eta}}\left(\sin^2\left[s_{0}\ln({a}/{a_{+})}\right]+\sinh^2\eta\right)
\nonumber\\
&+\frac{2A_wB_\eta}{e^{2\eta}}\frac{\sin\left[s_{0}\ln({a}/{a_{+})}\right]}{(1+e^{-2\eta})^{-1}}\left(\frac{r_{0}}{a}\right)^{s_{1}}
\nonumber\\
&~~~~~+\frac{B_{\eta}^2}{e^{4\eta}}\left(\frac{r_{0}}{a}\right)^{2s_{1}}\bigg]{k^{2\lambda+4}a^{2\lambda+5}},
\label{BosonicD3aposweak} 
\end{align}
where $A_w$ and $B_\eta$ relate to the coefficients $A$ and $B$ in Eq.~(\ref{ABosonicD3aposweak}) via
Eq.~(\ref{D3rate}), and $s_0\ln(a/a_+)=s_0\ln(a/r_0)+\Phi+\pi/2$. 
Since dissociation is the time-reversed process of recombination, the coefficients $A_{w}$ and $B_{\eta}$ are also
expected to be related to the ones for recombination in Eq.~(\ref{BosonicK3aposweak}). 

For repulsive systems, dissociation of weakly bound molecules is determined only considering contributions from 
pathways (I) and (II) in Eq.~(\ref{D3aposweak}). In this case, the corresponding amplitudes are given by
\begin{align}
&|A_{{\rm I}}^{\alpha\beta}|^2=A^2(ka)^{2\lambda+4}, 
\nonumber\\
&|A_{{\rm II}}^{\alpha\beta}|^2=B^2e^{-4\eta}\left(\frac{r_0}{a}\right)^{4p_0}(ka)^{2\lambda+4}, 
\end{align}
and $\phi_{\rm I}-\phi_{\rm II}=0$, which results in the rate for dissociation of weakly bound molecules to be:
\begin{align}
D^{w}_{3}(a>0)&=\frac{\hbar }{\mu_{ad}}
\bigg[A_{w}^2+B_\eta^2e^{-4\eta}\left(\frac{r_{0}}{a}\right)^{4p_{0}}
\nonumber\\
&+2A_{w}B_{\eta}e^{-2\eta}\left(\frac{r_{0}}{a}\right)^{2p_{0}}\bigg]k^{2\lambda+4}a^{2\lambda+5}.
\label{FermionicD3aposweak}
\end{align}
And as with recombination [Eq.~(\ref{FermionicK3aposweakKM})], the interference 
effects associated with KM states \cite{kartavtsev2007JPB,endo2011FBS,endo2012PRA,kartavtsev2014YF} 
can be introduced for the dissociation of weakly bound molecules.
In this case, we obtain
\begin{align}
D^{w}_{3\delta}(a>0)\approx&\frac{\hbar }{\mu_{ad}} \bigg[
4A_w^2
\left[1-\frac{(1-e^{-4\eta})}{2}\left(\frac{r_0}{a}\right)^{2p_0}\right]
\cos^2\Phi_{\delta}
\nonumber\\
&+A_w^2\frac{(1-e^{-4\eta})}{2}\left(\frac{r_{0}}{a}\right)^{4p_{0}}\bigg]{k^{2\lambda+4}a^{2\lambda+5}}.
\label{FermionicD3aposweakKM} 
\end{align}

The main difference between the results above and ones for recombination into weakly bound molecules is
the additional $ak^4$ dependence of $D^w_{3}$. In ultracold quantum gases
with $a\gg r_{0}$, molecules are extremely weakly bound and both collision-induced dissociation and atom-molecule relaxation
can cause molecular losses. The difference, however, is that while dissociation does not lead to
loss of atoms because its final products have a negligible kinetic energy, atom-molecule relaxation does
lead to loss of both atoms and molecules. In this case, the amount of energy released is too large,
causing both atom and molecules to escape from typical traps.
The influence of the Efimov physics on atom-molecule relaxation will be discussed in the next section.

\subsubsection{Atom-molecule relaxation} \label{VRSection}

In strongly interacting ultracold atom-molecule gas mixtures, the major collisional process that 
determines stability is atom-molecule relaxation, $XY^*+Z\rightarrow XY+Z$. 
The present study only concerns the collisional properties of weakly bound $s$-wave molecules, where universal properties
associated with Efimov physics can be derived. 
Collisional properties of deeply bound molecular states are insensitive to variations of $a$ and strongly dependent
on the details of interatomic interactions and, therefore, cannot be described within our model.
Nevertheless, the general definition for the atom-molecule relaxation rate is given by \cite{dincao2008PRA}
\begin{equation}
\beta_{\rm rel} = \sum_{f}4\pi\hbar\frac{(2J+1)}{\mu_{ad} k_{ad}}{|T_{fi}|^2}, 
\label{Vrelrate}
\end{equation}
where $\mu_{ad}$ is the atom-molecule reduced mass and $f$ is the label corresponding to all possible final states associated 
with deeply bound molecular channels. For relaxation of weakly bound $s$-wave molecules, the relevant wave vector associated to 
the initial collision channel is $k_{ad}^2=2\mu_{ad}(E-E^*_{v0})/\hbar^2$, where $E_{v0}^*\propto-1/a^2$, corresponding to the binding energy of the
weakly bound molecular state. The energy and scattering dependence for relaxation of weakly bound molecules can be determined solely 
from the properties of the initial collision channel $\beta$ of Figs.~\ref{SchPot}(a) and (c).
For simplicity, in what follows, we will denote $k=k_{ad}$. 

Within our model, the probability for relaxation of weakly bound molecules is determined from pathways (I) and (II) in Fig.~\ref{AllPaths}(e). 
Pathway (II) includes an intermediate transition to the energetically closed channel $\alpha$ of Figs.~\ref{SchPot}(a) and (c), 
and is, in general, less important. These pathways define the probability for relaxation of weakly bound molecules as
\begin{align}
|T_{\gamma\beta}|^2&=|A_{\rm I}^{\gamma\beta}|^2+|A_{\rm II}^{\gamma\beta}|^2+2|A_{\rm I}^{\gamma\beta}||A_{\rm II}^{\gamma\beta}|\cos(\phi_{\rm I}-\phi_{\rm II}),
\label{Vrelaposweak}
\end{align}
where the corresponding amplitudes can be derived from the pathways in Fig.~\ref{AllPaths}(e) as,
\begin{align}
&|A_{\rm I}^{\gamma\beta}|^2=P^{(\beta)}_{r_{c}\rightarrow a}P^{(\beta)}_{a\rightarrow r_{0}}P^{(\gamma)}_{r_{0}\rightarrow\infty},
\nonumber\\
&|A_{\rm II}^{\gamma\beta}|^2=P^{(\beta)}_{r_{c}\rightarrow a}P^{(\alpha)}_{a\rightarrow r_{0}}P^{(\gamma)}_{r_{0}\rightarrow\infty},
\label{AVrelaposweak}
\end{align}
where $r_{c}=(l+1/2)/k$ is the classical turning point for relaxation, determined from Eq.~(\ref{bc}), including
the Langer correction \cite{berry1966Proc.Phys.Soc.}. 
Note that in Eq.~(\ref{AVrelaposweak}),
since there is no tunneling in the final (exit) channel between $R=r_0$ and $R=\infty$,
the corresponding probability will be simply constant.
The energy dependence for relaxation can be determined using the tunneling probability given by 
Eq.~(\ref{TransProb}). In that case, similar to recombination and dissociation, tunneling in the regions from $r_{c}$ to $a$ 
completely determines the energy dependence of relaxation, regardless of the category into which the system falls,
\begin{equation}
P^{(\beta)}_{r_{c}\rightarrow a} \:{\propto}\:
(ka)^{2l+1}.
\label{TPbetax} 
\end{equation}   
The scaling with $a$ obtained from Eq.~(\ref{TPbetax}) 
is strongly modified by the three-body interaction in the region $r_{0}\ll R\ll a$, 
as determined by the Efimov physics characteristic of each category.

\paragraph{Relaxation for attractive systems.}
For attractive systems [see Figs.~\ref{SchPot}(a) and Table~\ref{Class}], relaxation 
is expected to exhibit resonant features related to the Efimov states formed at the collision threshold. 
In fact, similar to the $a<0$ recombination in attractive systems [see Eqs.~(\ref{K3anegdeepRes}) and (\ref{AK3anegdeepRes})]
the physics encapsulated in pathway (I) in Fig.~\ref{AllPaths}(e) is not
enough to describe resonant effects. Resonant transmission effects should include the multiple rattles within the range $r_0\ll R\ll |a|$, 
i.e., within to the region where the effective potential is attractive and supporting a resonant state.
Therefore, resonant transmission allows a substantial part of the wave function 
to reach the region where the coupling for relaxation peaks, $R\approx r_{0}$, enhancing the inelastic transitions. 
To account for the proper resonant effects, we include all important pathways by defining
\begin{align}
|T_{\gamma\beta}|^2=&\sum_{j,k}|A_{\rm I,j}^{\gamma\beta}||A_{{\rm I},k}^{\gamma\beta}|\cos(\phi_{{\rm I},j}-\phi_{{\rm I},k})
\nonumber\\
&+|A_{{\rm II}}^{\gamma\beta}|^2+2|A_{{\rm II}}^{\gamma\beta}|\sum_{j}|A_{\rm I,j}^{\gamma\beta}|\cos(\phi_{{\rm I},j}-\phi_{{\rm II}}),
\label{VrelaposweakRes}
\end{align}
with corresponding amplitudes determined from
\begin{align}
&|A_{{\rm I},j}^{\gamma\beta}|^2=A^2(1-e^{-4\eta})(e^{-4\eta})^{j-1}(ka)^{2l+1}, 
\nonumber\\
&|A_{{\rm II}}^{\gamma\beta}|^2=B^2(1-e^{-4\eta})\left(\frac{r_0}{a}\right)^{2s_1}(ka)^{2l+1}, 
\label{AVrelanegdeepRes}
\end{align}
where $j=1,2,...,\infty$. The corresponding WKB phases are given by
$\phi_{{\rm I},j}-\phi_{{\rm I},k}=2(j-k)[s_0\ln(a/r_0)+\Phi]$ and $\phi_{{\rm I},j}-\phi_{\rm II}=(2j-1)[s_0\ln(a/r_0)+\Phi]$. Replacing Eq.~(\ref{AVrelanegdeepRes})
into Eq.~(\ref{VrelaposweakRes}) and performing the summations analytically
yield the relaxation rate as
\begin{align} 
\beta^{w}_{\rm rel}(a>&0)=\frac{\hbar }{\mu_{ad}}\bigg[\frac{A_{d}^2}{2}\frac{\sinh2\eta}{\sin^2\left[s_0\ln({a}/{a_*})\right]+\sinh^2\eta}
\nonumber\\
&+\frac{A_{d}B_{\eta}}{(1-e^{-2\eta})^{-1}}\frac{\sinh2\eta\cos\left[s_0\ln({a}/{a_*})\right]}{\sin^2\left[s_0\ln({a}/{a_*})\right]+\sinh^2\eta}\left(\frac{r_{0}}{a}\right)^{s_{1}}
\nonumber\\
&~~~~~~+B_{\eta}^2\left(\frac{r_{0}}{a}\right)^{2s_{1}}\bigg]k^{2l}a^{2l+1},
\label{BosonicVrelaposweak}
\end{align}
where $s_0\ln(a/a_*)=s_0\ln(a/r_0)+\Phi$. Here, the three-body parameter $a_*$ is introduced and encapsulates the short range 
behavior of the three-body interactions. 
The first term displays the expected resonant enhancements each time $a$ is increased by $e^{\pi/s_{0}}$, creating a new Efimov state. 
The second term corresponds to interference effects between pathways (I) and (II). 
Within our model it is clear that $A_{d}$ should be a universal number, and that all dependence of relaxation 
on the details of the interactions is incorporated in the $a_{*}$, $\eta$, and $B_{\eta}$ parameters.

The above expression for the relaxation rate is a generalization of the result for $BBB$ ($J^\pi=0^+$ and $l=0$) 
systems in Ref.~\cite{braaten2004PRA}, where it was found that $A_d^2\approx20.3(4/3)$. This result can also be
derived from Ref.~\cite{efimov1979SJNP} by letting the transition probability at small distances be less than unity. 
In both cases, the linear dependence on $a$ was obtained indirectly by dimensional and physical arguments, while 
it follows naturally from Eq.~(\ref{TPbetax}) in the present analysis. In the present, simple 
approach, the mass dependence for relaxation is manifested only through the coefficient $s_{0}$ 
[see Fig.~\ref{CoeffsBBXFFX}(a), for instance]. In a more rigorous approach (Ref.~\cite{helfrich2010PRA}), 
the values for the universal constant $A_{d}$ were calculated for heteronuclear $BBX$ ($J^{\pi}=0^+$) systems in 
terms of the mass ratio.

\begin{figure}[htbp]
\begin{center}
\includegraphics[width=3.4in,angle=0,clip=true]{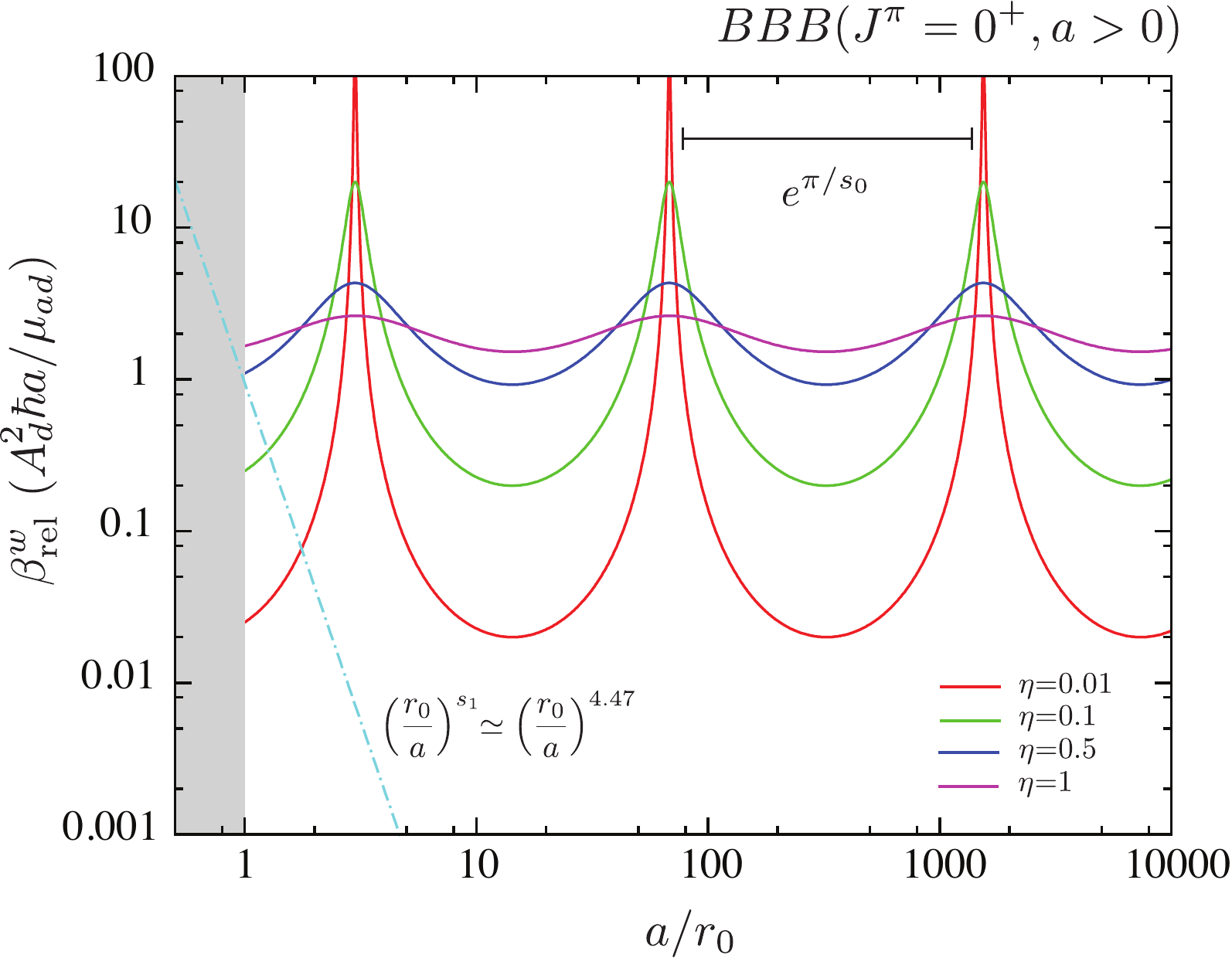}
\caption{Relaxation rate for $BBB$ ($J^\pi=0^+$, $l=0$) systems displaying the usual log-periodic structure associated 
with the Efimov effect. Resonances in relaxation reach a maximum contrast in the regime of small decay ($\eta\ll1$) 
and tend to be washed out in the limit of strong decay ($\eta\gg1$).}\label{VrelAtt}
\end{center}
\end{figure}

The scattering length dependence of the relaxation rate for $BBB$ ($J^\pi=0^+$, $l=0$) systems, displaying the usual 
log-periodic structure associated with the Efimov effect, is illustrated in Fig.~\ref{VrelAtt}. Here, similar to $a<0$ 
recombination [see Fig.~\ref{K3Att}(b)], resonances in relaxation reach a maximum contrast in the regime of small decay ($\eta\ll1$) 
and tend to be washed out in the limit of strong decay ($\eta\gg1$). Note that we only include the leading-order 
term of Eq.~(\ref{BosonicVrelaposweak}) and plot, separately, its next-to-leading-order term in Fig.~\ref{VrelAtt}. This
next-to-leading-order term, $(r_0/a)^{s_1}\simeq(r_0/a)^{4.47}$, is strongly suppressed for large $a$. 
However, for $BBX$ ($J^\pi=0^+$) systems with small mass ratios, where $s_1\simeq2$ [see Fig.~\ref{CoeffsBBXFFX}(a)], 
such terms can be more important. 

\paragraph{Relaxation for repulsive systems.}
\label{VrelSecCat}
For repulsive systems, no resonant enhancements of relaxation are, in general, expected to occur since these systems 
do not exhibit the Efimov effect. Nevertheless, Efimov physics still has a strong impact on 
the scattering length dependence of the relaxation of weakly bound molecules. 
In contrast with attractive systems, where the three-body attractive $1/R^2$ effective interaction leads to
resonant effects, the main physical phenomena that controls the scattering length dependence 
of relaxation in repulsive three-body systems is tunneling through the repulsive $1/R^2$ effective
potential in the region $r_{0}\ll R\ll a$. In our model, all the pathways relevant for relaxation [Fig.~\ref{AllPaths}(e)] 
suffer from such tunneling effects. Moreover, since the spatial extent of this Efimov potential increases with $a$, it is 
reasonable to expect that relaxation should be suppressed in the large $a$ limit. 
This suppression, first predicted in Ref.~\cite{petrov2004PRLb} in the context of two-spin Fermi gases,
has been observed in several experiments
~\cite{greiner2003NT,jochim2003Sci,strecker2003PRL,cubizolles2003PRL,jochim2003PRL,zwierlein2003PRL,bartenstein2004PRLb,
regal2004PRL,bourdel2004PRL,zirbel2008PRL,spiegelhalder2009PRL,khramov2012PRA,wu2012PRL,laurent2017PRL}
and has proved to be an extremely beneficial property that allows for long molecular lifetimes. 
However, Efimov physics has not been traditionally recognized
as the main mechanism controlling the collisional properties of weakly bound molecules.
Rather, the suppression of relaxation is often associated with Pauli blocking~\cite{petrov2004PRLb}, which
prevents two identical fermions from coexisting at short distances (where inelastic transitions occur). 
Although the Pauli blocking picture is not necessarily misleading, it omits the importance of the Efimov physics
and does not explain cases in which no fermions are involved in the collision process~\cite{dincao2008PRL}.

We derive the explicit scattering length dependence of relaxation of weakly bound molecules for repulsive three-body 
systems from Eq.~(\ref{Vrelaposweak}), with amplitudes [Eq.~(\ref{AVrelaposweak})] given by
\begin{align}
&|A_{{\rm I}}^{\gamma\beta}|^2=A^2(1-e^{-4\eta})\left(\frac{r_0}{a}\right)^{2p_0}(ka)^{2l+1}, 
\nonumber\\
&|A_{{\rm II}}^{\gamma\beta}|^2=B^2(1-e^{-4\eta})\left(\frac{r_0}{a}\right)^{2p_1}(ka)^{2l+1}, 
\end{align}
and $\phi_{\rm I}-\phi_{\rm II}=0$. In this case, the relaxation of weakly bound $s$-wave molecules is
given by
\begin{align}
\beta^{w}_{\rm rel}(a>0)=\frac{\hbar }{\mu_{ad}}&(1-e^{-4\eta})
\bigg[A_{\eta}^2\left(\frac{r_0}{a}\right)^{2p_0}+B_\eta^2\left(\frac{r_{0}}{a}\right)^{2p_{1}}
\nonumber\\
&~~+2A_{\eta}B_{\eta}\left(\frac{r_{0}}{a}\right)^{p_{0}+p_1}\bigg]k^{2\lambda}a^{2l+1},
\label{FermionicVrelaposweak}
\end{align}
where amplitudes $A_{\eta}$ and $B_{\eta}$ will depend on the details of the 
interatomic interactions; they originate from the contributions
of pathways (I) and (II) in Fig.~\ref{AllPaths}(e), respectively. 
The dominant contribution scales as $a^{2l+1-2p_{0}}$, 
implying that the suppression of relaxation occurs whenever $2l+1<2p_{0}$, and depends crucially on the 
strength of the Efimov repulsive potential, $p_{0}$ [see Eq.~(\ref{effermionic})].
As a result, we conclude that the suppression of relaxation is a direct consequence of Efimov physics.
For fermionic $FFF'$ ($J^{\pi}=0^+$, $l=0$) systems, for instance, we find that relaxation
scales as $a^{-3.332}$ (see Table~\ref{TabI} for the value for $p_0$). This result agrees with the prediction of 
Petrov {\it et al.}~\cite{petrov2004PRLb} (see also Refs.~\cite{dincao2005PRL,dincao2008PRA}) and 
is consistent with various experiments. 
For bosonic  $BBB'$ ($J^{\pi}=0^+$, $l=0$) systems, where only the $BB$ interaction is resonant (see Table~\ref{TabI}), 
relaxation scales as $a^{-1}$, serving as evidence that a Pauli-blocking mechanism is not the only factor
that determines the suppression of the relaxation rate.
In fact, the $a^{-1}$ suppression of relaxation
holds for {\em any} system of three dissimilar particles with a single resonant pair, $XYZ$($a_{XY}$), regardless of the mass 
ratios. For such cases, the suppression of $XY^*+Z$ collisions has also been observed experimentally
\cite{zirbel2008PRL,spiegelhalder2009PRL,khramov2012PRA,wu2012PRL,laurent2017PRL}. 

\begin{figure}[htbp]
\begin{center}
\includegraphics[width=3.2in,angle=0,clip=true]{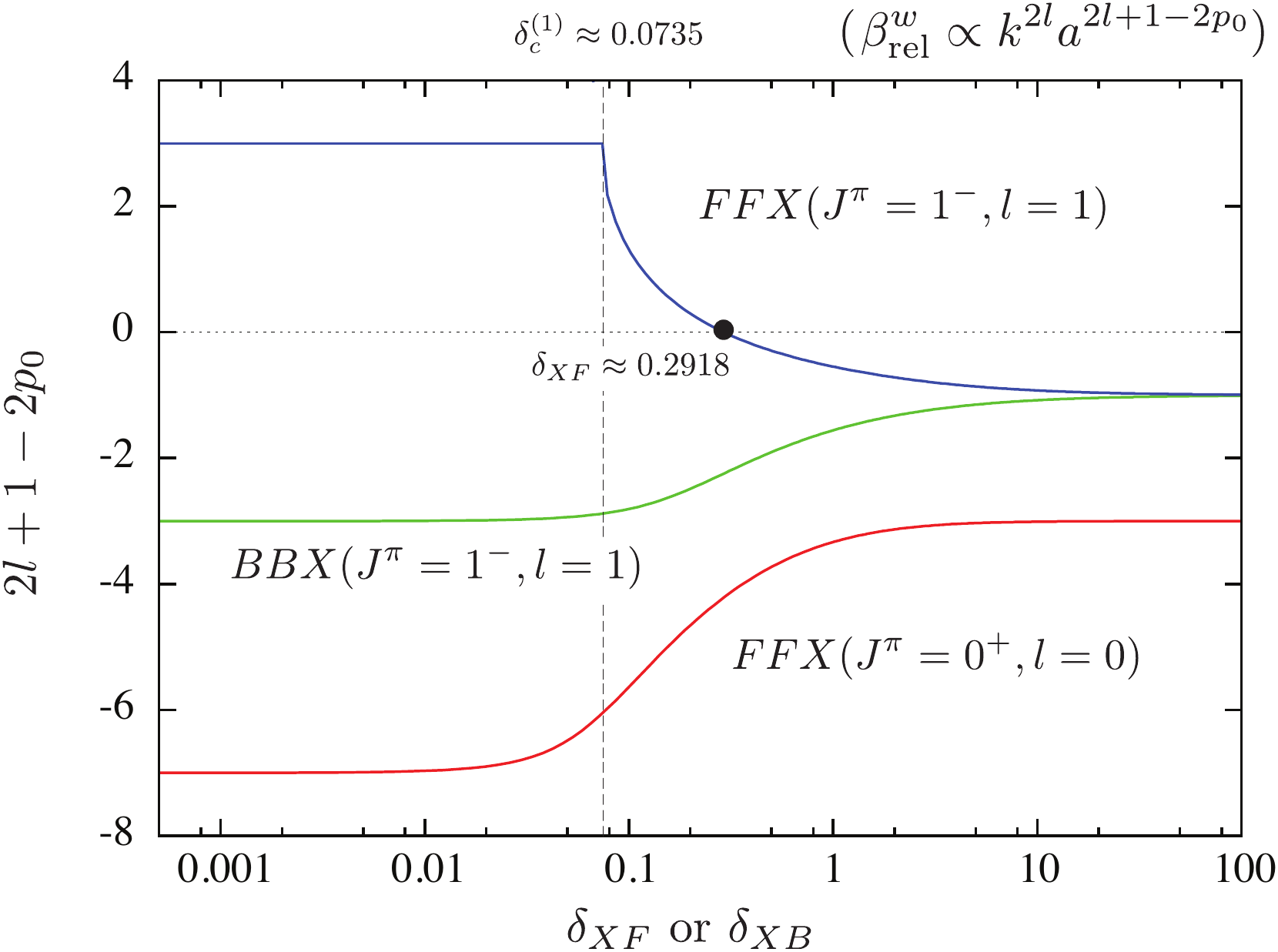}
\caption{Coefficient $2l+1-2p_0$ that determines the scattering length dependence for 
the relaxation of weakly bound molecules (\ref{FermionicVrelaposweak}) for different
values of the relevant mass ratios. For $FFX$ ($J^\pi=0^+$) and $BBX$ ($J^\pi=1^-$) 
systems, the suppression occurs for all mass ratios. For $FFX$ ($J^\pi=1^-$) systems, 
the suppression of relaxation only occurs for $\delta_{XF}\gtrsim0.2918$.}\label{VrelRep}
\end{center}
\end{figure}

Therefore, the result in Eq.~(\ref{FermionicVrelaposweak}) predicts a suppression of relaxation for {\em any} three-body repulsive system in 
which $2l+1<2p_{0}$ and that this suppression is not unique to fermionic $FFX$ systems. 
The same suppression can also happen, for instance, for high partial wave contributions in repulsive $BBX$ ($J^\pi\ne0$) systems.
This is illustrated in Fig.~\ref{VrelRep}, where the coefficient $2l+1-2p_0$, determining the scattering length dependence for 
the relaxation of weakly bound molecules (\ref{FermionicVrelaposweak}), is shown for $BBX$ and $FFX$ systems 
as a function of the relevant mass ratios. The coefficient $p_0$ is determined from Eqs.~(\ref{bc2ibBX}) and (\ref{bc2if}). 
For $FFX$ ($J^\pi=0^+$) and $BBX$ ($J^\pi=1^-$) systems, the suppression occurs for all mass ratios, but manifests
more strongly in the limit of small mass ratios. However, for $FFX$ ($J^\pi=1^-$) systems, the suppression of relaxation only occurs for 
$\delta_{XF}\gtrsim0.2918$, i.e., for mass ratios in which $2l+1-2p_0<0$. For $\delta_{XF}<0.2918$, relaxation increases 
with $a$, reaching its strongest dependence for $\delta_{XF}\leq\delta_{c}\approx0.0735$, when the system becomes attractive
and relaxation becomes proportional to $k^{2}a^{3}$ and displays resonant effects
associated with the formation of Efimov states [see Eq.~(\ref{BosonicVrelaposweak})].

In our model, the effects of universal KM three-body states \cite{kartavtsev2007JPB,endo2011FBS,endo2012PRA,kartavtsev2014YF}
(see Section \ref{Spectrum}) can also be incorporated.
For $BBX$ ($J^\pi>0$-even) and $FFX$ ($J^\pi$-odd) systems with $\delta>\delta_{c}$, the mass ratio
dependence of such states can lead to resonant effects in atom-molecule
collisions \cite{endo2011FBS}. 
The leading-order term of relaxation can be obtained following the same derivation
for Eq.~(\ref{BosonicVrelaposweak}), replacing $s_0\ln(a/a_*)\rightarrow\Psi_\delta$, and 
$e^{-4\eta}\rightarrow e^{-4\eta_0}=1-(1-e^{-4\eta})(r_0/a)^{2p_0}$, where $\Psi_{\delta}$ is a universal mass-ratio-dependent 
phase \cite{endo2011FBS}. By expanding the results for small 
$\eta_0\approx(1/4)(1-e^{-4\eta})(r_0/a)^{2p_0}$, one obtains
\begin{align} 
\beta^{w}_{{\rm rel}\delta}(a>0)=\frac{\hbar }{\mu_{ad}}&\bigg[
\frac{{A_{d}^2}\frac{(1-e^{-4\eta})}{8}(\frac{r_{0}}{a})^{2p_{0}}}{\sin^2\Psi_\delta+[\frac{(1-e^{-4\eta})}{4}(\frac{r_{0}}{a})^{2p_{0}}]^2}
\bigg]k^{2l}a^{2l+1}.
\label{FermionicVrelaposweakKM}
\end{align}
As a result, atom-molecule resonances occur whenever $\Psi_\delta$ is an integer multiple of $\pi$.
Note that the dependence on $r_0/a$ above derives from the existence of the Efimov repulsive barrier [Eq.~(\ref{effermionic})]
within the range $r_0\ll R \ll a$ that suppresses the probability of inelastic transitions. This suppression indicates that, in the limit of large $a$, 
KM states should be extremely long-lived.

%%%%%%%%%%%%%%%%%%%%%%%%%%%%%%%%%%%%%%%%%%%%%%%%%%%%
\subsection{Three-body elastic collisions}
\label{ElasticCol}

In this section, we analyze properties of three-body elastic processes relevant for ultracold quantum gases.
While the determination of three-body inelastic collisional properties is crucial for the understanding of the lifetime and 
stability of ultracold gases, three-body elastic collisions play a different role. Three-body elastic processes help to characterize
the many-body behavior of the system, in a way similar to two-body elastic processes. 
For instance, from the mean-field point of view, large and positive values of the two-body scattering length 
are generally associated with a collective regime characterized by strongly repulsive (mean-field) interactions.
In contrast, for large and negative scattering lengths the mean-field interactions now become strongly attractive. 
Therefore, understanding the physics controlling three-body elastic properties, whether or not they are affected
by the presence of three-body states, is of crucial importance.
Here, we use our pathway analysis to provide a clear physical picture for the relevant three-body elastic processes 
in strongly interacting ultracold gases. Specifically, we examine {\em three-body elastic scattering} (describing the collision between 
three free atoms) {\em atom-molecule elastic scattering}. 

For atom-molecule collisions, the description of elastic properties is essentially equivalent to 
a two-body process. Here, however, the asymptotic form of the effective potential for the case of 
weakly bound $s$-wave molecules is given by Eq.~(\ref{bc}) (with $l'=0$). In this case, 
the relative atom-molecule angular momentum $l$ defines the partial wave contributions for 
elastic scattering. As a result, the $l$-wave atom-molecule scattering length can be written as
\begin{align}  
a_{ad}^{(l)}&=-\lim_{k_{ad}\rightarrow0}{\rm Re}\left[\frac{\tan\delta^{(l)}_{ad}(k_{ad})}{k_{ad}^{2l+1}}\right],
\label{Aad}
\end{align}
where $k_{ad}^2=2\mu_{ad}(E-E^*_{v0})/\hbar^2$, with $E_{v0}^*\propto-1/a^2$,
is the wave vector associated with the initial atom-molecule collision channel, and $\delta_{ad}$ is
the corresponding phase-shift. (Note that although we refer to the above quantity as a scattering {\em length},
its actual unit is length$^{2l+1}$. Note also that, in the presence of inelastic processes, $\delta_{ad}$ is a complex 
quantity with its real and imaginary parts defining the elastic and inelastic properties of the collision 
process \cite{hutson2007NJP}, respectively.)
Similarly, for three-body elastic scattering, each three-body continuum state will define a partial-wave contribution.
In this case, however, the effective potential representing the collision of three free particles
is given asymptotically by Eq.~(\ref{ch}) and $l_{\rm eff}=\lambda+3/2$ will characterize the effective 
hyperangular momentum for the collision process. 
As a result, we define the three-body scattering length as
\begin{align} 
A_{3b}^{(\lambda)}&=-\lim_{k\rightarrow0}{\rm Re}\left[\frac{\tan\delta_{3b}^{(\lambda)}(k)}{k^{2\lambda+4}}\right],
\label{A3B}
\end{align}
where $k^2=2\mu E/\hbar^2$ is the wave vector associated with the initial collision channel, and $\delta_{3b}$ is
the corresponding three-body phase-shift.

In the next sections we determine the explicit dependence of the atom-molecule and three-body 
scattering lengths in terms of the two-body scattering length.
The analysis of the relevant collision pathways will, once again, reveal the pervasive influence of 
Efimov physics.

\subsubsection{Pathways for elastic collisions}

\begin{figure}[htbp]
\begin{center}
\includegraphics[width=3.3in,angle=0,clip=true]{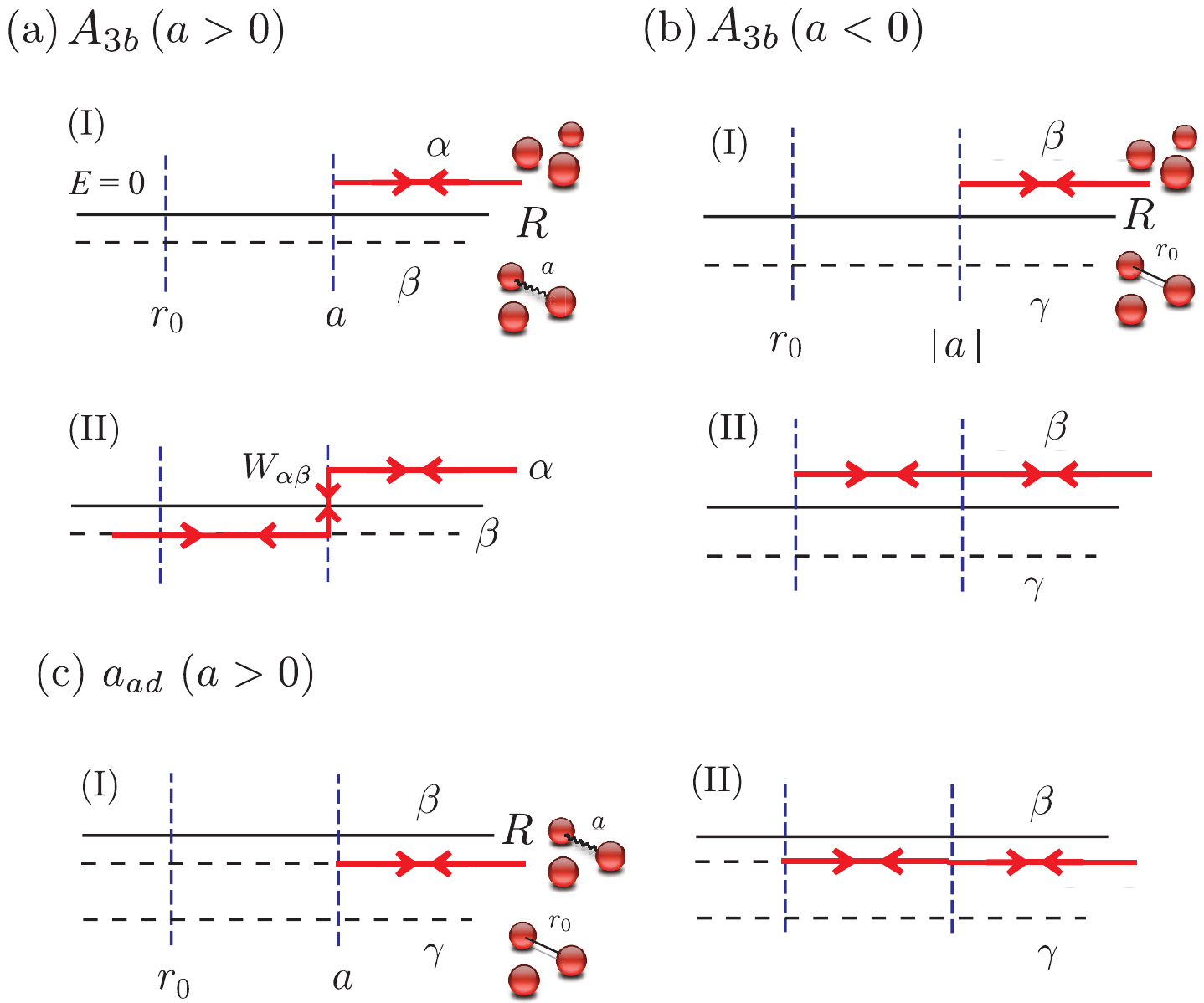}
\caption{Relevant pathways for three-body elastic scattering for (a) $a>0$ and (b)
$a<0$ as well as for (c) $a>0$ atom-molecule elastic scattering. In (a) and (c) channels $\alpha$ and $\beta$
correspond to the lowest continuum and the weakly bound molecular channels, respectively [see Figs.~\ref{SchPot}(a) and \ref{SchPot}(c)].
In (b), channels $\beta$ and $\gamma$ correspond to the lowest continuum and a deeply bound
molecular channel [see Figs.~\ref{SchPot}(b) and \ref{SchPot}(d)].}\label{AllPathsElastic}
\end{center}
\end{figure}

The dominant pathways for elastic processes considered in our analysis
are shown schematically in Fig.~\ref{AllPathsElastic}. As in inelastic processes (Section~\ref{InelasticCol}),
the pathways for elastic processes are independent of the category the three-body system belongs to, i.e., 
they do not depend on whether the three-body system has an attractive or repulsive $1/R^2$ effective
potential in the range $r_{0}\ll R \ll |a|$ (see Fig.~\ref{SchPot}). 
The relative importance of each pathway will, of course, depend on the specific form of the potentials,
and whether they allow for interference and resonant effects.
We show the pathways for three-body elastic processes
with $a>0$ and $a<0$ in Figs.~\ref{AllPathsElastic}(a) and (b), respectively. 
For $a>0$, while pathway (I) originates 
from a simple reflection near $R=a$, the pathway (II) involves an inelastic transition near $R=a$
between the initial three-body continuum channel $\alpha$ and the weakly bound molecular 
channel $\beta$. For $a<0$, pathway (I) also originates from a simple reflection near $R=|a|$,
while, now, pathway (II) involves a transmission through $R=|a|$, and special care should be taken
if three-body states exist, to allow for resonant transmission effects. In this case, one should also consider 
pathways similar to pathway (II) in Fig.~\ref{AllPathsElastic}(b) that accounts for multiple reflections 
between $R=r_0$ and $R=|a|$, which are characteristic of resonant phenomena in the semi-classical picture. 
Here, however, we properly address such resonant effects whenever necessary. For elastic atom-molecule collisions, 
the relevant pathways are shown in Fig.~\ref{AllPathsElastic}(c) and are in complete analogy to the pathways
for three-body elastic processes in Fig.~\ref{AllPathsElastic}(b), including the possibility for
resonant transmission effects whenever a three-body state occurs at the atom-molecule threshold.

To determine the three-body and atom-molecule scattering lengths in Eqs.~(\ref{Aad}) and (\ref{A3B}),
respectively, the elastic $T$-matrix element can be written simply as the sum of the
contributions for each pathway, i.e.,
\begin{align}
{\rm Re}[T_{}]&=-{\rm Re}\left[\sum_{j}|A_j^{}|e^{i\phi_{j}}\right]
=-\sum_{j}|A_j^{}|\cos\phi_{j},\label{GenT}
\end{align}
where $j=$ I, II, III, etc., and amplitudes $|A_j|$ determined
from the tunneling probability in Eq.~(\ref{TransProb}). 
The fact that, in the low-energy limit, the $T$-matrix element is simply $T=(e^{2i\delta}-1)/2i\approx\delta$, 
allows us to directly relate the amplitude and phases in Eq.~(\ref{GenT}) to the phase-shift, $\delta$.
[Note that the overall minus sign in the Eq.~(\ref{GenT}) was chosen to provide the correct sign for the 
phase-shift in our analysis.]
As in inelastic processes (Section~\ref{InelasticCol}), we define the scattering parameters 
in terms of constants whose specific values can be determined 
though numerical calculations or more sophisticated analytical approaches.

\subsubsection{Three-body elastic scattering} \label{A3bSection}

The problem of determining the three-body elastic properties, i.e., properties of three colliding particles 
at low energy, is a fundamental and highly nontrivial problem 
\cite{amado1970PRL,gerjuoy1970JPB,adhikari1971PRL,amado1971PRD,gerjuoy1971PTBSLA,gibson1972PRA,
adhikari1973PRD,matsuyama1991NPA,newton1972AP,newton1974JMP,newton1976PRA,potapov1977PRAa,potapov1977PRAb}. 
More recently, this problem has been revitalized due to the interest in understanding high-order three-body 
correlations in ultracold Bose gases in the strongly interacting regime \cite{braaten1999EPJB,braaten2002PRL,kohler2002PRL,
bulgac2002PRL,bedaque2003PRA,tan2008PRA,STanDAMOP2010,colussi2014PRL,colussi2016JPB},
where Efimov physics plays an important role.
The main difficulty in studying three-body elastic processes is the necessity for determining scattering contributions that originate from 
multiple scattering events where two atoms interact, while the third only spectates \cite{amado1970PRL}. 
These contributions must be subtracted out to determine scattering events that 
are truly of a three-body nature \cite{gerjuoy1971PTBSLA}, i.e., collision events in which all three atoms participate.
In our present formulation, as well as in Efimov's original analysis in Ref.~\cite{efimov1979SJNP}, 
we neglect such contributions, bypassing the issue of removing multiple scattering contributions by accounting 
for interactions only in the regions where all three atoms interact, i.e., the region in which the interactions are characterized by 
the Efimov potentials (\ref{efbosonic}) and (\ref{effermionic}) within the range $r_{0}\ll R\ll |a|$. In fact, within the adiabatic hyperspherical 
representation, it can be shown \cite{dincao2016Prep} that spectator multiple-scattering
contributions are manifested through high-order corrections to the asymptotic potentials in Eq.~(\ref{ch}), for $R\gg|a|$. 
Evidently, an accurate determination of the three-body scattering length [Eq.~(\ref{A3B})] requires a more rigorous 
approach such as the ones in Refs.~\cite{braaten2002PRL,tan2008PRA,STanDAMOP2010}.
Nevertheless, our simple formulation is capable of determining the $a$ dependence of the three-body scattering 
length while elucidating the role of Efimov physics in the collision process. 

For $a>0$, the relevant pathways for three-body elastic collisions are given 
in Fig.~\ref{AllPathsElastic}(a). In this case, the elastic phase-shift, which is associated to the relevant $T$-matrix 
for elastic scattering in the limit of $k\rightarrow0$, is simply determined by
\begin{align}
%{\rm Re}[T_{\alpha\alpha}]&=  {\rm Re}[\tan\delta_{3b}]=\nonumber\\
{\rm Re}[T_{\alpha\alpha}]&(\approx{\rm Re}[\delta_{3b}])=\nonumber\\
&=-|A_{{\rm I}}^{\alpha\alpha}|\cos\phi_{\rm I}-|A_{{\rm II}}^{\alpha\alpha}|\cos\phi_{\rm II},
\label{T2A3Bapos}
\end{align}
where $\phi_{\rm I}$ and $\phi_{\rm II}$ are the WKB phases in Eq.~(\ref{PhaseWKB}), corresponding
to pathways (I) and (II) in Fig.~\ref{AllPathsElastic}(a), respectively, with amplitudes determined from
\begin{align}
&|A_{{\rm I}}^{\alpha\alpha}|^2=P^{(\alpha)}_{r_{c}\rightarrow a}P^{(\alpha)}_{a\rightarrow r_{c}}P^{(\alpha)}_{r_c\rightarrow\infty},
\nonumber\\
&|A_{{\rm II}}^{\alpha\alpha}|^2=P^{(\alpha)}_{r_{c}\rightarrow a}P^{(\beta)}_{a\rightarrow r_{0}}
P^{(\beta)}_{r_{0}\rightarrow a}P^{(\alpha)}_{a\rightarrow r_{c}}P^{(\alpha)}_{r_c\rightarrow\infty}. \label{AA3Bapos}
\end{align}
For $a<0$, the entrance channel now corresponds to channel $\beta$ in Fig.~\ref{AllPathsElastic}(b), and the 
corresponding elastic phase-shift is given by
\begin{align}
%{\rm Re}[T_{\beta\beta}]&={\rm Re}[\tan\delta_{3b}]=\nonumber\\
{\rm Re}[T_{\beta\beta}]&(\approx{\rm Re}[\delta_{3b}])=\nonumber\\
&=-|A_{{\rm I}}^{\beta\beta}|\cos\phi_{\rm I}-|A_{{\rm II}}^{\beta\beta}|\cos\phi_{\rm II},
\label{T2A3Baneg}
\end{align}
with WKB phases $\phi_{\rm I}$ and $\phi_{\rm II}$ [Eq.~(\ref{PhaseWKB})] and amplitudes determined from
\begin{align}
&|A_{{\rm I}}^{\beta\beta}|^2=P^{(\beta)}_{r_{c}\rightarrow |a|}P^{(\beta)}_{|a|\rightarrow r_{c}}P^{(\beta)}_{r_c\rightarrow\infty},
\nonumber\\
&|A_{{\rm II}}^{\beta\beta}|^2=P^{(\beta)}_{r_{c}\rightarrow |a|}
P^{(\beta)}_{|a|\rightarrow r_{0}}
P^{(\beta)}_{r_{0}\rightarrow |a|}P^{(\beta)}_{|a|\rightarrow r_{c}}P^{(\beta)}_{r_c\rightarrow\infty}, \label{AA3Baneg}
\end{align}
corresponding to pathways (I) and (II) in Fig.~\ref{AllPathsElastic}(b), respectively.
Note that in Eqs.~(\ref{AA3Bapos}) and (\ref{AA3Baneg}), since there is no tunneling in the final (exit) channel between $R=r_c$ and $R=\infty$,
the corresponding probabilities will be simply a constant. 

At ultracold energies, the energy dependence of the elastic phase-shift in Eqs.~(\ref{T2A3Bapos}) and (\ref{T2A3Baneg})
originates solely from the terms describing tunneling from the classical turning point $r_{c}=(\lambda+2)/k$
to $|a|$. From the tunneling probability in Eqs.~(\ref{TransProb}) and the asymptotic form of the potentials (\ref{ch}), 
we obtain
\begin{equation}
P^{(\alpha)}_{r_{c}\rightarrow a}
\:{\propto}\:
(ka)^{2\lambda+4} ~~\mbox{and}~~
P^{(\beta)}_{r_{c}\rightarrow |a|}
\:{\propto}\: (k|a|)^{2\lambda+4},
\end{equation}
ensuring that the scattering length in Eq.~(\ref{A3B}) is constant in the limit of $k\rightarrow0$, 
and proportional to $a^{2\lambda+4}$. However, the other terms in Eqs.~(\ref{AA3Bapos}) and (\ref{AA3Baneg}) 
originating from pathway (II) will depend critically on the attractive or repulsive character of the three-body 
interactions within the region $r_0\ll R\ll|a|$. In fact, these are the terms through which the Efimov physics is 
introduced, and we will discuss them according to the category into which the three-body system falls.

\begin{figure}[htbp]
\begin{center}
\includegraphics[width=3.4in,angle=0,clip=true]{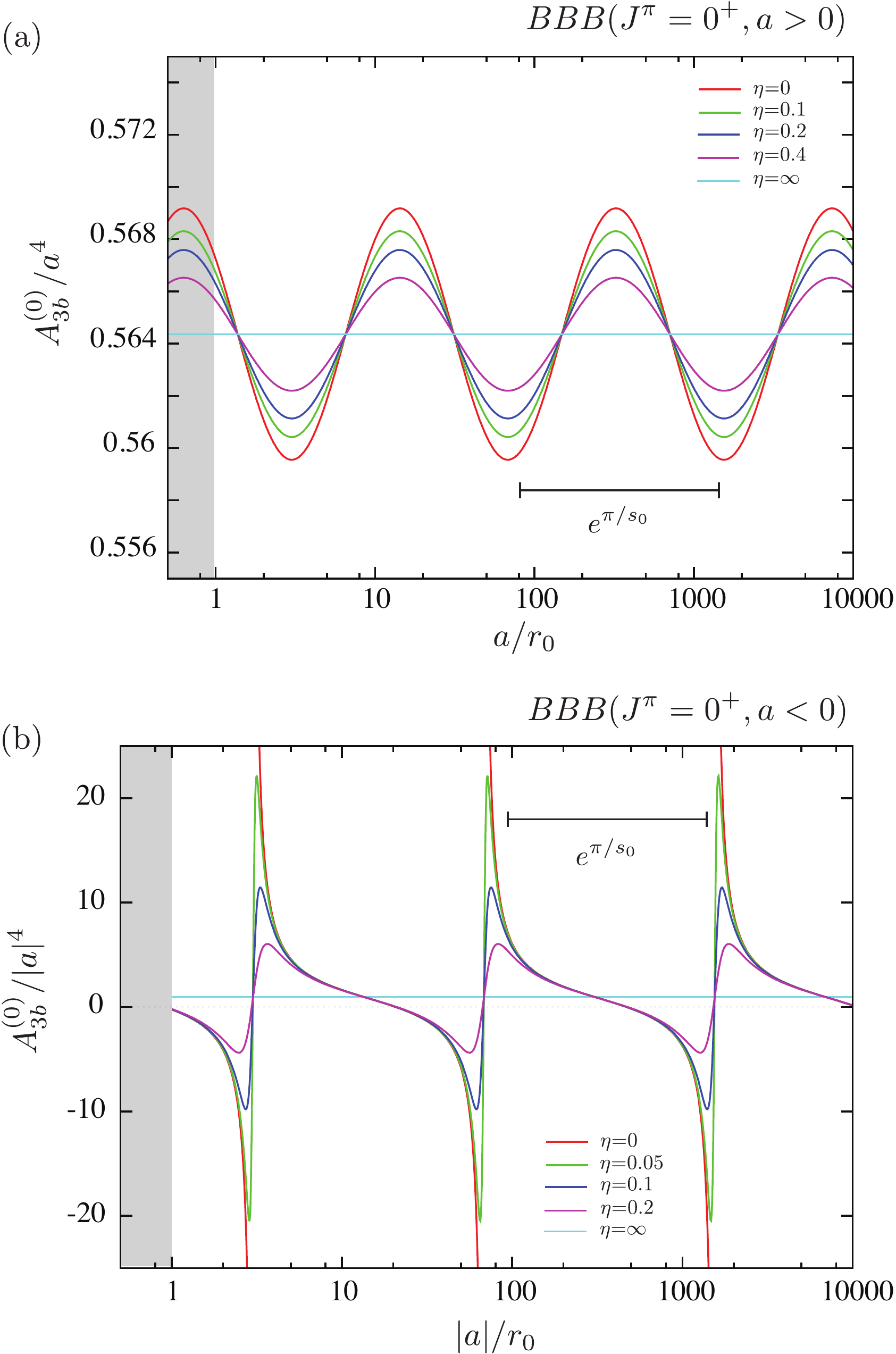}
\caption{Scattering length dependence of the three-body scattering length $A_{3b}$ for
$BBB$ ($J^\pi=0^+$, $\lambda=0$) \cite{braaten2002PRL,STanDAMOP2010} displaying the 
expected log-periodic features associated with the Efimov effect. (a) For $a>0$, $A_{3b}$ displays
interference effects manifested by the small amplitude oscillation, while for $a<0$, (b), $A_{3b}$ displays
resonant effects due to the formation of Efimov states. In the limit of strong decay ($\eta\gg1$), the Efimov
features are strongly suppressed, and $A_{3b}$ is given by its background value.
}\label{A3bAtt}
\end{center}
\end{figure}

\paragraph{Three-body elastic scattering for attractive systems.}
For three-body attractive systems, we have already shown that 
specific signatures in the inelastic scattering occur in relation to the formation of Efimov states. 
For elastic scattering, Efimov states should also lead to specific signatures. 
For $a>0$, the signatures of the Efimov effect in elastic scattering originate from the same
interference effect occurring for recombination into weakly bound molecules [see Eq.~(\ref{BosonicK3aposweak})]. 
Here, the amplitudes for elastic scattering from pathways (I) and (II) in Eq.~(\ref{AA3Bapos}) are given by
\begin{align}
&|A_{{\rm I}}^{\alpha\alpha}|^2=A^4(ka)^{4\lambda+8},
\nonumber\\
&|A_{{\rm II}}^{\alpha\alpha}|^2=B^4e^{-4\eta}(ka)^{4\lambda+8},\label{AA3bapos}
\end{align}
which include the probability of decay to deeply bound states via the contribution from pathway (II) in Fig.~\ref{AllPathsElastic}(a).
Since no phase is accumulated in pathway (I), we set $\phi_{\rm I}=0$, while for pathway (II) we obtain: $\phi_{\rm II}=2[s_0\ln(a/r_0)+\Phi]-\pi/2$, 
according to Eq.~(\ref{PhaseWKB}). In Eq.~(\ref{AA3bapos}) the coefficients $A$ and $B$ are expected to be universal and related to
the reflection and transmission probability near $R=a$.
Therefore, replacing Eq.~(\ref{AA3bapos}) and corresponding phases in Eq.~(\ref{T2A3Bapos}),
allows us to determine the three-body scattering length in Eq.~(\ref{A3B}) as
\begin{align}
A_{3b}^{(\lambda)}=&\bigg[\left(A^2-\frac{B^2}{e^{2\eta}}\right)
+\frac{2B^2}{e^{2\eta}}\sin^{2}[s_{0}\ln(a/a_+)-\frac{\pi}{4}]\bigg]a^{2\lambda+4},
\label{A3BBosonicApos}
\end{align}
with the three-body parameter $a_+$ defined identically as we showed in the recombination into weakly
bound molecules in Eq.~(\ref{BosonicK3aposweak}). Note, however, that the result of Eq.~(\ref{A3BBosonicApos}) is $\pi/4$ out-of-phase from
Eq.~(\ref{BosonicK3aposweak}). Our result in Eq.~(\ref{A3BBosonicApos})
is general and displays interference effects whenever $a$ increases by the geometric factor, $e^{\pi/s_0}$,
for all three-body attractive systems. In the limit of weak decay ($\eta\ll1$),
Eq.~(\ref{A3BBosonicApos}) is consistent with the results from Refs.~\cite{braaten2002PRL,STanDAMOP2010} 
for $BBB$ ($J^\pi=0^+$ and $\lambda=0$) systems, with $(A^2-B^2)\approx 1.22(4\pi-3\sqrt{3})/(2\sqrt{3}\pi)$, 
and $2B^2\approx0.021(4\pi-3\sqrt{3})/(2\sqrt{3}\pi)$. 
Figure \ref{A3bAtt}(a) shows the scattering length dependence of the three-body scattering length $A_{3b}$ ($a>0$) 
for $BBB$ ($J^\pi=0^+$, $\lambda=0$) systems \cite{braaten2002PRL,STanDAMOP2010}. This dependence displays the 
expected log-periodic features associated with the Efimov effect, but with a very small amplitude (associated
due to the small value for $B^2$ for identical bosons). In fact, as one can see in Fig.~\ref{A3bAtt}(a),
in the limit of strong decay ($\eta\gg1$), the sinusoidal term in Eq.~(\ref{A3BBosonicApos}) is completely 
suppressed. 
Although here we don't determine the explicit values for the coefficients $A$ and $B$ in Eq.~(\ref{A3BBosonicApos}), one would
in general expect that $A^2-B^2>0$ since the value of $B^2$ depends on the (normally weakly) inelastic transition 
probability near $R=a$, associated with the pathway (II) in Fig.~\ref{AllPathsElastic}(a). In this case,
from Eq.~(\ref{A3BBosonicApos}) we expect that, in general, $A_{3b}>0$, regardless of the value of $\eta$. 

For $a<0$, our pathway analysis [see pathways (I) and (II) in Fig.~\ref{AllPathsElastic}(b)] indicates 
that the main physical mechanism that controls the elastic process is resonant scattering associated with the 
formation of Efimov states in the entrance channel $\beta$. 
In this case, which is similar to the resonant effects in inelastic processes [Eq.~(\ref{BosonicK3anegdeep})],
the contribution from pathway (II) in Eq.~(\ref{T2A3Baneg}) must be generalized to
account for the multiple rattles within the range $r_0\ll R\ll |a|$, characteristic of the resonant state.
This effect can be simply incorporated into our analysis by writing Eq.~(\ref{T2A3Baneg}) as
\begin{align}
%{\rm Re}[T_{\beta\beta}]&={\rm Re}[\tan\delta_{3b}]=\nonumber\\
{\rm Re}[T_{\beta\beta}]&(\approx{\rm Re}[\delta_{3b}])=\nonumber\\
&=-|A_{{\rm I}}^{\beta\beta}|\cos\phi_{\rm I}-\sum_j|A_{{\rm II},j}^{\beta\beta}|\cos\phi_{{\rm II},j},
\label{T2A3BanegRes}
\end{align}
with amplitudes determined from
\begin{align}
&|A_{{\rm I}}^{\beta\beta}|^2=A^4(k|a|)^{4\lambda+8},
\nonumber\\
&|A_{{\rm II},j}^{\beta\beta}|^2=B^4(e^{-4\eta})^j(k|a|)^{4\lambda+8},\label{AmpRes}
\end{align}
where $j=1,2,...,\infty$, and corresponding WKB phases
$\phi_{\rm I}=0$ and $\phi_{{\rm II},j}=(2j)[s_0\ln(|a|/r_0)+\Phi]-\pi/2$.
The coefficients $A$ and $B$ in Eq.~(\ref{AmpRes}) are expected to be universal, since both related to the physics near $R=|a|$.
Replacing the Eq.~(\ref{AmpRes}) into Eq.~(\ref{T2A3BanegRes}), performing the summations 
analytically and substituting the results into Eq.~(\ref{A3B}), one finally obtains
\begin{align}
A_{3b}^{(\lambda)}=
\left[A^2+\frac{B^2}{4}\frac{\sin[2s_{0}\ln(|a/a_-|)]}
{\sin^2[s_{0}\ln(|a/a_-|)]+\sinh^2\eta}\right]|a|^{2\lambda+4},
\label{A3BBosonicAneg}
\end{align}
where $s_0\ln(|a/a_-|)=s_0\ln(|a|/r_0)+\Phi$. The first term in Eq.~(\ref{A3BBosonicAneg}) originates from
the contribution of pathway (I) and corresponds to a hard-sphere (of radius $R=|a|$) scattering. The second term,
originating from the pathway (II), corresponds to resonant transmission effects.
The expression in Eq.~(\ref{A3BBosonicAneg}) is consistent with the results from Refs.~\cite{braaten2002PRL,STanDAMOP2010} 
for $BBB$ ($J^\pi=0^+$ and $\lambda=0$) systems, where it was found that $A^2\approx 1.23(4\pi-3\sqrt{3})/(2\sqrt{3}\pi)$ 
and $B^2/4\approx3.16(4\pi-3\sqrt{3})/(2\sqrt{3}\pi)$, obtained in the absence of inelastic decay ($\eta=0$).
In this regime, $A_{3b}$ diverges exactly at values $a=a_-(e^{\pi/s_0})^n$, where $n=$1, 2, 3, ..., corresponding to the
value of $a$ in which an Efimov state becomes bound. 
Figure \ref{A3bAtt}(b) shows the scattering length dependence of the three-body scattering length $A_{3b}$ ($a<0$) 
for $BBB$ ($J^\pi=0^+$, $\lambda=0$) systems \cite{braaten2002PRL,STanDAMOP2010}. It displays the 
expected log-periodic resonant features associated with the Efimov effect. 
Note, however, for any nonzero inelastic decay ($\eta\ne0$), 
$A_{3b}$ does not diverge on resonance. On resonance, it rather assumes its background value
(proportional to $A^2$), in agreement with the analysis of Ref.~\cite{hutson2007NJP}. 
Note also that, in the limit of strong decay ($\eta\gg1$), the second term in Eq.~(\ref{A3BBosonicAneg}) vanishes, and 
$A_{3b}$ is given by its background value for all values of $a$.

\paragraph{Three-body elastic scattering for repulsive systems.}
For the three-body repulsive systems the elastic properties can be determined following the 
same procedure described in the previous section. For $a>0$, the amplitudes [Eq.~(\ref{AA3Bapos})] for elastic scattering from 
pathways (I) and (II) in Eq.~(\ref{T2A3Bapos}) are given by
\begin{align}
&|A_{{\rm I}}^{\alpha\alpha}|^2=A^4(ka)^{4\lambda+8},
\nonumber\\
&|A_{{\rm II}}^{\alpha\alpha}|^2=B^4e^{-4\eta}\left(\frac{r_0}{a}\right)^{4p_0}(ka)^{4\lambda+8},
\end{align}
and we set $\phi_{\rm I}=\phi_{\rm II}=0$, since all potentials are repulsive and, consequently, no phase is accumulated.
In this case, the three-body scattering length is determined from Eq.~(\ref{A3B}) to be
\begin{align}
A_{3b}^{(\lambda)}=\left[A^2+B^2e^{-2\eta}\left(\frac{r_0}{a}\right)^{2p_0}\right]a^{2\lambda+4},
\label{A3BFermionicApos}
\end{align}
with both $A$ and $B$ being universal constants dependent on the physics near $R=a$. The inclusion 
of interference effects related to KM states follows an analogous procedure to the derivation
of Eq.~(\ref{A3BBosonicApos}). But now, we replace $s_0\ln(a/a_+)\rightarrow\Phi_\delta+\pi/2$, and 
$e^{-4\eta}\rightarrow e^{-4\eta_0}=1-(1-e^{-4\eta})(r_0/a)^{2p_0}$. Then, expanding for small values of
$\eta_0$, we obtain
\begin{align}
A_{3b\delta}^{(\lambda)}&=\bigg[\left(A^2-B^2\left[1-\frac{1-e^{-4\eta}}{2}\left(\frac{r_0}{a}\right)^{2p_0}\right]\right)
\nonumber\\
&+2B^2\left[1-\frac{1-e^{-4\eta}}{2}\left(\frac{r_0}{a}\right)^{2p_0}\right]\cos^2(\Phi_\delta-\pi/4)\bigg]a^{2\lambda+4},
\label{A3BFermionicAposKM}
\end{align}
where $\Phi_\delta$ is the same universal, mass-ratio dependent phase appearing in Eq.~(\ref{FermionicK3aposweakKM}).
For $a<0$, the amplitudes [Eq.~(\ref{AA3Baneg})] for elastic scattering from pathways (I) and (II) in Eq.~(\ref{T2A3Baneg}) 
are given by
\begin{align}
&|A_{{\rm I}}^{\beta\beta}|^2=A^4(k|a|)^{4\lambda+8},
\nonumber\\
&|A_{{\rm II}}^{\beta\beta}|^2=B^4e^{-4\eta}\left(\frac{r_0}{|a|}\right)^{4p_0}(k|a|)^{4\lambda+8},
\end{align}
with phases $\phi_{\rm I}=\phi_{\rm II}=0$. In this case, we find the three-body scattering length to be
\begin{align}
A_{3b}^{(\lambda)}&=\left[A^2+B^2e^{-2\eta}\left(\frac{r_0}{|a|}\right)^{2p_0}\right]|a|^{2\lambda+4}.
\label{A3BFermionicAneg}
\end{align}
Note that since the second terms in Eqs.~(\ref{A3BFermionicApos}), (\ref{A3BFermionicAposKM}), and (\ref{A3BFermionicAneg}),
are strongly suppressed for large values of $a$, the three-body elastic scattering problem for repulsive
systems can be simply seen as a problem of scattering from a {\em hard-hypersphere} of hyperradius $|a|$.

\subsubsection{Atom-molecule elastic scattering} \label{AadSection}

The problem of determining the elastic properties of collisions involving weakly bound dimers has been the subject
of various works where both atom-dimer \cite{petrov2003PRA,braaten2004PRA,mora2004PRL,dincao2008PRA,helfrich2009EPL,
levinsen2009PRL,helfrich2010PRA,helfrich2011JPB,levinsen2011EPJD} and molecule-molecule
\cite{petrov2004PRLb,petrov2005PRA,marcelis2008PRA,dincao2009PRLb,dincao2009PRA} elastic
collisions have been studied. In this section we only focus on the properties of elastic atom-molecule collisions, 
although some of the ideas discussed here can be applied for elastic molecule-molecule collisions~\cite{rittenhouse2011JPB}.

According to the pathways shown in Fig.~\ref{AllPathsElastic}(c), the entrance channel corresponds to channel 
$\beta$, and the corresponding atom-molecule elastic phase-shift is given by
\begin{align}
%{\rm Re}[T_{\beta\beta}]&={\rm Re}[\tan\delta_{ad}]=\nonumber\\
{\rm Re}[T_{\beta\beta}]&(\approx{\rm Re}[\delta_{ad}])=\nonumber\\
&=-|A_{{\rm I}}^{\beta\beta}|\cos\phi_{\rm I}-|A_{{\rm II}}^{\beta\beta}|\cos\phi_{\rm II},
\label{T2Aadapos}
\end{align}
where WKB phases $\phi_{\rm I}$ and $\phi_{\rm II}$ are determined from Eq.~(\ref{PhaseWKB}) and amplitudes are given by
\begin{align}
&|A_{{\rm I}}^{\beta\beta}|^2=P^{(\beta)}_{r_{c}\rightarrow a}P^{(\beta)}_{a\rightarrow r_{c}}P^{(\beta)}_{r_c\rightarrow\infty},
\nonumber\\
&|A_{{\rm II}}^{\beta\beta}|^2=P^{(\beta)}_{r_{c}\rightarrow a}
P^{(\beta)}_{a\rightarrow r_{0}}
P^{(\beta)}_{r_{0}\rightarrow a}P^{(\beta)}_{a\rightarrow r_{c}}P^{(\beta)}_{r_c\rightarrow\infty}. \label{AAadapos}
\end{align}
The amplitudes for both pathways (I) and (II) in Eq.~(\ref{AAadapos}) have contributions from tunneling from the classical turning point $r_{c}=(l+1/2)/k$ to $R=a$, 
determining the low-energy dependence of the phase-shift
\begin{equation}
P^{(\beta)}_{r_{c}\rightarrow a}\:{\propto}\:
(ka)^{2l+1}.
\end{equation}
The other terms in Eq.~(\ref{AAadapos}), however, depend critically on the attractive or repulsive character of the three-body 
interactions controlled by the Efimov physics. 

\paragraph{Atom-molecule elastic scattering for attractive systems.}

\begin{figure}[htbp]
\begin{center}
\includegraphics[width=3.4in,angle=0,clip=true]{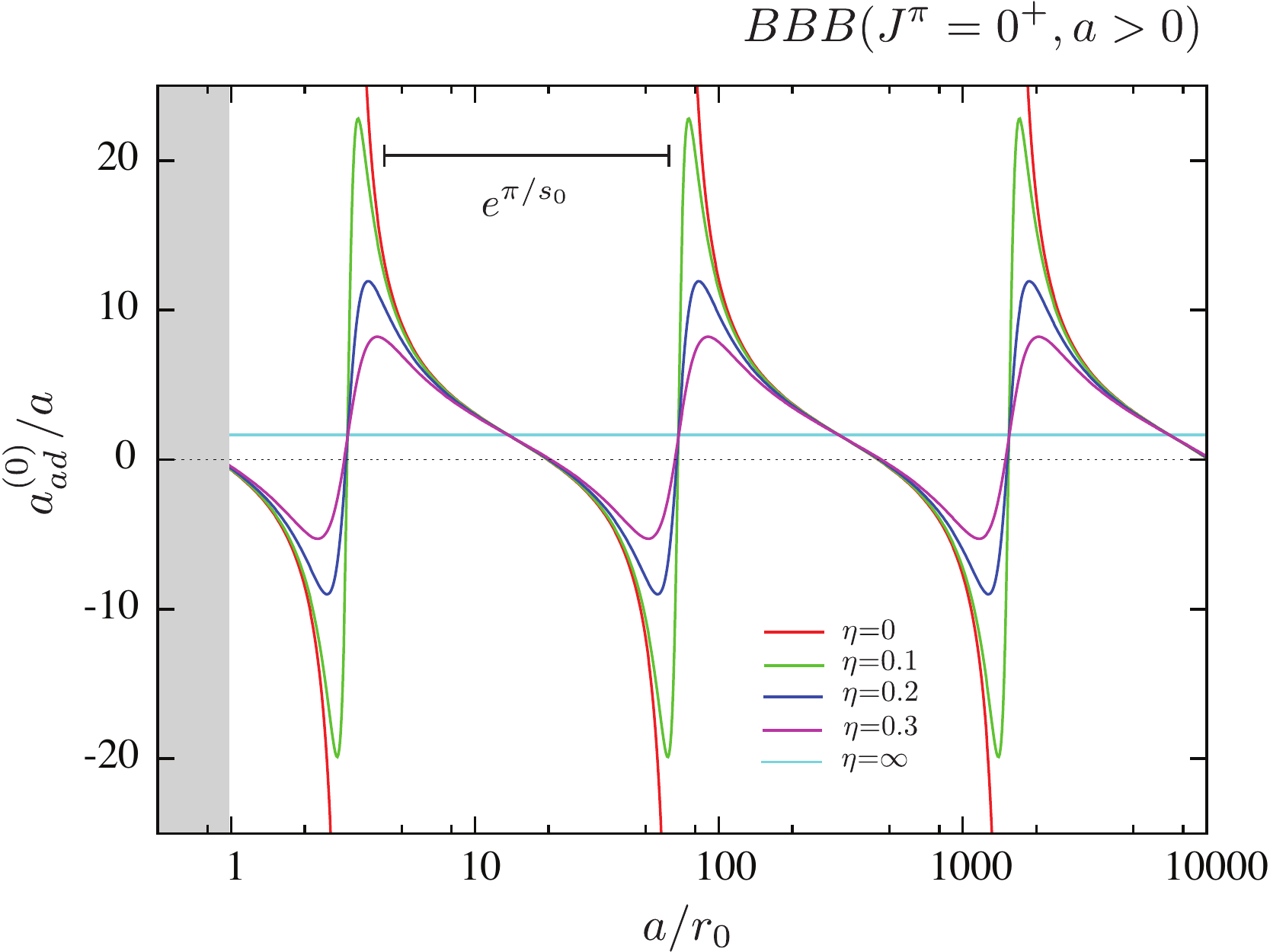}
\caption{Scattering length dependence of the atom-molecule
scattering length, $a_{ad}$, for $BBB$ ($J^\pi=0^+$, $l=0$) systems \cite{braaten2004PRA} displaying the log-periodic resonant structure associated 
with the formation of Efimov states at the collision threshold.
In the absence of inelastic decay ($\eta=0$), $a_{ad}$ diverges at values $a=a_*(e^{\pi/s_0})^n$ ($n=$1, 2, 3, etc),
while for $\eta\ne0$ decay to deeply bound molecules suppresses the poles in $a_{ad}$ \cite{hutson2007NJP}. 
}\label{AadAtt}
\end{center}
\end{figure}

For attractive systems, the main expected features in elastic atom-molecule collisions are resonant effects
due to the appearance of Efimov states at the atom-molecule threshold [see Fig.~\ref{SchPot}(a)], i.e,
corresponding to the entrance channel $\beta$ in Fig.~\ref{AllPathsElastic}(c). 
While the contribution to Eq.~(\ref{T2Aadapos}) from pathway (I) corresponds to a hard-sphere (of radius $R=a$) scattering,
the resonant contribution originates from pathway (II) and is obtained by rewriting Eq.~(\ref{T2Aadapos}) to include the multiple
rattles within the range $r_0\ll R\ll a$, i.e.,
\begin{align}
%{\rm Re}[T_{\beta\beta}]&={\rm Re}[\tan\delta_{ad}]=\nonumber\\
{\rm Re}[T_{\beta\beta}]&(\approx{\rm Re}[\delta_{ad}])=\nonumber\\
&=-|A_{{\rm I}}^{\beta\beta}|\cos\phi_{\rm I}-\sum_j|A_{{\rm II},j}^{\beta\beta}|\cos\phi_{{\rm II},j},
\label{T2AadaposRes}
\end{align}
where the amplitudes are determined from
\begin{align}
&|A_{{\rm I}}^{\beta\beta}|^2=A^4(ka)^{4l+2},
\nonumber\\
&|A_{{\rm II}}^{\beta\beta}|^2=B^4(e^{-4\eta})^j(ka)^{4l+2},\label{Ampad}
\end{align}
with corresponding WKB phases $\phi_{\rm I}=0$, and $\phi_{{\rm II},j}=(2j)[s_0\ln(a/r_0)+\Phi]-\pi/2$.
In this case, substituting the amplitudes in Eq.~(\ref{Ampad}) into Eq.~(\ref{T2AadaposRes}), we obtain the $l$-wave atom-molecule 
scattering length [Eq.~(\ref{Aad})] as
\begin{align}
a_{ad}^{(l)}=
\left[A^2+\frac{B^2}{4}\frac{\sin[2s_{0}\ln(a/a_*)]}
{\sin^2[s_{0}\ln(a/a_*)]+\sinh^2\eta}\right]a^{2l+1},
\label{ScalAttractive}
\end{align}
where $s_0\ln(a/a_*)=s_0\ln(a/r_0)+\Phi$. Both coefficients $A$ and $B$ are expected to be universal, as they are determined by the physics 
near $R=|a|$. 
Equation (\ref{ScalAttractive}) is consistent with Ref.~\cite{braaten2004PRA,helfrich2010PRA,helfrich2011JPB}. 
For $BBB$ systems ($J^\pi=0^+$ and $l=0$), the coefficients $A$ and $B$ were determined numerically to be $A^2\approx1.46$ and
$B^2/4\approx2.15$ in Ref.~\cite{braaten2004PRA}; for heteronuclear $BBX$ systems values for such universal constants 
can be found in Ref.~\cite{helfrich2010PRA}. Figure \ref{AadAtt} shows the $a$ dependence of the atom-molecule
scattering length for $BBB$ ($J^\pi=0^+$, $l=0$) systems \cite{braaten2004PRA} displaying the log-periodic resonant structure associated 
with the formation of Efimov states at the collision threshold.
As expected, in the absence of inelastic decay ($\eta=0$) $a_{ad}$ diverges at values $a=a_*(e^{\pi/s_0})^n$ ($n=$1, 2, 3, etc.),
corresponding to the values of $a$ in which an Efimov state becomes bound. For $\eta\ne0$, however, decay to deeply
bound molecules suppresses the poles in $a_{ad}$, with $a_{ad}$ assuming its background value, proportional to $A^2$ on 
resonance \cite{hutson2007NJP}. 

\paragraph{Atom-molecule elastic scattering for repulsive systems.}
For three-body repulsive systems, since the repulsive interaction holds for $r_{0}\ll R\ll a$, 
the expectation is that the atom-molecule scattering length is determined mostly from a hard-sphere problem 
with radius $a$. Within our model, hard-sphere scattering corresponds to the contributions in Eq.~(\ref{T2Aadapos}) 
associated with the pathway (I) in Fig.~\ref{AllPathsElastic}(c), while pathway (II) leads to a correction to the hard-sphere scattering. 
In this case, the amplitudes for each pathway are given by
\begin{align}
&|A_{{\rm I}}^{\beta\beta}|^2=A^4(ka)^{4l+2},
\nonumber\\
&|A_{{\rm II}}^{\beta\beta}|^2=B^4e^{-4\eta}\left(\frac{r_0}{a}\right)^{4p_0}(ka)^{4l+2},
\end{align}
with corresponding WKB phases $\phi_{\rm I}=\phi_{{\rm II}}=0$, since no phase is
accumulated within the pathways. From the results above, along with Eq.~(\ref{T2Aadapos}),
the atom-molecule scattering length [Eq.~(\ref{Aad})] can be determined to be
\begin{align}
a_{ad}^{(l)}=
\left[A^2+B^2e^{-2\eta}\left(\frac{r_0}{a}\right)^{2p_0}\right]a^{2l+1},
\label{ScalRepulsive}
\end{align}
with $A$ and $B$ being universal constants. This result is consistent with the findings for
$FFF'$ systems ($J^{\pi}=0^+$ and $l=0$) \cite{petrov2003PRA,dincao2008PRA}, 
where $A^2\approx1.2$, and for $FFX$ systems \cite{petrov2003PRA}. 
Although our approach only determines the overall dependence of the atom-molecule scattering length
on $a$, the precise value of the universal constants are extremely important.
In particular, for $FFF'$ systems the values for atom-molecule \cite{petrov2003PRA,dincao2008PRA}
and molecule-molecule \cite{petrov2004PRLb,dincao2009PRA} scattering lengths are important for
describing the BEC-BCS crossover physics \cite{eagles1969PR,leggett1980JPC}.

For the cases in which universal KM states \cite{kartavtsev2007JPB,endo2011FBS,endo2012PRA,kartavtsev2014YF}
can occur (see Section \ref{Spectrum}), the derivation of the atom-dimer scattering length closely follows
the derivation of Eq.~(\ref{ScalAttractive}), but replaces $s_0\ln(a/a_*)\rightarrow\Psi_\delta$, where $\Psi_\delta$
is a universal mass-ratio dependent phase \cite{endo2011FBS}, and 
$e^{-4\eta}\rightarrow e^{-4\eta_0}=1-(1-e^{-4\eta})(r_0/a)^{2p_0}$. Expanding the results for small 
$\eta_0\approx(1/4)(1-e^{-4\eta})(r_0/a)^{2p_0}$, one finally obtains
\begin{align}
a_{ad\delta}^{(l)}=
\left[A^2+\frac{B^2}{4}\frac{\sin(2\Psi_\delta)}
{\sin^2\Psi_\delta+[\frac{(1-e^{-4\eta})}{4}(\frac{r_{0}}{a})^{2p_{0}}]^2}\right]a^{2l+1}.
\label{ScalRepulsiveKM}
\end{align}
In this case, the atom-molecule scattering length also displays resonant effects whenever a KM state 
is formed, in agreement with Ref.~\cite{endo2011FBS}.
The dependence on $r_0/a$ in Eq.~(\ref{ScalRepulsiveKM}), however, leads to narrow resonant features in $a_{ad}$, which 
is also consistent with the predicted long lifetime of KM states \cite{kartavtsev2007JPB,endo2011FBS,endo2012PRA,kartavtsev2014YF}.

.pdf% !TEX root = ./TutorialJPB.tex

%%%%%%%%%%%%%%%%%%%%%%%%%%%%%%%%%%%%%%%%%%%%%%%%%%%%
%\section{Efimov physics in ultracold quantum gases} \label{EfimovPhysics}
\section{Three-body collisions in ultracold gases} \label{EfimovPhysics}

The analysis of three-body scattering processes presented in the previous sections
has benefited from our ability to classify three-body systems according to
their characteristic attractive and repulsive interactions. This classification allows us to state that systems 
falling into the same category have essentially the same scattering properties.
Scattering observables for three-body attractive systems display clear signatures
of Efimov effects via the occurrence of interference and resonant effects
whenever the scattering length changes by the characteristic Efimov geometric factor, $e^{\pi/s_0}$.
In contrast, for repulsive systems the characteristic signature of Efimov physics
is the (relative or absolute) suppression of various inelastic processes. 

In the context of ultracold gases, among all possible contributions to the
collision processes, only a few (if not only one) are expected to dominate because of
their corresponding low-energy behavior, as dictated by Wigner's threshold laws. 
These dominant scattering processes will, ultimately, determine the stability and lifetime
of the ultracold gas. 
Therefore, it is crucial to build up an overall picture to evaluate {\em how} and {\em when} 
Efimov physics can be experimentally accessible and whether its appearance will be detrimental or beneficial to the 
system's stability. 
Moreover, for any finite energy system, one also must understand the regime in which
finite temperature effects become important.
In other words, one needs to understand the breakdown of the {\em threshold regime}, which ensures 
the applicability and usefulness of the Wigner threshold laws. Consequently, in this section, we 
provide a summary of the relevant three-body scattering rates and an analysis of the dominant scattering 
contributions and the collisional stability of various types of ultracold gases. We also provide
an analysis of the notion of a threshold regime and finite temperature effects on three-body scattering observables.

From the perspective of the experimental observability of Efimov physics in ultracold gases, signatures of Efimov 
physics for three-body repulsive systems can be easily identified through the suppression of atomic 
and molecular losses. In contrast, the observation of signatures of Efimov physics for three-body attractive  
systems is more delicate, since they are accompanied with generally strong losses, and they are more sensitive 
to finite temperature effects. 
For such cases, the observation of geometric scaling for Efimov states, i.e., 
the observation of multiple Efimov features in atomic and molecular losses, depend strongly
on the experimental ability to keep the gas within the threshold regime \cite{dincao2004PRL,dincao2006PRA,dincao2009JPB}. 
In fact, at any finite temperature, the threshold regime persists only within a range of 
scattering lengths, which define a maximum value of the scattering length in which the interference and resonant effects 
associated with Efimov states can be observed \cite{dincao2004PRL}. 
Outside the threshold regime, the system enters into the so-called {\em unitarity regime}, 
where scattering rates becomes independent of $a$, and all Efimov features are 
simply washed away \cite{dincao2004PRL}. Therefore, finite temperature effects sets a limit on the number of consecutive features that can
be experimentally accessible \cite{dincao2006PRA}.

\begin{table*}[htbp]
\caption{Energy and scattering length scaling laws for three-body inelastic
  rates for systems in which only one scattering length (indicated in the Table) is resonant.
  For a given value of $J^\pi$, the attractive and repulsive character of the three-body systems is indicated by the subscripts A and R, respectively, added to the
  value of $J^\pi$.
  Note that for  $BBX$ ($J^\pi=2^+$) and $FFX$ ($J^\pi=1^-$) systems, the rates are given for
  $\delta<\delta_{c}$ and $\delta>\delta_{c}$, respectively, emphasizing the change 
  in the scattering properties due to the appearance of the Efimov Effect for $\delta<\delta_{c}$. (For $J^{\pi}=1^{-}$ we 
  have $\delta_{c}\approx0.073492$, and for $J^{\pi}=2^{+}$ we have $\delta_{c}\approx0.025887$.) 
  Boldface type indicates the leading contribution at threshold.}
\label{TabRates}
{\begin{ruledtabular}
\begin{tabular}{ccccccccc}
& & 
& \multicolumn{1}{c}{\underline{\hspace{0.3in}$\beta^{w}_{\rm rel}$\hspace{0.3in}}} 
&   
& \multicolumn{1}{c}{\underline{\hspace{0.3in}$K^{w}_{3}$\hspace{0.3in}}}
& \multicolumn{2}{l}{\underline{\hspace{0.6in}$K^{d}_{3}$\hspace{0.8in}}}
& {\underline{\hspace{0.3in}$D^{w}_{3}$\hspace{0.3in}}}  \\ 
System & $J^{\pi}_{\rm C}$ & $l$ & $a>0$ & $\lambda_{\rm m}$ & $a>0$ & $a>0$ & $a<0$ & $a>0$ \\ \hline
$BBB$($a_{BB}$)
   & $0^+_{\rm A}$ 
   	   & 0
           & \boldmath$P_{ad}^d(a)${$a$}
   	   & 0
           & \boldmath$M_{3b}^{w}(a)${$a^{4}$}  
           & \boldmath{$a^{4}$}    
           & \boldmath$P_{3b}^{d}(a)${$|a|^{4}$}  
           & \boldmath$M_{3b}^{w}(a)${$k^4$$a^{5}$}  \\  [0.00in]
   & $1^-_{\rm R}$
   	   & 1
	   & $k^2$$a^{-2.728}$ 
   	   & 3
           & $k^6$$a^{10}$ 
           & $k^6$$a^{4.272}$      
           & $k^6$$|a|^{4.272}$  
           & $k^{10}$$a^{11}$\\     [0.00in]
   & $2^+_{\rm R}$ 
   	   & 1
	   & $k^4$$a^{-0.647}$ 
   	   & 2
           & $k^4$$a^{8}$ 
           & $k^4$$a^{2.353}$ 
           & $k^4$$|a|^{2.353}$  
           & $k^8$$a^{9}$ \\ [0.05in]
$BBB'$($a_{BB'}$)
   & $0^+_{\rm A}$ 
   	   & 0
           & \boldmath$P_{ad}^{d}(a)$$a$ 
   	   & 0
           & \boldmath$M_{3b}^{w}(a)$$a^{4}$ 
           & \boldmath$a^{4}$  
           & \boldmath$P_{3b}^{d}(a)$$|a|^{4}$  
           & \boldmath$M_{3b}^{w}(a)$$a^{5}$ \\ [0.00in]
   & $1^-_{\rm R}$ 
   	   & 1
           & $k^2$$a^{-1.558}$ 
   	   & 1
           & $k^2$$a^{6}$  
           & $k^2$$a^{1.443}$   
           & $k^2$$|a|^{1.443}$ 
           & $k^6$$a^{7}$    \\  [0.00in]
   & $2^+_{\rm R}$ 
   	   & 2
           & $k^4$$a^{-0.815}$ 
   	   & 2
           & $k^4$$a^{8}$  
           & $k^4$$a^{2.815}$   
           & $k^4$$|a|^{2.815}$ 
           & $k^8$$a^{9}$    \\ [0.05in]
$FFF'$($a_{FF'}$)
   & $0^+_{\rm R}$ 
   	   & 0
           & \boldmath{$a^{-3.332}$}
   	   & 2
           & $k^4$$a^{8}$  
           & $k^4$$a^{3.668}$ 
           & $k^4$$|a|^{3.668}$ 
           & $k^8$$a^{9}$  \\[0.00in]
   & $1^-_{\rm R}$ 
   	   & 1
           & $k^2$$a^{-0.546}$ 
   	   & 1
           & \boldmath$k^2${$a^{6}$}  
           & \boldmath$k^2${$a^{2.455}$} 
           & \boldmath$k^2${$|a|^{2.455}$} 
           & \boldmath$k^6$$a^{7}$  \\ [0.00in]
   & $2^+_{\rm R}$
   	   & 2
           & $k^4$$a^{-1.210}$ 
   	   & 2
           & $k^4$$a^{8}$  
           & $k^4$$a^{1.790}$
           & $k^4$$|a|^{1.790}$
           & $k^8$$a^{9}$  \\ [0.05in]
$BBX$($a_{BX}$)
   & $0^+_{\rm A}$
   	   & 0
           & \boldmath{$P_{ad}^{d}(a)$$a$} 
   	   & 0
           & \boldmath{$M_{3b}^{w}(a)$$a^{4}$} 
           & \boldmath{$a^{4}$} 
           & \boldmath{$P_{3b}^{d}(a)$$|a|^{4}$}   
           & \boldmath{$M_{3b}^{w}(a)$$k^4$$a^{5}$} \\[0.00in]
   & $1^-_{\rm R}$ 
   	   & 1
           & $k^2$$a^{3-2p_{0}}$ 
   	   & 1
           & $k^2$$a^{6}$ 
           & $k^2$$a^{6-2p_{0}}$  
           & $k^2$$|a|^{6-2p_{0}}$ 
           & $k^6$$a^{7}$ \\ [0.025in]
   & $2^+_{\rm A}$
   	   & 2
           & $P_{ad}^{d}(a)$$k^4$$a^{5}$
   	   & 2
           & $M_{3b}^{w}(a)$$k^4$$a^{8}$
           & $k^4$$a^{8}$  
           & $P_{3b}^{d}(a)$$k^4$$|a|^{8}$
           & $M_{3b}^{w}(a)$$k^8$$a^{9}$ \\ [0.00in]
   & $2^+_{\rm R}$
   	   & 2
           & $k^4$$a^{5-2p_{0}}$ 
   	   & 2
           & $k^4$$a^{8}$ 
           & $k^4$$a^{8-2p_{0}}$  
           & $k^4$$|a|^{8-2p_{0}}$ 
           & $k^8$$a^{9}$ \\ [0.05in]
$FFX$($a_{FX}$)
   & $0^+_{\rm R}$
   	   & 0
           & \boldmath{$a^{1-2p_{0}}$} 
   	   & 2
           & $k^4$$a^{8}$ 
           & $k^4$$a^{8-2p_0}$ 
           & $k^4$$|a|^{8-2p_0}$ 
           & $k^4$$a^{9}$ \\[0.00in]
   & $1^-_{\rm A}$ 
   	   & 1
           & $P_{ad}^{d}(a)$$k^2$$a^{3}$
   	   & 1
           & \boldmath{$M_{3b}^{w}(a)$$k^2$$a^{6}$}
           & \boldmath$k^2${$a^{6}$}
           & \boldmath{$P_{3b}^{d}(a)$$k^2$$|a|^{6}$}
           & \boldmath{$M_{3b}^{w}(a)$$k^6$$a^{7}$} \\[0.00in]
   & $1^-_{\rm R}$
   	   & 1
           & $k^2$$a^{3-2p_{0}}$
   	   & 1
           & \boldmath$k^2${$a^{6}$}
           & \boldmath$k^2${$a^{6-2p_{0}}$}
           & \boldmath$k^2${$|a|^{6-2p_{0}}$}
           & \boldmath$k^6${$a^{7}$} \\[0.00in]
   & $2^+_{\rm R}$ 
   	   & 2
           & $k^4$$a^{5-2p_{0}}$ 
   	   & 2
           & $k^4$$a^{8}$  
           & $k^4$$a^{8-2p_0}$
           & $k^4$$|a|^{8-2p_0}$ 
           & $k^8$$a^{9}$  \\  [0.05in]
$BBX$($a_{BB}$)
   & $0^+_{\rm R}$ 
   	   & 0
           & \boldmath{$a^{-1}$} 
   	   & 0
           & \boldmath{$a^{4}$} 
           & \boldmath{$a^{2}$} 
           & \boldmath{$|a|^{2}$} 
           & \boldmath$k^4$$a^{5}$  \\[0.00in]
[$XYZ$($a_{XY}$)]           
   & $1^-_{\rm R}$ 
   	   & 1
           & $k^2$$a^{-1}$
   	   & 1
           & $k^2$$a^{6}$ 
           & $k^2$$a^{2}$
           & $k^2$$|a|^{2}$
           & $k^6$$a^{7}$  \\[0.00in]
   & $2^+_{\rm R}$
   	   & 2
           & $k^4$$a^{-1}$ 
   	   & 2
           & $k^4$$a^{8}$  
           & $k^4$$a^{2}$
           & $k^4$$|a|^{2}$
           & $k^8$$a^{9}$  \\
\end{tabular}
\end{ruledtabular}}
\end{table*}

%%%%%%%%%%%%%%%%%%%%%%%%%%%%%%%%%%%%%%%%%%%%%%%%%%%%
\subsection{Scaling and threshold laws: Summary} 
\label{SummaryScalingLaws}

To gain a more comprehensive view of the relevant inelastic scattering processes in ultracold quantum
gases, we summarize in Table~\ref{TabRates} our results for the energy and scattering length dependence for
three-body recombination, $K_{3}$ (Section \ref{K3Section}), collision-induced dissociation, $D_{3}$ 
(Section \ref{D3Section}), and atom-molecule relaxation, $\beta_{\rm rel}$ (Section \ref{VRSection}).
The results in Table~\ref{TabRates} cover the inelastic processes for all three-body systems in which only one type 
of pair interaction is resonant, with the relevant scattering length for each system also indicated. 
(Results for cases in which more than one scattering length is relevant can be found in Ref.~\cite{dincao2009PRL}.)
Table \ref{TabRates} will serve as a guide for various discussions following this section. As a result, here, we carefully 
describe how these results are structured, and their corresponding physical relevance.

The boldface type in Table~\ref{TabRates} indicates the dominant partial wave contribution for each scattering process, as 
determined by its energy dependence, i.e., by Wigner's threshold laws. Note that the dominant contribution does not necessarily 
correspond to the lowest $J$. Although this is always the case for atom-molecule relaxation, for recombination, 
the dominant contribution depends essentially on $\lambda=\lambda_{\rm m}$, i.e., the minimum value of $\lambda$ determined by the 
permutational symmetry and total angular momentum $J^{\pi}$ (see Table~\ref{TabI}). 
By including results for different partial waves, Table~\ref{TabRates} indicates how strong should be the contribution from the next leading 
term for finite temperatures. 

In Table~\ref{TabRates} we only show the main power-law behavior for each rate. We have added a label to the 
the total angular momentum, $J^{\pi}$, to indicate whether the corresponding three-body system has attractive or 
repulsive Efimov interactions, by marking it with the subscripts A and R, respectively.
For three-body attractive systems, the terms that are responsible for Efimov features in the inelastic rates 
are given in Table~\ref{TabRates} by
\begin{align}
&M_{3b}^{w}\left(a\right)=4A_w^2e^{-2\eta}[\sin^2[s_{0}\ln(a/a_{+})]+\sinh^2\eta]\frac{\hbar}{\mu},\label{M3b}\\
&P_{3b}^{d}\left(a\right)=\frac{A_d^2}{2}\frac{\sinh2\eta}{\sin^2[s_{0}\ln(a/a_{-})]+\sinh^2\eta}\frac{\hbar}{\mu},\label{P3b}
\end{align}
describing the interference phenomena in recombination and dissociation involving weakly bound molecules ($a>0$), and 
resonant effects in recombination into deeply bound states ($a<0$), respectively (see Sections \ref{K3Section} and \ref{D3Section}). 
For atom-molecule relaxation, the resonant effects are encapsulated in the term (see Section \ref{VRSection})
\begin{align}
P_{ad}^{d}\left(a\right)=\frac{A_d^2}{2}\frac{\sinh2\eta}{\sin^2[s_{0}\ln(a/a_{*})]+\sinh^2\eta}\frac{\hbar}{\mu_{ad}}.\label{Pad}
\end{align}
The universal nature of the constants $A_w$, $C_d$, and $A_d$ in Eqs.~(\ref{M3b})-(\ref{Pad}) is discussed 
in Sections~\ref{K3Section}-\ref{VRSection}, respectively.
Equations~(\ref{M3b})-(\ref{Pad}) also explicitly show the geometric-scaling properties characteristic of the Efimov effect 
via their log-periodic dependence on $a$. Together they show that an Efimov feature (a minimum or a resonant enhancement
due to the formation of a Efimov state) should occur whenever $a\rightarrow e^{\pi/s_0}a$. The values of $a$ in which
such features occur are determined by the three-body parameters $a_+$, $a_-$ and $a_*$, which encapsulate the short-range
properties of the three-body interactions---see Section \ref{Universality} for a discussion on the universality of such three-body parameters.
As one can see from Table~\ref{TabRates}, the inelastic rates associated with three-body repulsive systems are typically characterized
by their suppression as $a$ increases. Note that for repulsive systems, while the atom-dimer relaxation rate vanishes for large $a$,
recombination into deeply bound is enhanced, but with a weaker $a$ dependence, comparable to recombination into weakly bound molecules.
Note also that for $BBX$ and $FFX$ repulsive systems, the scattering length dependence of the inelastic rates depends on the specific value 
of the mass ratio, $\delta$, through its dependence on the parameter $p_0$.
Recall that for $BBX$ and $FFX$ systems, for values of $\delta$ smaller than a critical value, $\delta<\delta_{c}$, the system is 
characterized by an attractive interaction (the Efimov effect occurs), while for $\delta>\delta_{c}$ a repulsive interaction prevails, and the 
rates depend on the strength of such repulsion, $p_0$  [see Figs.\ref{CoeffsBBXFFX}(a)--(h)]. 

Similarly, Table~\ref{TabElastic} summarizes our results for the elastic three-body scattering length,
$A_{3b}$ (Section \ref{A3bSection}), and the atom-molecule scattering length, $a_{ad}$ (Section \ref{AadSection}).
For those cases, the interference and resonant phenomena in three-body attractive systems, associated with
the Efimov effect, are encapsulated in the terms 
\begin{align}
&I_{3b}^w\left(a\right)=(A^2-\frac{B^2}{e^{2\eta}})+\frac{2B^2}{e^{2\eta}}\sin^{2}[s_{0}\ln(a/a_+)-\frac{\pi}{4}],\label{A3bPos}\\
&R_{3b}^{d}\left(a\right)=A^2+\frac{B^2}{4}\frac{\sin[2s_{0}\ln(|a/a_-|)]}
{\sin^2[s_{0}\ln(|a/a_-|)]+\sinh^2\eta},\label{A3bNeg}
\end{align}
for the three-body scattering length, $A_{3b}$, while the resonant effects in the atom-molecule scattering length are
encapsulated through 
\begin{align}
&R_{ad}^{d}\left(a\right)=A^2+\frac{B^2}{4}\frac{\sin[2s_{0}\ln(a/a_*)]}
{\sin^2[s_{0}\ln(a/a_*)]+\sinh^2\eta}.\label{AadPos}
\end{align}
The nature of the universal constants above is discussed in Sections~\ref{A3bSection} and \ref{AadSection}.
Note that the only cases in Table \ref{TabElastic} in which a negative value of the scattering length 
can be obtained are the ones in which a three-body state is formed at the collision threshold.
In the case of attractive systems, formation of Efimov states leads to resonant effects parametrized by 
Eqs.~(\ref{A3bNeg}) and (\ref{AadPos}). 
For repulsive systems, although not explicitly shown in Table~\ref{TabElastic}, a negative value of the atom-molecule scattering length
is also possible for some values of the mass ratio due to the presence of KM states \cite{kartavtsev2007JPB,endo2011FBS,endo2012PRA,kartavtsev2014YF} 
for $BBX$ ($J^\pi>0$-even) and $FFX$ ($J^\pi$-odd) systems, as explained in Section~\ref{AadSection}.
Although we do not explore the relevance of the three-body elastic parameters $A_{3b}$ and $a_{ad}$, it is well-known
that, from the mean-field point of view, both the magnitude and sign of the scattering lengths are important
in determining the character of the three-body interactions in a many-body system. 
For large and positive values for the elastic scattering parameters, mean-field interactions are expected to be 
strong and repulsive, while for large and negative values, mean-field interactions are strong and attractive. 
Therefore, the results in Table~\ref{TabElastic} can offer further insights on the impact of three-body
elastic processes for the collective behavior of ultracold quantum gases, an analysis beyond the scope of the
present study.

\subsection{Collisional stability of ultracold gases}

Using the results in Table~\ref{TabRates}, we now analyze the collisional properties of ultracold quantum 
gases and identify the important processes determining their lifetimes and stabilities. We do this here, however,
only to provide an overall look at such issues and characterize the Efimov physics accessible in different 
quantum gases. We first analyze the relevant scattering processes according to the structure of
the gas, i.e., whether it forms a pure atomic sample, an atom-dimer mixture, or a pure molecular sample. 
We then perform the analysis on the lifetime and stability of such samples, which strongly depend on their 
intrinsic quantum statistics.

\paragraph{Atomic and molecular loss processes.}
For {\em pure atomic} samples, recombination represents the main collisional process leading to 
atomic losses. However, for large values of $a>0$, while recombination to deeply bound molecules ($K_3^d$) will
unequivocally lead to atomic losses (both atom and diatomic molecules will have enough kinetic energy to escape from 
the trap), weakly bound molecules formed via recombination ($K_3^w$) can still remain trapped. In this case, 
both relaxation ($\beta_{\rm rel}^w$) and dissociation ($D_3^w$) of weakly bound molecules can start contributing to the collisional 
and dynamical properties of the system. Note that although relaxation will lead to atomic and molecular losses, 
dissociation does not necessarily leads to losses. Weakly bound molecules formed by recombination can have enough kinetic energy
to be dissociated due to collisions with other atoms (or molecules), thus leading to an increase in the number of atoms.
In the case of {\em atom-molecule} samples, if the temperature is lower than the molecular binding energy, relaxation (due to collisions of
other atoms or molecules) is the main process controlling the molecular lifetime. The stability of the atomic component of 
this mixture, however, will be controlled by both relaxation and recombination.
Evidently, if the temperature is larger than the molecular binding energy, or if recombination can lead to the formation
of trappable weakly bound molecules, dissociation will also play a role, greatly complicating the dynamics of the system.
In {\em pure molecular} gases, molecule-molecule collisions (not covered in Table~\ref{TabRates}) are the main collisional 
process controlling the stability of such samples, and whether such molecules are composite bosons or fermions have 
important implications for their collisional properties due the Wigner threshold law. For bosonic molecules relaxation is
constant at threshold, while for fermionic molecules, relaxation is suppressed due to their $p$-wave character. 
Moreover, if any of the possible three-body subsystems of the molecule-molecule four-body compound belongs to the attractive 
or repulsive class of three-body systems, inelastic molecule-molecule collisions can display resonant effects or be suppressed
\cite{petrov2004PRLb,petrov2005PRA,marcelis2008PRA,dincao2009PRLb,dincao2009PRA}.

\paragraph{Collisional stability of ultracold quantum gases.}
%Bose gases
In {\em ultracold Bose gases} (single component), the main collisional behavior is determined from the $J^\pi=0^+$ partial-wave
contribution for $BBB$ systems. Efimov resonant and interference features can be observed in atomic and molecular losses 
caused by recombination, dissociation and relaxation, with rates given according to Table~\ref{TabRates}. 
%Bose gas mixtures
A similar situation is found in {\em ultracold Bose gas mixtures} with two components, where the main collisional behavior is also determined 
from the $J^\pi=0^+$ partial-wave contribution. In this case, however, the main collisional behavior originates from two different systems, 
$BBb$ and $Bbb$, where $B$ and $b$ denote two dissimilar bosons. If the interspecies interactions are resonant,
the Efimov effect occurs for both systems. As a result, a more complicated structure of interference and resonances
can be expected to occur, since $BBb$ and $Bbb$ systems can have different
geometric-scaling parameters, $e^{\pi/s_0}$ (see Fig.~\ref{S0MassDep}). 
Moreover, depending upon the mass ratio, $\delta_{bB}<\delta_c\approx0.025887$, the contributions from $BBb$ collisions
can also display Efimov features, associated with $J^\pi=2^+$ Efimov states.
However, if the density of the $B$ and $b$ species is substantially different,
only one of the processes will dominate the collisional behavior of the system. For instance, if the density of $B$ atoms is much 
higher than the density of $b$ atoms, one can expect $BBb$ processes to be the leading contribution to atomic and molecular 
losses. 
Whether or not the Bose gas is made out of a single or two components, in both cases the dominant contribution for recombination 
and relaxation is constant at low energies and proportional to $a^4$ and $a$, respectively---see Table~\ref{TabRates}. This dependence 
of the loss rates on $a$ should lead to a drastic reduction of the system's lifetime as $a$ increases. 
As a result, bosonic gases are intrinsically unstable in the resonant regime.

%Fermi gas 
In contrast, {\em ultracold Fermi gases} (single component) are far more stable. In such systems, however, the interactions can only be
tuned via $p$-wave interactions. In this case, although recombination is still enhanced for large values of the $p$-wave scattering
length \cite{suno2003PRL,suno2003NJP,regal2003PRLb,ticknor2004PRA,zhang2004PRA,jona-lasinio2008PRA}, their energy 
dependence (controlled by the Wigner threshold law \cite{esry2002PRA}) still ensures its suppression. Although in molecular 
samples, collisions between $p$-wave molecules with other atoms and molecules are not suppressed at low energies, 
they remain constant for large values of the $p$-wave scattering length \cite{dincao2008PRA}.
%Fermi gas mixtures
In the case of {\em ultracold Fermi gas mixtures} with two components, strong $s$-wave interactions are allowed between the 
components. For pure atomic samples, the main collisional behavior is determined from the $J^\pi=1^-$ partial-wave contribution for recombination 
and dissociation, while for atom-molecule mixtures, the collisional behavior is determined from the $J^\pi=0^+$ partial-wave for relaxation (see Table~\ref{TabRates}). 
Similar to Bose gas mixtures, atomic and molecular losses in a two-component Fermi gas can originate from two different three-body systems, 
$FFf$ and $Fff$, where $F$ and $f$ denote two dissimilar fermions.
For this case, whenever $\delta_{fF}<\delta_c\approx0.073492$, one should expect the contributions from $FFf$ collisions
to display Efimov features---associated with $J^\pi=1^-$ Efimov states--- that can be observable through atomic losses associated to
three-body recombination. 
For values of the mass ratios in which none of the systems display the Efimov effect, the signature of Efimov physics characteristic for 
this case is the suppression of recombination into deeply bound molecular states, which comes from the $a^{6-2p_0}$ dependence
on recombination (see Table~\ref{TabRates}). For specific values of $p_0$, see Fig.~\ref{CoeffsBBXFFX}(e).
In general, in Fermi gas mixtures one should expect long lifetimes in both pure atomic samples 
and atom-molecule mixtures, even in the large $a$ regime. 
In atomic samples, recombination is suppressed due to its $k^2$ energy dependence, while in atom-molecule mixtures, the 
suppression occurs due the $a^{1-2p_0}$ dependence on relaxation, which is a direct result of the Efimov physics (see Table~\ref{TabRates}).

%Bose-Fermi gas mixtures
In {\em ultracold Bose-Fermi mixtures} (two components), as with the other gas mixtures, two different inelastic processes
can contribute, one involving two bosons, $BBF$, and the other involving two fermions, $BFF$. 
The dominant collisional behavior, however, is solely determined from the $J^\pi=0^+$ partial-wave contribution for $BBF$ collisions, since 
contributions for $BFF$ are suppressed because of its energy and/or scattering length dependence. 
For instance, while $J^\pi=0^+$ recombination for $BBF$ systems is constant in the low energy limit (displaying Efimov features), 
recombination for a $BFF$ system is determined by the $J^\pi=1^-$ contribution, which vanishes as $k^2$ at threshold 
(see Table~\ref{TabRates}). In the case of relaxation, the relaxation rate for both systems is constant at low energies 
(determined by the $J^\pi=0^+$ contribution). However, relaxation for $BFF$ is suppressed in the limit of large scattering lengths as 
$a^{1-2p_0}$, while relaxation for a $BBF$ systems scales linearly with $a$ and displays resonant enhancements due to the formation of
Efimov states (see Table~\ref{TabRates}). We also note that, for mass ratios $\delta_{BF}<\delta_c\approx0.073492$, the
Efimov effect can also occur for $BFF$ systems where resonant and interference effects due to $J^\pi=1^-$ Efimov states 
can also occur and might still be visible if the temperature of the gas is not too low. 
Because the dominant collisional behavior is controlled by the $BBF$ system, Bose-Fermi 
mixtures are expected, in general, to be short-lived. 
However, for a mixture of $BF$ molecules with {\em only} $F$ atoms, relaxation is suppressed due to its $a^{1-2p_0}$ dependence
(see Table~\ref{TabRates}), which results in a long-lived mixture for large values of $a$. Specific values
for the scaling of relaxation for different Bose-Fermi mixtures can be found in Ref.~\cite{dincao2006PRAb}.

Evidently, the determination of collisional properties for multicomponent ultracold gases,
when multiple scattering lengths can be tuned to large values simultaneously, is more complicated and 
depends strongly on the quantum statistics of the gas.
Nevertheless, this is a tractable problem, and we have addressed most of the relevant issues
in the analysis presented in Refs.~\cite{dincao2008PRL,dincao2009PRL}, where a similar
approach to the one presented in Section \ref{Collisions} has been applied to determine
the relevant scattering rates. As shown in these works, a multitude of interesting effects 
can lead to unique ways of probing Efimov physics and controlling the collisional
properties of the system.

\begin{table}[htbp]
\caption{Scattering length dependence of the elastic atom-molecule scattering length, $a_{ad}$,
  and three-body scattering length $A_{3b}$. The attractive and repulsive character of the three-body 
  systems is indicated by A and R, respectively. Boldface type indicates the leading contribution at threshold.}
\label{TabElastic}
{
\begin{ruledtabular}
\begin{tabular}{ccccccc}
& & 
& \multicolumn{1}{c}{\underline{\hspace{0.15in}$a_{ad}$\hspace{0.15in}}} 
&   
& \multicolumn{2}{c}{\underline{\hspace{0.45in}$A_{3b}$\hspace{0.45in}}}
  \\ 
System & $J^{\pi}_{\rm C}$ & $l$ & $a>0$ & $\lambda_{\rm m}$ & $a>0$ & $a<0$  \\ \hline
$BBB$($a_{BB}$)
   & $0^+_{\rm A}$
   	   & 0
           & \boldmath$R_{ad}^d(a)${$a$} 
   	   & 0
           & \boldmath$I_{3b}^{w}(a)${$a^{4}$}  
           & \boldmath$R_{3}^{d}(a)${$a^{4}$} \\  [0.00in]
   & $1^-_{\rm R}$
   	   & 1
	   & $a^{3}$ 
   	   & 3
           & $a^{10}$ 
           & $a^{10}$ \\ [0.00in]
   & $2^+_{\rm R}$ 
   	   & 2
	   & $a^{5}$ 
   	   & 2
           & $a^{8}$ 
           & $a^{8}$ \\ [0.05in]
$BBB'$($a_{BB'}$)
   & $0^+_{\rm A}$ 
   	   & 0
           & \boldmath$R_{ad}^{d}(a)$$a$ 
   	   & 0
           & \boldmath$I_{3b}^{w}(a)$$a^{4}$ 
           & \boldmath$R_{3}^{d}(a)$$a^{4}$ \\ [0.00in]
   & $1^-_{\rm R}$ 
   	   & 1
           & $a^{3}$ 
   	   & 1
           & $a^{6}$  
           & $a^{6}$ \\  [0.00in]
   & $2^+_{\rm R}$
   	   & 2
           & $a^{5}$ 
   	   & 2
           & $a^{8}$  
           & $a^{8}$ \\ [0.05in]
$FFF'$($a_{FF'}$)
   & $0^+_{\rm R}$
   	   & 0
           & \boldmath{$a$}
   	   & 2
           & $a^{8}$  
           & $a^{8}$ \\[0.00in]
   & $1^-_{\rm R}$ 
   	   & 1
           & $a^{3}$ 
   	   & 1
           & \boldmath{$a^{6}$}  
           & \boldmath{$a^{6}$} \\ [0.00in]
   & $2^+_{\rm R}$
   	   & 2
           & $a^{5}$ 
   	   & 2
           & $a^{8}$  
           & $a^{8}$ \\ [0.05in]
$BBX$($a_{BX}$)
   & $0^+_{\rm A}$
   	   & 0
           & \boldmath{$R_{ad}^{d}(a)$$a$} 
   	   & 0
           & \boldmath{$I_{3b}^{w}(a)$$a^{4}$} 
           & \boldmath{$R_{3}^{d}(a)$$a^{4}$} \\ [0.00in]
   & $1^-_{\rm R}$
   	   & 1
           & $a^{3}$ 
   	   & 1
           & $a^{6}$ 
           & $a^{6}$ \\ [0.025in]
   & $2^+_{\rm A}$
   	   & 2
           & $R_{ad}^{d}(a)$$a^{5}$
   	   & 2
           & $I_{3b}^{w}(a)$$a^{8}$ 
           & $R_{3b}^{d}(a)$$a^{8}$ \\ [0.00in]
   & $2^+_{\rm R}$
   	   & 2
           & $a^{5}$ 
   	   & 2
           & $a^{8}$ 
           & $a^{8}$ \\ [0.05in]
$FFX$($a_{FX}$)
   & $0^+_{\rm R}$
   	   & 0
           & \boldmath{$a$} 
   	   & 2
           & $a^{8}$ 
           & $a^{8}$ \\ [0.00in]
   & $1^-_{\rm A}$
   	   & 1
           & $R_{ad}^{d}(a)$$a^{3}$
   	   & 1
           & \boldmath{$I_{3b}^{w}(a)$$a^{6}$}
           & \boldmath{$R_{3b}^{d}(a)$$a^{6}$} \\ [0.00in]
   & $1^-_{\rm R}$
   	   & 1
           & $a^{3}$
   	   & 1
           & \boldmath{$a^{6}$}
           & \boldmath{$a^{6}$} \\ [0.00in]
   & $2^+_{\rm R}$
   	   & 2
           & $a^{5}$ 
   	   & 2
           & $a^{8}$  
           & $a^{8}$ \\  [0.05in]
$BBX$($a_{BB}$)
   & $0^+_{\rm R}$
   	   & 0
           & \boldmath{$a$} 
   	   & 0
           & \boldmath{$a^{4}$} 
           & \boldmath{$a^{4}$} \\ [0.00in]
[$XYZ$($a_{XY}$)]           
   & $1^-_{\rm R}$
   	   & 1
           & $a^{3}$
   	   & 1
           & $a^{6}$ 
           & $a^{6}$ \\ [0.00in]
   & $2^+_{\rm R}$
   	   & 2
           & $a^{5}$ 
   	   & 2
           & $a^{8}$  
           & $a^{8}$ \\
\end{tabular}
\end{ruledtabular}}
\end{table}

%%%%%%%%%%%%%%%%%%%%%%%%%%%%%%%%%%%%%%%%%%%%%%%%%%%%
\subsection{Threshold regime and finite energy effects} 
\label{ThresholdRegime}

In the previous two sections, the characterization of the collisional properties of ultracold quantum 
gases was accomplished by identifying the dominant contribution for the inelastic rates in the low-energy
limit without explicitly quantifying what sets up this regime. 
In fact, the analysis in the previous sections is only meaningful provided that the temperature of the system represents 
the smallest energy scale in the problem. 
This condition relates to the validity of the Wigner's threshold laws, which determines the low-energy behavior of scattering observables.
Therefore, whenever the temperature (or collision energy) is found to be the smallest energy scale in
the system, it is said that the system is in the {\em threshold regime} and the collisional rates obey Wigner's threshold laws. 
If the condition for the threshold regime is not satisfied, finite energy effects in the three-body 
scattering rates have been shown to be extremely important \cite{dincao2004PRL,jonsell2006EPL,braaten2007PRA,
braaten2007PRAb,yamashita2007PLA,braaten2008PRA,wang2010PRL,wang2011NJP,petrov2015pra}.
In this case, not only are the low-energy expressions for the three-body rates not valid, 
but also other higher partial-wave contributions are expected to substantially contribute to the total rates. 
Moreover, as shown in Ref.~\cite{dincao2004PRL}, at {\em any} finite temperature, the condition for the threshold
regime is always violated for values of $a$ exceeding a certain critical value, $a_c$ (characterized below).
Therefore, it is crucial to understand the physics behind such breakdown of the threshold regime.

The violation of the threshold regime shown in Ref.~\cite{dincao2004PRL} was found for $|a|\gtrsim a_c$
as a result of the appearance of a new energy scale. 
For example, for $a>0$, this energy scale is set by the binding energy of the weakly bound molecular state, 
while for $a<0$ the energy scale is set by the height of a potential barrier.
Both this height and binding energy are proportional to $1/a^2$, and become increasingly small as $a$ 
becomes large. Therefore, for three-body collisions, the condition $k|a|\ll1$ specifies the range of energies, 
or, equivalently, the range of scattering lengths, $|a|\ll a_c=1/k$, where the system 
is in the threshold regime. In this case, the three-body scattering observables can 
be quantitatively described by their dominant contribution, as shown in Table~\ref{TabRates}.
For three-body repulsive systems the violation of the threshold regime only implies a change in behavior for the 
collision rates with $a$ beyond $a_c$. However, for three-body attractive systems, $a_c$ also defines the 
maximum value of the scattering length in which interference and resonant effects associated
to Efimov states can be observed. As a result, the condition for the threshold regime sets a limit 
on the number of consecutive Efimov features that can be experimentally accessible for a 
given temperature (see discussion in Ref.~\cite{dincao2006PRA}). 

For values of $|a|$ beyond $a_c$ the transition probabilities $|T_{fi}|^2$ in Eqs.~(\ref{K3rate}), (\ref{D3rate}) 
and (\ref{Vrelrate}) become constant, therefore drastically changing the energy and scattering length dependence of 
the inelastic scattering rates \cite{dincao2004PRL}. In fact, in this {\em unitarity regime} (since $|T_{fi}|^2$ approaches unit value), 
the rates become independent of $a$.
For recombination (\ref{K3rate}), dissociation (\ref{D3rate}), and relaxation (\ref{Vrelrate}), the unitarity-limited values for
the rates (valid for $|a|\gg a_c$ or, equivalently, $k|a|\gg1$) are given, respectively, by
\begin{align}
K_3^{u}(E)&=\frac{8n!\pi^2\hbar^5(2J+1)}{\mu^3E^{2}}(1-e^{-4\eta}),\label{K3u}\\
D_3^{u}(E)&=\frac{4\pi\hbar^2(2J+1)}{2^{1/2}\mu_{ad}^{3/2}E^{1/2}}(1-e^{-4\eta}),\label{D3u}\\
 \beta_{\rm rel}^{u}(E)&=\frac{4\pi\hbar^2(2J+1)}{2^{1/2}\mu_{ad}^{3/2}E^{1/2}}(1-e^{-4\eta}),\label{Betau}
\end{align}
where $n$ is the number of identical particles and $\eta$ the usual three-body inelasticity parameter \cite{braaten2006PRep}. 
The $\eta$-dependent term in Eq.~(\ref{K3u}) was first obtained in Refs.~\cite{rem2013PRL,petrov2015pra}.
Although Eqs.~(\ref{K3u})-(\ref{Betau}) give a limiting value for the inelastic collision rates in the unitarity regime,
they are only part of the story for attractive three-body systems. 
In the unitarity regime, Efimov states lead to oscillations in the inelastic three-body rates as the energy changes, 
with the number of oscillations determined by the number of Efimov states for a given value of $a$ \cite{wang2011NJP,wang2010PRL} 
(see also Refs.~\cite{wang2013Adv} and \cite{shu2016ARX}) . 

\begin{figure*}[htbp]
\begin{center}
\includegraphics[width=6.8in,angle=0,clip=true]{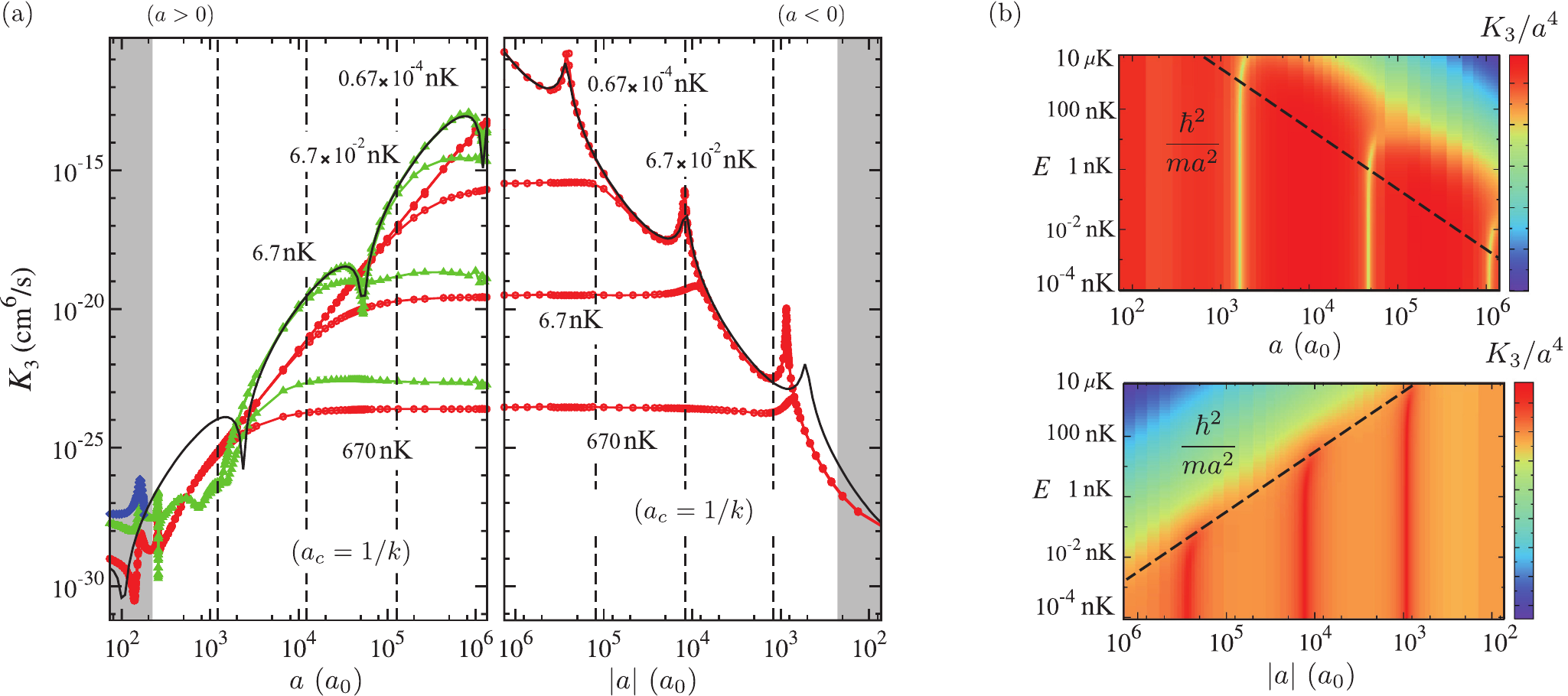}
 \caption{(a) Comparison of the finite-energy and finite-range numerical calculations for $J^\pi=0^+$ recombination
of Cs atoms to the analytical results from Table~\ref{TabRates} (black solid lines).
For $a>0$, green and red curves (with symbols) represent recombination into weakly and deeply bound molecules, respectively,
while for $a<0$ recombination only occurs to deeply bound molecular states. 
For $a>0$ there exists an additional $d$-wave resonance feature (blue curve) which appears for values of $a$ on the order of the
range of the interaction. The gray shaded areas indicate the region $a$ in which finite-range effects
are expected to be important. Finite energy effects are evident for $|a|>a_{c}$, where recombination becomes 
independent of $a$~\cite{dincao2004PRL}. 
(b) Contour plot of $K_{3}/a^{4}$ demonstrating the importance of the fundamental 
energy scale, proportional to $1/a^2$, in determining the threshold regime. 
} \label{Validity}
\end{center}
\end{figure*}

The validity, and eventual breakdown, of the threshold regime can be illustrated by comparisons between the universal 
formulas from Table~\ref{TabRates} and the fully numerical calculations. This comparison is done in Fig.~\ref{Validity}(a), where black solid lines 
represents the analytical, zero-range, and zero-energy results for recombination for $BBB$ ($J^\pi=0^+$) systems from Table \ref{TabRates}.
The numerical results [lines with symbols in Fig.~\ref{Validity}(a)] were obtained for Cs atoms in Ref.~\cite{dincao2009JPB},
using a finite-range two-body interaction supporting up to three diatomic molecular states. The results display three interference 
minima ($a>0$) and three resonances ($a<0$).
For $a>0$ recombination can lead to weakly bound and deeply bound 
molecules (green and red curves, respectively), while for $a<0$ recombination can only proceed to deeply bound 
molecules (see Ref.~\cite{dincao2009JPB}). 
Note that for $a>0$ there exists an additional $d$-wave resonance feature that appears when the scattering length is 
of the order of the range of the interaction, $r_{0}=100a_0$, where $a_0$ is the Bohr radius [see blue curve in Fig.~\ref{Validity}(a)]. 
In fact, the gray-shaded area in Fig.~\ref{Validity}(a) indicates the region in which $|a|\lesssim r_{0}$, 
emphasizing the region in $a$ in which finite-range effects not incorporated within the
zero-range results [black solid lines in Fig.~\ref{Validity}(a)] can play an important rule.
One clear consequence of such corrections is that the predicted geometric scaling between scattering
features, $e^{\pi/s_0}\approx22.7$, can only be quantitatively verified for those deep in the $|a|\gg r_0$ regime, even in the
zero energy limit.

For the lowest collision energy in Fig.~\ref{Validity}(a) ($E/k_B=0.67\times10^{-4}$nK), the critical scattering 
length, $a_c=1/(mE)^{1/2}$ (where $m$ is the atomic mass) is about $a_{c}\approx4.4\times10^6a_0$, thus
ensuring the validity of the zero-energy limit results of Table~\ref{TabRates} 
for the entire range of $a$ displayed in Fig.~\ref{Validity}(a). 
As the collision energy increases toward more accessible values,
this range gets substantially smaller. For $E/k_B=6.7\times10^{-2}$nK, $6.7$nK and $670$nK, the corresponding
values for $a_c$ are $1.3\times10^5a_0$, $1.3\times10^4a_0$, and $1.3\times10^3a_0$, as indicated 
by the vertical dashed lines in Fig.~\ref{Validity}(a). 
For a given collision energy, recombination becomes independent of $a$ 
for values of $|a|\gtrsim a_{c}$, i.e., when the system enters into the unitarity regime~\cite{dincao2004PRL}, 
with values consistent with the unitarity-limited rate in Eq.~(\ref{K3u}). 
In the unitarity regime, all Efimov features (minima and resonances) in recombination are completely washed out,
illustrating the fact that the number of observable Efimov features will depend on the energy [Fig.~\ref{Validity}(a)].
One can, however, estimate the number of Efimov features that are observable in loss rates for a given temperature.
Accordingly to Eq.~(\ref{NumStates}) and the discussion in Section \ref{Spectrum}, this number can be estimated 
by \cite{dincao2006PRA}
\begin{equation}
N\approx \frac{s_{0}}{\pi}\ln(a_c/r_{0})=\frac{s_{0}}{\pi}\ln(1/kr_{0}).
\end{equation}
This result emphasizes that in systems with large values of $s_0$, such as those with two heavy and one light atoms,
the number of experimentally accessible Efimov features will be larger for a given temperature, 
thus allowing for the observation of multiple features \cite{dincao2006PRA}. 
However, a series of universal relationships that connect different Efimov scattering features offer a path 
around the limitations related to the finite temperature effects discussed here, while still providing
evidence of the Efimov effect (see Section \ref{EfimovUniversality}).

A more clear demonstration of the importance of the $\hbar^2/ma^2$ energy scale in determining the range of validity of the 
threshold regime can be seen in Figure~\ref{Validity}(b), where a contour plot of $K_{3}/a^{4}$ is shown
for both $a>0$ and $a<0$. [The dashed line in Fig.~\ref{Validity}(b) is the corresponding $\hbar^2/ma^2$ energy scale.] 
Figure~\ref{Validity}(b) also illustrates the energy dependence of the position of the minima and resonances in 
recombination. Even before reaching the unitarity regime, $|a|\gg a_{c}$, the finite energy effects already 
can modify such positions and therefore impact the observation of geometric scaling. 
Note that passing the range of validity of the threshold regime, higher partial waves
can also contribute, and can even dominate the total rate. This fact was shown in Ref.~\cite{dincao2004PRL}
by calculating the $J^\pi=2^+$ contribution for recombination of three identical bosons (the next leading-order contribution
for recombination in Table~\ref{TabRates}). Other examples can be found in the literature 
(see, for instance, Refs.~\cite{suno2002PRA,suno2008PRA,suno2009PRA}).
Note also that although our analysis indicates that $\hbar^2/ma^2$ is the important 
energy scale controlling the validity of the threshold regime, changes this rule can be expected if
some other smaller energy scale is introduced in the problem.

Taking account the discussions here, it is clear that whenever $|a|\gtrsim a_c$ finite energy effects 
and contributions from higher-$J^{\pi}$ partial waves must 
be taken into account to accurately determine the scattering rates.
In the context of ultracold quantum gases, however, thermal effects must also to be considered. In a gas of ultracold 
atoms and/or molecules, a thermal distribution of energies, rather than in a well-defined
energy state is found. In the thermal regime (i.e., in the nondegenerate regime), atoms and molecules should 
follow a Boltzmann distribution of energies, and three-body scattering rates should be properly thermally 
averaged accordingly to this distribution. For a system at temperature $T$, the proper thermally averaged 
recombination, dissociation, and relaxation rates are given by \cite{dincao2004PRL}:
\begin{align}
\langle K_3\rangle&=\frac{1}{2 (k_BT)^3}\int_0^\infty K_3E^2e^{-E/k_BT}dE,\label{K3T}\\
\langle D_3\rangle&=\frac{1}{\pi^{1/2} (k_BT)^{3/2}}\int_0^\infty D_3E^{1/2}e^{-E/k_BT}dE,\label{D3T}\\
\langle \beta_{\rm rel}\rangle&=\frac{1}{\pi^{1/2} (k_BT)^{3/2}}\int_0^\infty \beta_{\rm rel}E^{1/2}e^{-E/k_BT}dE,\label{BetaT}
\end{align}
respectively. Within the threshold regime, $k|a|\ll1$, the effect of thermal averaging is simply determined based
on the threshold law for each scattering observable: $K_3\propto k^{2\lambda}$, $D_3\propto k^{2\lambda+4}$
and $\beta_{\rm rel}\propto k^{2l}$ (see Section \ref{Collisions}). Using Eqs.~(\ref{K3T})-(\ref{BetaT}), 
the thermally averaged rates are found to be
\begin{align}
\langle K_3\rangle&\approx\frac{\Gamma(\lambda_{\rm m}+3)}{2}K_3(k_BT),\\
\langle D_3\rangle&\approx\frac{\Gamma(\lambda_{\rm m}+\frac{7}{2})}{2}D_3(k_BT),\\
\langle \beta_{\rm rel}\rangle&\approx\frac{\Gamma(l+\frac{3}{2})}{2}\beta_{\rm rel}(k_BT),
\end{align}
where $\lambda_{\rm m}$ and $l$ are given in Table~\ref{TabRates}.
Therefore, except for some proportionality constants, the thermally averaged rates
are obtained simply by substituting $E$ with $k_BT$. For $BBB$ systems, since $J^\pi=0^+$ 
($\lambda_{\rm m}=0$ and $l=0$) is the dominant contribution at threshold, thermal 
averaging has no effect in recombination, but averaging does enhance dissociation by a factor $15\sqrt{\pi}/16\approx1.66$
and reduces relaxation by a factor $\sqrt{\pi}/4\approx0.44$. Now, whenever the system is in the unitarity regime, 
$k|a|\gg1$, the thermally averaged rates can also be determined simply from their energy dependence in
Eqs.~(\ref{K3u})-(\ref{Betau}). In this case, we obtain the thermally averaged, unitarity-limited rates as
\begin{align}
\langle K_3^{u}\rangle&\approx \frac{1}{2}K_3^u(k_BT),\\
\langle D_3^{u}\rangle&\approx\frac{1}{\pi^{1/2}}D_3^u(k_BT),\\
\langle \beta_{\rm rel}^{u}\rangle&\approx\frac{1}{\pi^{1/2}}\beta_{\rm rel}^u(k_BT).
\end{align}
Again, apart from a proportionality constant, the thermally averaged rates are obtained simply by substituting 
$E$ with $k_BT$ in Eqs.~(\ref{K3u})-(\ref{Betau}). In all cases, however, thermal averaging leads to a reduction of
the unitarity-limited rates.

% !TEX root = ./TutorialJPB.tex

%%%%%%%%%%%%%%%%%%%%%%%%%%%%%%%%%%%%%%%%%%%%%%%%%%%%
\section{Universality in few-body systems} \label{EfimovUniversality}

To some extent, everything we have discussed so far is a result of the universality of
few-body systems. The fact that all three-body systems with resonant $s$-wave interactions
can be classified into only two classes, the nature and origin of Efimov states and other few-body 
states, the scattering length dependence of the collisional processes and various other properties 
are all independent of the actual form of the interparticle interactions. All that is assumed is that $s$-wave 
interactions are strong and that they have a finite-range character.
Evidently, since most of the universal properties of few-body systems are derived from the use
of zero-range interactions, more realistic systems will display finite-range corrections to the 
universality when the interactions are not strong enough, i.e, when the condition $|a|\gg r_0$ 
is not fully satisfied [see discussion of Fig.~\ref{Validity}(a) in Section~\ref{ThresholdRegime}]. 
In fact, understanding finite-range corrections and determining whether they
are themselves universal is a crucial point in further developing our knowledge 
of the underlying universal phenomena. Understanding finite-range effects is especially important considering that under 
accessible experimental conditions, few-body observables will likely be sensitive 
to finite-range corrections.

In this Section, we analyze a set of important universal relations among different few-body 
parameters derived from Refs.~\cite{braaten2006PRep,gogolin2008PRL,helfrich2010PRA,helfrich2011JPB},
from which we can estimate the effect of finite-range interactions.
Such universal relations connect the different Efimov interferences and resonances 
as well as the energy of Efimov states in an elegant and unifying way. At the same time, they offer 
a path that mitigates finite-temperature effects (see Section \ref{ThresholdRegime}) by allowing for the 
exploration of the universality between Efimov features associated with the low-lying states, i.e., without requiring 
extremely large values of $a$. 
In that case, however, understanding finite-range effects in such universal relations is of critical importance.
We conclude this section by briefly discussing the recent developments, both experimentally and theoretically,
that led to the establishment of the universality of both three-body and four-body parameters. This universality represents a unique 
characteristic of atomic systems that can open up ways
to expand our knowledge of the Efimov universal phenomena in ultracold quantum gases.

%%%%%%%%%%%%%%%%%%%%%%%%%%%%%%%%%%%%%%%%%%%%%%%%%%%%
\subsection{Universal relations for Efimov features} 
\label{UniversalRelations}

From the perspective of the experimental observation of the Efimov effect in ultracold
quantum gases, one direct way to clearly identify such phenomena is through the
observation of multiple consecutive Efimov features in atomic and molecular losses
following the characteristic geometric scaling, $e^{\pi/s_0}$. As shown in the previous section, 
however, this approach requires tuning the scattering length to very large values and extremely 
low temperatures to keep the system within the threshold regime. 
There are, however, other properties one can identify that also relate to the Efimov effect 
that may circumvent the requirement to keep the system under such extreme conditions. 
In this section, we describe the universal relations between Efimov features from different scattering observables. 
The existence of a universal relationship between these features was first determined by Braaten and Hammer in 
Ref.~\cite{braaten2006PRep}, and later expanded in Ref.~\cite{gogolin2008PRL}, for the case of three identical 
bosons and then extended to heteronuclear systems in Ref.~\cite{helfrich2010PRA,helfrich2011JPB}. 
The observation of a single Efimov feature in any two observables and the verification of 
their corresponding universal relations, serves as a unique way to demonstrate the physics implied by the 
Efimov effect.
 
\begin{figure}[htbp]
\begin{center}
\includegraphics[width=3.4in,angle=0,clip=true]{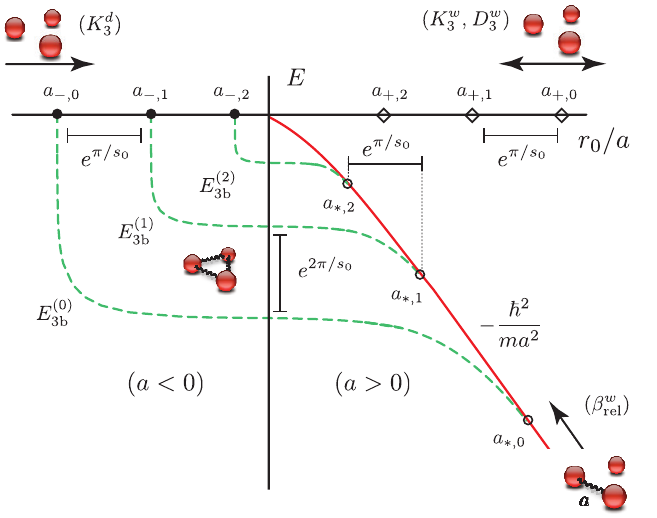}
\caption{Illustration of the Efimov energy spectrum and corresponding three-body parameters, $a_+$, $a_-$ and $a_*$,  
relevant for three-body scattering. For $a<0$ Efimov states are formed at scattering lengths $a_{-,i}$ while for $a>0$ they are formed
at $a_{*,i}$, causing resonant effects in recombination and relaxation, respectively, at such values. 
For $a>0$, Efimov physics is manifested via destructive interference effects leading to minima in recombination for scattering lengths 
$a_{+,i}$. The position of these resonant and interference features are universally related according to
 Eq.~(\ref{UnivRel}).}\label{EfimovSpectrum}
\end{center}
\end{figure}

The key underlying concept explored by the universal relations between Efimov features is that the three-body physics 
depends on a single {\em three-body parameter} that incorporates the short distance 
behavior of the three-body interactions~\cite{efimov1970PLB,efimov1971SJNP,efimov1972JETPL,efimov1973NPA}. 
This concept implies that the observation of a single Efimov feature (or, equivalently, the determination of a single three-body 
parameter) would be sufficient to derive all the properties of the system. As we discussed in Sections \ref{InelasticCol}
and \ref{ElasticCol}, the relevant three-body 
parameters for scattering observables are $a_+$, $a_-$ and $a_*$, determining, respectively, the values of the scattering 
length in which interference minima ($a>0$) and resonances ($a<0$) occur in three-body recombination [Eqs.~(\ref{BosonicK3aposweak}) 
and (\ref{BosonicK3anegdeep})] and the resonances in atom-molecule relaxation ($a>0$) [Eq.~(\ref{BosonicVrelaposweak})].
Figure~\ref{EfimovSpectrum} illustrates the Efimov energy spectrum, indicating the positions of the Efimov features
in scattering observables. Efimov states, with energy $E_{\rm 3b}$, are labeled by the index $i$
(with $i=0,1,2,...,\infty$) and the additional index on the $a_+$, $a_-$ and $a_*$ parameters indicates their 
Efimov family parentage. [According to the usual geometric scaling properties of the problem,
each value corresponds to: $a_{+,i}=a_+(e^{\pi/s_0})^i$, $a_{-,i}=a_-(e^{\pi/s_0})^i$, and $a_{*,i}=a_*(e^{\pi/s_0})^i$.] 
One can generically define the {\em universal relations} between the Efimov features as
\begin{align}
a_{\alpha,i}/a_{\beta,j}&=\theta^{\alpha}_{\beta}~(e^{\pi/s_{0}})^{i-j},\label{UnivRel}
\end{align}
where $\theta^{\alpha}_{\beta}=a_{\alpha}/a_{\beta}$, with $\alpha$ and $\beta$ assuming the values ``$-$", ``$+$" and ``$*$", 
and $i$ and $j$ running over the index labeling the Efimov states. 
Note that universal relations also exist between the scattering three-body parameters ($a_+$, $a_-$ and $a_*$) and the
energy of the Efimov states at $a=\pm\infty$ \cite{braaten2006PRep,gogolin2008PRL}.
The coefficients $\theta$ in Eq.~(\ref{UnivRel}) define the {\em fundamental ratios}. Evidently, whenever
the universal relation in Eq.~(\ref{UnivRel}) is applied to features belonging to the
same scattering observable (i.e., when $\alpha=\beta$), we obtain $\theta^\alpha_\alpha=1$,
and the universal relation simply refers to the geometric-scaling property of Efimov physics.
The fundamental ratios for Efimov features from different scattering observables are universal and deeply rooted 
in the universal properties of Efimov physics. As shown in Refs.~\cite{braaten2006PRep,gogolin2008PRL}, for the case of 
three identical bosons, their values are:
\begin{align}
\theta^{+}_{-}&=a_{+}/a_{-}\approx-0.209914,\nonumber\\ 
\theta^{*}_{-}&=a_{*}/a_{-}\approx-0.046938,\nonumber\\
\theta^{*}_{+}&=a_{*}/a_{+}\approx0.223604. \label{thetavalues}
\end{align}
From Eq.~(\ref{UnivRel}), it is clear that the determination of a {\em single} Efimov feature 
will determine {\em all} other scattering features. The universal relations in Eq.~(\ref{thetavalues}) are extremely useful 
and important, considering that experiments are likely to be limited by the observation of a small number of consecutive 
features; the ability to determine their relationship to other features offers strong evidence of the universal behavior 
related to Efimov physics.

\begin{figure}[htbp]
\begin{center}
\includegraphics[width=2.8in,angle=0,clip=true]{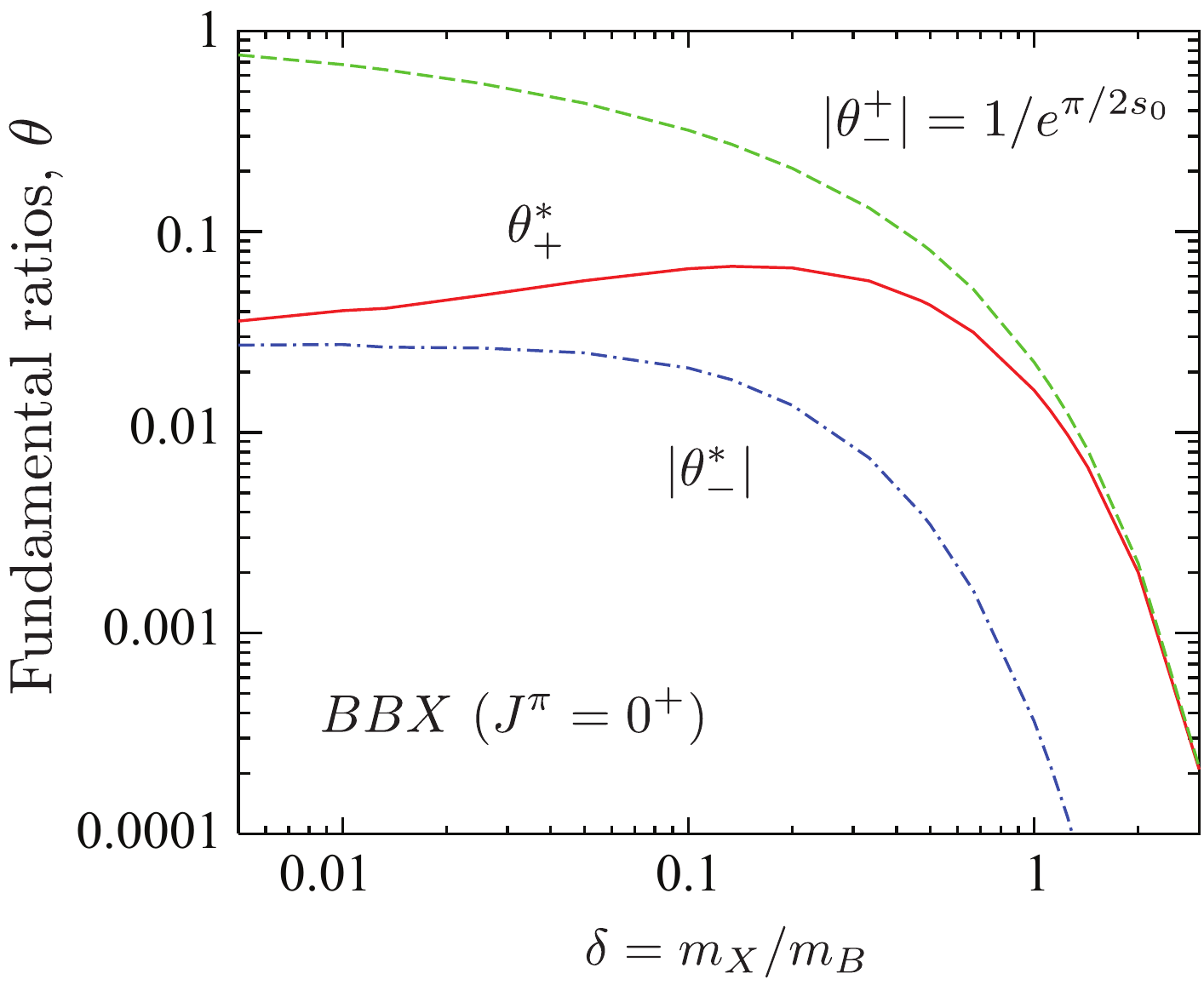}
\caption{Mass ratio dependence of the value of the the fundamental ratios 
$\theta^{+}_{-}$, $\theta^{*}_{-}$ and $\theta^{*}_{+}$ from Ref.~\cite{helfrich2010PRA}, specifying 
the relative position of the Efimov features in three-body scattering observables for $BBX$ ($J^\pi=0^+$).}\label{RatiosPhases}
\end{center}
\end{figure}

For heteronuclear $BBX$ systems with resonant interspecies interactions, the fundamental ratios 
$\theta^{+}_{-}$, $\theta^{*}_{-}$ and $\theta^{*}_{+}$ were determined for the $J^\pi=0^+$ state in terms of 
their mass ratio $\delta_{XB}=m_{X}/m_{B}$ dependence in Ref.~\cite{helfrich2010PRA}. 
Results are shown in Fig.~\ref{RatiosPhases}. The ratio $\theta^{*}_{-}$ has also been
calculated in Ref.~\cite{helfrich2011JPB} for $FFX$ ($J^\pi=1^-$) systems for values of the mass ratio $\delta_{XF}=m_{X}/m_{F}$ 
smaller than the critical value $\delta_c\approx0.073492$.
According to the analysis of Ref.~\cite{helfrich2010PRA}, the ratio $|\theta^{+}_{-}|=1/e^{\pi/2s_{0}}$, thus
depending on the mass ratios only through $s_{0}$ and valid for other systems. This remarkably simple result
implies that the maxima in $K_{3}^{w}/a^4$ $(a>0)$ (i.e., the maximum between two interference minima)
and $K_{3}^{d}/a^4$ $(a<0)$ are symmetric across the resonance ($|a|\rightarrow\infty$). Moreover,
since $\theta^{+}_{-}$ can be determined analytically, and because $\theta^{*}_{-}=\theta^{+}_{-}\theta^{*}_{+}$,
knowing one of the remaining two fundamental ratios is enough to determine the other. 

Note that, in principle, the universal relations in Eq.~(\ref{UnivRel}) could only be expected to be valid if they are obtained 
from Efimov features that belong to the same two-body resonance, since different resonances would imply different
short-range physics. This expectation was shown explicitly in Ref.~\cite{dincao2009JPB}, where recombination was numerically calculated 
for three identical bosons interacting via a finite-range potential. However, more recent developments have shown
that the specific form of the interactions between neutral atoms leads to a novel type of universality 
in which case the universal relations obtained from Efimov features that belong to different two-body resonances can 
indeed be expected to satisfy Eq.~(\ref{UnivRel}). (We discuss such aspects in the next section.)
It is also important to emphasize that the universal relations in Eq.~(\ref{thetavalues}) were 
obtained assuming a zero-range interaction model. As a consequence, finite-range corrections to these results can 
be expected for the relations between Efimov features associated to low-lying states. 
In fact, strong deviations from the geometric scaling between consecutive 
Efimov features, $e^{\pi/s_0}$, were found whenever the features occur for scattering lengths not deep inside 
the universal regime ($|a|\gg r_{0}$) \cite{dincao2009JPB}. The existence of such deviations are important to 
be recognized by any experimentalist seeking to verify 
Efimov features. Moreover, understanding the physics controlling the deviations and whether they are 
themselves universal, are of crucial importance. Much progress has been achieved recently. 
We explore some of this progress in the next section.

%%%%%%%%%%%%%%%%%%%%%%%%%%%%%%%%%%%%%%%%%%%%%%%%%%%%
\subsection{Universality of the three-body parameter} \label{Universality}

In the recent years, an unexpected development has changed one of most fundamental aspects
of Efimov physics conceived at the time of its original prediction 
\cite{efimov1970PLB,efimov1971SJNP,efimov1972JETPL,efimov1973NPA}: 
Although the universal physics related to the Efimov effect depends on a single three-body
parameter, this three-body parameter is itself expected to be {\em nonuniversal}. In fact, since 
the three-body parameter depends on the short-range behavior of the three-body interactions,
this concept {\em does} make sense---different atomic species or different two-body resonances should 
imply different three-body interactions at short distances, especially considering the importance of 
short-ranged nonadditive three-body forces \cite{soldan2003PRA,epelbaum2002PRC}.
Nevertheless, this scenario has proven to be not true for ultracold atoms. This fact was first shown
from the experimental observations in $^{133}$Cs \cite{berninger2011PRL} where the observed value for 
the three-body parameter $a_-$ was the {\em same} (within a 15\% margin) for different 
two-body resonances. However, the theory at that point essentially expected no correlation between the results. 
Moreover, if the value for $a_-$ is recast in terms of the van der Waals length, $r_{\rm vdW}$, the observations 
in all other available atomic species also led to similar results, $a_-\approx-10r_{\rm vdW}$ 
(see Refs.~\cite{wang2013Adv,ferlaino2011FBS} for a summary of such experimental findings). 
This new universal picture for the three-body parameter was then confirmed and interpreted theoretically 
as intimately related to the van der Waals forces in atomic systems 
\cite{wang2012PRL,dincao2013FBS,mestrom2016ARX,wang2012PRLb,wang2015PRL,naidon2014PRA,
naidon2014PRL,schmidt2012EPJB,blume2015FBS,
wang2014NTP,li2016PRA,giannakeas2016ARX}.

\begin{figure}[htbp]
\begin{center}
\includegraphics[width=3.4in,angle=0,clip=true]{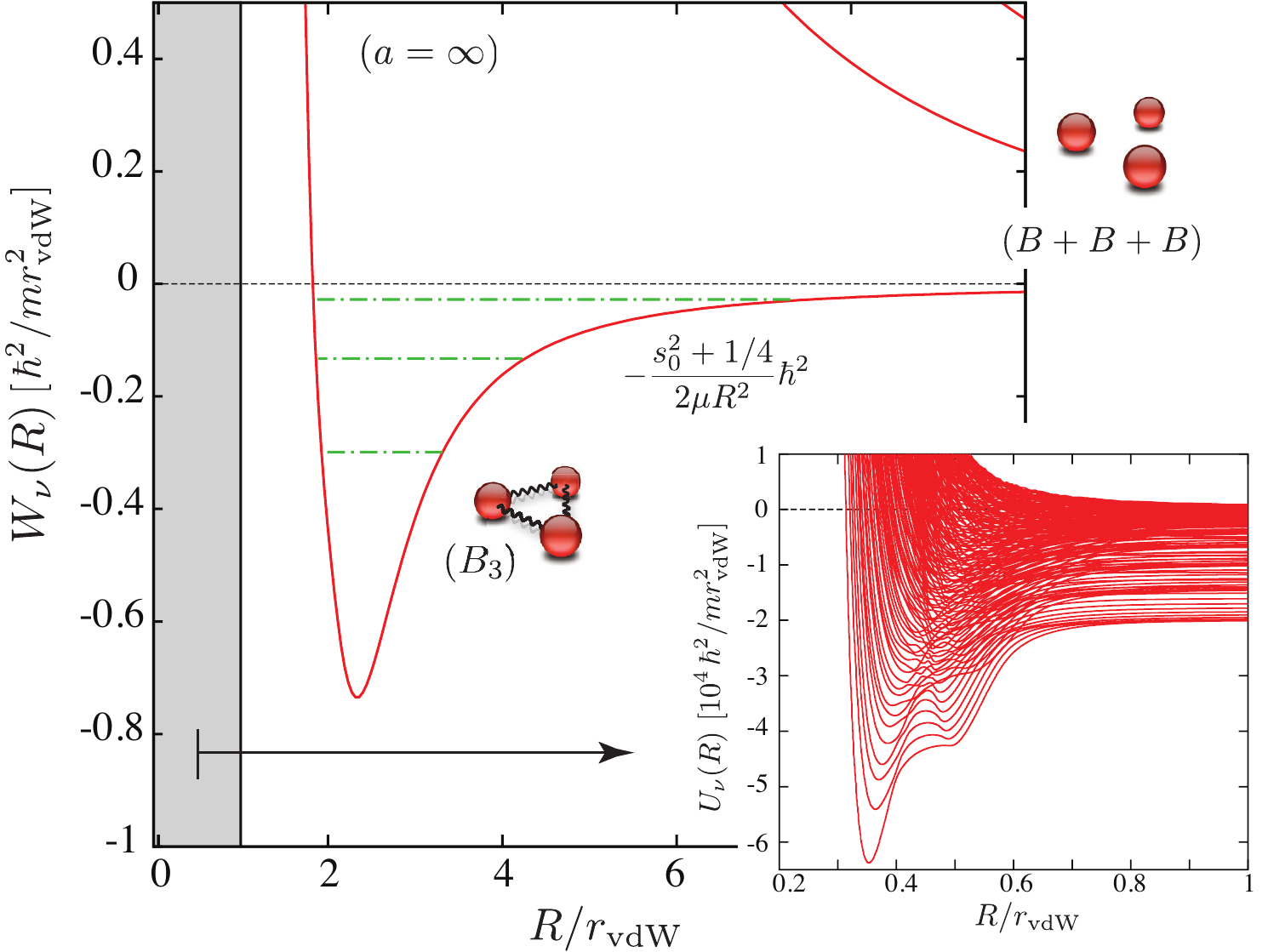}
\caption{Three-body effective potentials (in van der Waals units) for the $BBB$ ($J^\pi=0^+$) system 
with atoms interacting via a Lennard-Jones potential tuned to give $a=\infty$ and no deeply bound states
\cite{wang2012PRL,dincao2013FBS,mestrom2016ARX}. The relevant attractive $1/R^2$ effective potential for Efimov physics 
is repulsive for distances $R\lesssim2r_{\rm vdW}$, preventing atoms from probing the details of the interaction (see inset). 
The universality of this result was systematically tested for systems supporting multiple diatomic states and 
different model interactions \cite{wang2012PRL,dincao2013FBS}.} \label{PotScaLInfty}
\end{center}
\end{figure}

In Refs.~\cite{wang2012PRL,dincao2013FBS,mestrom2016ARX,wang2012PRLb,wang2015PRL,naidon2014PRA,naidon2014PRL} 
the origin of the universality of the three-body parameter was found to be rooted in the sudden increase of kinetic energy as atoms 
approach each other and, consequently, leads to the suppression of the probability of finding atoms at short distances.
When two atoms approach each other, the steep change in the $-C_6/r^6$ interaction 
(where $C_6$ is the usual van der Waals dispersion coefficient) leads to a sudden increase 
of the local two-body kinetic energy that, consequently, suppresses the amplitude of the wave function for 
$r<r_{\rm vdW}$ (see Ref.~\cite{dincao2013FBS}). Intuitively, on the three-body level, this increase of kinetic energy 
should now be manifested as a repulsive barrier for $R<r_{\rm vdW}$. The three-body 
effective potentials (in van der Waals units) for three-identical bosons \cite{wang2012PRL,dincao2013FBS,mestrom2016ARX}
are shown in Fig.~\ref{PotScaLInfty}. The atoms interact via a Lennard-Jones potential tuned to give $a=\infty$ and no deeply 
bound states. As one can see, the relevant channel for Efimov physics 
(i.e., the attractive $1/R^2$ effective potential) is, in fact, repulsive for distances $R\lesssim2r_{\rm vdW}$, thus
consistent with the increase of the local two-body kinetic energy. 
A more detailed study in Ref.~\cite{naidon2014PRA} has provided further insights into the origin of the 
universality of the three-body parameter by showing that, because of the high kinetic-energy cost, the three-body system is subject to a
geometric deformation. This deformation effectively prevents three atoms from coming within distances smaller than $r_{\rm vdW}$
and, consequently, leads to a universal three-body parameter (see more details in Ref. \cite{naidon2017RPP}). 
The universality of the three-body parameter has also been explored for three-body systems with interactions other 
than van der Waals, leading to interesting prospects in nuclear physics and solid-state physics \cite{naidon2014PRL}.

The universality for ultracold atoms was systematically studied for systems supporting multiple diatomic states and 
different model interactions \cite{wang2012PRL,dincao2013FBS}.
Physically, this potential barrier prevents all three atoms, colliding at ultracold energies, from approaching at
distances smaller than $R\lesssim2r_{\rm vdW}$. As a result, the atoms can not probe the details of the interactions, 
thus leading to a universal three-body parameter.
At higher collision energies ($E\gg\hbar^2/mr_{\rm vdW}^2$) or collision processes involving a deeply bound diatomic state, however, the 
system can, in fact, probe the short-range behavior of the interactions, and the universality of three-body observables
should not be expected. The inset of Fig.~\ref{PotScaLInfty} shows a typical short-range behavior for a three-body 
system with many diatomic states \cite{wang2012PRL}. The figure illustrates the complexity that is avoided at ultracold energies 
by the existence of the potential barrier at $R\simeq2r_{\rm vdW}$.

The repulsive interaction in Fig.~\ref{PotScaLInfty} also persists for finite values of $a$ and translates into 
the universality of all three-body parameters $a_+$, $a_-$, and $a_*$ \cite{wang2012PRL,mestrom2016ARX}.
In fact, universality holds for scattering lengths of 
the order of $r_{\rm vdW}$, 
%JPD
%but it is also influenced by effects associated to the van der Waals
%universality \cite{gao2000PRA,wang2012PRA}. 
%\textcolor{red}
{in which case effects associated to the van der Waals
universality \cite{gao2000PRA,wang2012PRA} becomes important.}
The results for the relevant three-body 
parameters obtained from averaging the results of Ref.~\cite{mestrom2016ARX}, using a different 
number of diatomic states, are listed in Table \ref{TabComp}. 
The results from Ref.~\cite{mestrom2016ARX} deviate among themselves within less than 2\%, 
a strong indication of universality. Moreover, since Ref.~\cite{mestrom2016ARX} uses a finite-range potential,
this result indicates that finite-range effects should also be universal.
Note that the three-body parameter $a_{*,0}$, associated with the lowest Efimov state is absent in the table, since,
as shown in Ref.~\cite{mestrom2016ARX}, this state fails to cross the atom-molecule threshold because of
the existence of a variational principle \cite{bruch1973PRL,lee2007PRA} that prevents the energy of the three-body ground state
from exceeding that of the two-body system.

Now, to compare the results from Table~\ref{TabComp} to the zero-range results [Eqs.~(\ref{UnivRel}) and (\ref{thetavalues})],
we define the universal relation between three-body parameters as 
\begin{align}
a_{\alpha,i}/a_{\beta,j}&=\Theta^{\alpha\beta}_{i,j}~(e^{\pi/s_{0}})^{i-j},\label{UnivRelG}
\end{align}
where $\alpha$ and $\beta$ can assume
the values ``$-$", ``$+$" and ``$*$", while $i$ and $j$ run over the index labeling the Efimov state. 
For features obtained for large values of $a$, i.e., for large values of $i$ and $j$, finite-range effects should be
less important, and the fundamental ratio $\Theta$ should approach the values for the ratio $\theta$ obtained from 
the zero-range model [Eq.~(\ref{thetavalues})]. Therefore, Eq.~(\ref{UnivRelG}) allows us to determine the size of 
finite-range effects depending on the states analyzed.
The values for $\Theta$ obtained from the three-body parameters are listed in Table~\ref{TabComp}
and compared to those from the zero-range model for three-identical bosons. 
As one can see, the deviations are substantial. For a more comprehensive analysis on the
finite-range effects, see Ref.~\cite{mestrom2016ARX}. 

\begin{table}[htbp]
\caption{Values for the three-body parameters $a_-$, $a_*$ and $a_+$ for the lowest two Efimov scattering features in 
recombination and relaxation, obtained by averaging the results from Ref.~\cite{mestrom2016ARX} that were calculated from different 
numbers of diatomic states. From these results, we determine the fundamental ratios, $\Theta^{\alpha\beta}_{i,j}$ in Eq.~(\ref{UnivRelG})
and the corresponding geometric-scaling factor for consecutive Efimov features.
We compare these ratios to the zero-range results for three identical bosons: $e^{\pi/s_0}\approx22.694384$, 
$\theta^{+}_{-}\approx-0.209914$, $\theta^{*}_{-}\approx-0.046938$ and 
$\theta^{*}_{+}\approx0.223604$ \cite{braaten2006PRep,gogolin2008PRL}.}\label{TabComp}
\begin{ruledtabular}
\begin{tabular}{ccccccc}
 \multicolumn{4}{l}{Three-body parameters} &  \multicolumn{3}{l}{Geometric scaling}\\
 $(i)$ & ($0$) & ($1$) & ($2$) & \multicolumn{3}{l}{~~~$e^{\pi/s_0}\times\Theta^{\alpha\alpha}_{i,i+1}$}  \\
 \hline
 $a_{+,i}/r_{\rm vdW}$ & $1.4$ & $27.7$ & --- & \multicolumn{3}{l}{ 19.8 ($\Theta^{++}_{1,0}\approx0.87$)} \\
 $a_{-,i}/r_{\rm vdW}$ & $-9.77$ & $-163$ & --- & \multicolumn{3}{l}{16.7 ($\Theta^{--}_{1,0}\approx0.74$)} \\
 $a_{*,i}/r_{\rm vdW}$ & --- & $3.33$ & $159$ &  \multicolumn{3}{l}{47.8  ($\Theta^{**}_{2,1}\approx2.10$)}  \\ [0.05in]
\hline
 \multicolumn{7}{l}{Universal relations}\\
$(i,j)$ & $(0,0)$ & $(0,1)$ & $(1,0)$ & $(1,1)$ & $(2,0)$ & $(2,1)$ \\
 \hline
 $\Theta^{+-}_{i,j}$ & $-0.143$ & $-0.195$ & $-0.125$ & $-0.170$ & --- & --- \\
 $\Theta^{*-}_{i,j}$  & --- & --- & $-0.015$ & $-0.020$ & $-0.032$ & $-0.043$ \\
 $\Theta^{*+}_{i,j}$ & --- & --- & 0.105 & 0.120 & 0.221 & 0.253 \\
\end{tabular}
\end{ruledtabular}
\label{Tab3BP}
\end{table}
 
Understanding the nature of finite-range effects in the universal relations obtained for features associated with the low-lying 
Efimov states is of crucial importance, since those are the features most likely accessible to experiments.
In fact, much progress has been recently achieved
\cite{kievsky2013PRA,ji2012AP,ji2015PRA,platter2009PRA,hammer2007PRA,garrido2013PRA,kievsky2015PRA}. 
A key point to be better understood is the dependence of the three-body parameters on the width of the Feshbach resonance \cite{chin2010RMP}, 
which is closely related to the two-body effective range \cite{gogolin2008PRL,fedorov2001JPA,petrov2004PRL,wang2011PRAb}. 
In particular, the results from Table~\ref{TabComp} and the zero-range results in Eq.~(\ref{thetavalues}) are expected 
to be valid only for broad resonances. For narrow, or even intermediate, resonances, strong modifications for the universal ratios 
can be expected. A more comprehensive, and perhaps more useful, discussion of the effects of the width of the resonance 
can be found, for instance, in Refs.~ \cite{schmidt2012EPJB,fedorov2001JPA,petrov2004PRL,wang2011PRAb}. Here, our goal
is to simply emphasize the importance of this issue.

As we see here, in the adiabatic hyperspherical representation, the origin of the universality of the three-body parameter 
can be related to the existence of a repulsive barrier on the three-body potentials for $R\lesssim2r_{\rm vdW}$, which 
prevents atoms from probing the details of the interatomic interactions.
In retrospect, the existence of a three-body potential barrier had also been found in other types of three-body systems.
For example, in Refs.~\cite{wang2011PRL,wang2011PRLb}, it was shown that a similar repulsive interaction exists for a system of three 
strongly interacting dipoles, leading to a universal three-body parameter. 
In this case, however, the range of the repulsive three-body 
interaction was found to be controlled by the dipole length, $d_{\ell}$, characterizing the range of the interparticle interactions.
As a consequence of the universality in the problem, Ref.~\cite{wang2011PRL} found that the three-body parameters for three identical
bosonic dipoles ($a_+$, $a_-$, and $a_*$) depended on $d_{\ell}$, instead $r_{\rm vdW}$.
The key difference here is that the dipole length can largely exceed $r_{\rm vdW}$ for strongly dipolar atomic and molecular 
species \cite{lahaye2009RPP,aikawa2012PRL,lu2011PRL,lu2011PRLb,jin2011PT,jin2012CR}, and,
in the case of electronic dipolar interaction, $d_{\ell}$ can actually be tuned. 
From the three-body perspective, a large value of $d_{\ell}$ (implying a stronger
repulsive barrier) not only leads to the universality of the three-body parameters, but also implies long lifetimes for Efimov states, 
which can make ultracold dipolar gases an ideal system for exploring the Efimov effect, as shown (Ref.~\cite{wang2011PRL}). 

%%%%%%%%%%%%%%%%%%%%%%%%%%%%%%%%%%%%%%%%%%%%%%%%%%%%
\subsection{Four-boson universality}

The idea of extending the Efimov effect to four or more identical bosons was pursued almost immediately 
after Efimov's original prediction \cite{amado1973PRD}. 
Considering that a {\em true $N$-body Efimov effect} would consist of an $N$-body system possessing an infinity of bound 
states when an $(N-1)$-body bound state formed at the $N$-body breakup threshold, while still no other bound subsystem would exist, 
Amado and Greenwood~\cite{amado1973PRD} found that the Efimov effect for four or more identical bosons was not possible. 
The absence of an Efimov effect for four identical bosons, however, does not preclude the existence of
other Efimov-related states or other universal phenomena in such systems. 
In fact, much has been learned in recent years about the universal aspects of four 
boson systems, both theoretically  
\cite{platter2004PRA,hammer2007EPJAb,yamashita2006EPL,hanna2006PRA,stecher2009NTP,deltuva2011FBS,hadizadeh2011PRL,
stecher2010JPB,yamashita2010PRA,yan2015PRA,stecher2011PRL,gattobigio2011PRA,gattobigio2012PRA,gattobigio2014PRA,kievsky2014PRA,
mehta2009PRL,deltuva2012PRA,
dincao2009PRLb,deltuva2012PRAb,deltuva2011PRA,
deltuva2010PRAb,deltuva2011EPL}
and experimentally \cite{ferlaino2009PRL,zaccanti2009NTP,pollack2009Sci,ferlaino2011FBS,dyke2013PRA,zenesini2012NJP}.
For a more comprehensive review on the theoretical and experimental progress on the four-boson problem
see Ref.~\cite{wang2013Adv}.

The pioneering work of Platter, Hammer and Mei{\ss}ner \cite{platter2004PRA} (see also Ref.~\cite{hammer2007EPJAb})
found that four-boson systems can support weakly bound states that have an intrinsic connection 
to Efimov states. In particular, two weakly bound 
four-body states whose energies were a multiple of the energy of the lowest Efimov state at $a=\pm\infty$ were found, 
and that the same relationship should also persist for excited Efimov trimer states, with each of them having an associated pair of tetramer states
\cite{platter2004PRA,hammer2007EPJAb}.
Since the energies of the tetramer states depend {\em only} on the energy of the Efimov trimers, these studies brought in the 
notion that the four-body parameters should also depend on the three-body physics alone,
without the need of any additional parameter associated with the short-range, nonuniversal aspect of the interatomic interactions. 
Although there was some initial controversy \cite{yamashita2006EPL}, today the relationship between the energy of the ground and excited
tetramers with the energies of the Efimov trimers has been largely verified \cite{hanna2006PRA,stecher2009NTP,deltuva2011FBS,hadizadeh2011PRL,
stecher2010JPB,stecher2011PRL,gattobigio2011PRA,gattobigio2012PRA,gattobigio2014PRA,kievsky2014PRA},
in agreement with Ref.~\cite{platter2004PRA,hammer2007EPJAb}, including for the tetramers associated with excited Efimov states.
This relationship can be conveniently expressed by 
\begin{align}
E_{\rm 4b}^{(i,j)} &= c_{j}E^{(i)}_{\rm 3b},\nonumber\\
&= c_{j}E^{(0)}_{\rm 3b}/(e^{2\pi/s_0})^{i},\label{Energies4B}
\end{align} 
where $j=1$ and $2$ and $E_{\rm 3b}^{(n)}$ is the energy of the $n^{\rm th}$ Efimov trimer at $a=\pm\infty$. 
The constants in Eq.~(\ref{Energies4B}) were found to be 
$c_1=5.0$ and $c_2=1.01$ from Ref.~\cite{hammer2007EPJAb}, $c_1=5.008$ in Ref.~\cite{hanna2006PRA},
and $c_1=4.58$ and $c_2=1.01$ in Ref.~\cite{stecher2009NTP}. Such constants were later refined through more numerically 
accurate calculations in Ref.~\cite{deltuva2011FBS} to be $c_{1}=4.6108$ and $c_{2}=1.00228$ for states
in which finite-range effects are negligible, i.e., for four-boson states associated with highly excited Efimov states.
Note that these values should also be compared with the ones from more recent calculations in Refs.~\cite{stecher2010JPB,stecher2011PRL,hadizadeh2011PRL,gattobigio2011PRA,gattobigio2012PRA}. 
Note also that the small numerical discrepancy between these results can be understood in terms of different finite-range 
corrections, and are present in all calculations described here.

\begin{figure}[htbp]
\includegraphics[width=3.4in,angle=0,clip=true]{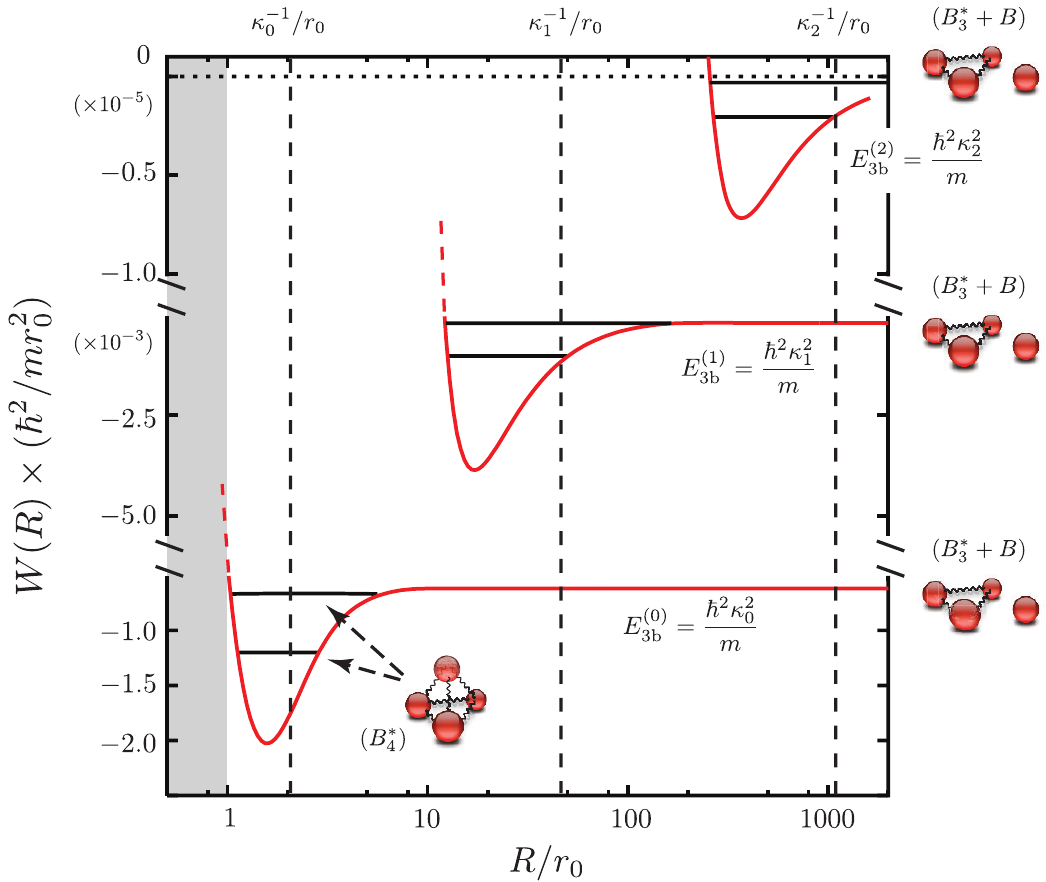}
\caption{Effective four-boson potentials for $|a|=\infty$, converging at large $R$ to the atom-trimer thresholds, $E_{\rm 3b}$. 
Horizontal solid black lines represent the energy of the two four-boson states associated to each trimer. Results from
Ref.~\cite{stecher2009NTP}.}  
\label{4BosonsFig}
\end{figure}

The origin of the universality of the four-boson energies [expressed through their relationship with the 
energies of Efimov trimers in Eq.~(\ref{Energies4B})] was explained in Ref.~\cite{stecher2009NTP} and further
extended for other four-boson scattering observables \cite{stecher2009NTP,mehta2009PRL,deltuva2012PRA,gattobigio2012PRA,
dincao2009PRLb,deltuva2012PRAb,deltuva2011PRA,deltuva2010PRAb,deltuva2011EPL}. 
Within the adiabatic hyperspherical representation in Ref.~\cite{stecher2009NTP}, the
four-body potentials displayed a repulsive barrier for values of the hyperradius $R$ that were comparable with the size of the
Efimov states. The four-boson hyperspherical potentials are shown in Fig.~\ref{4BosonsFig} for $|a|=\infty$.
In this case, the lowest hyperspherical potential supports the lowest two four-boson states and converges asymptotically ($R\rightarrow\infty$)
to the energy of the lowest Efimov trimer.
The other potentials support the four-boson states associated with the excited Efimov trimers, as indicated.
The location of the minimum of each potential and the barrier at shorter distances scale with the size of the respective Efimov trimer, $\kappa^{-1}$.
As a result, since the size of the Efimov states is larger compared to the range of the two-body interactions, $r_0$, the repulsive barrier
prevents atoms from approaching short distances to probe the nonuniversal details of the interactions. This result is similar to what was found
on the three-body level (see Section~\ref{Universality}), although it was identified a few years before \cite{stecher2009NTP}.

\begin{figure}[htbp]
\includegraphics[width=3.4in,angle=0,clip=true]{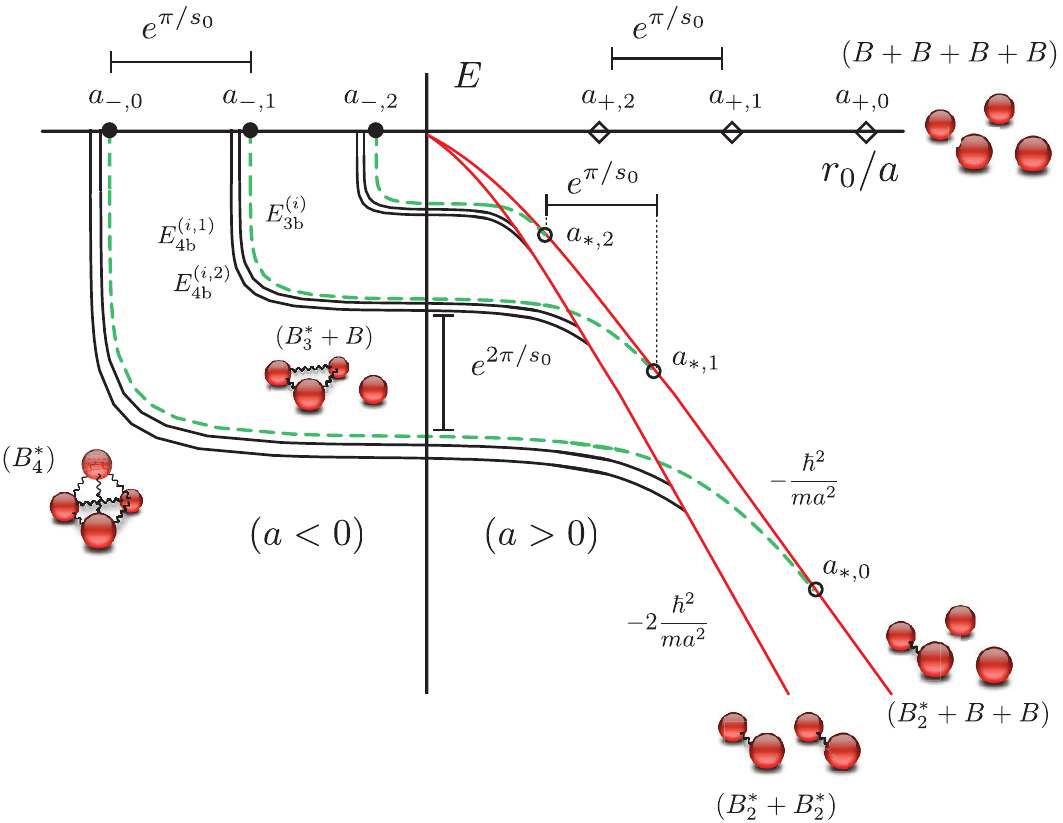}
\caption{General structure of four-boson states and their relation to Efimov states and corresponding three-body parameters
$a_+$, $a_-$ and $a_{*}$ (Fig.~\ref{EfimovSpectrum}). For each Efimov state (dashed-green curves), there exist two universal four-boson
states (solid back curves). Different families of four-boson states are related by the Efimov geometric factor, $e^{\pi/s_0}$.
For $a<0$, such states are formed in the collision threshold for four free atoms, while for $a>0$ these states merge into the dimer-dimer
threshold. A more detailed description of the four-boson energy spectrum (including states not shown in this figure) is given in 
Fig.~\ref{4BP} and the corresponding discussion in the text.}  
\label{4BE}
\end{figure}

\begin{figure}[htbp]
\includegraphics[width=3.4in,angle=0,clip=true]{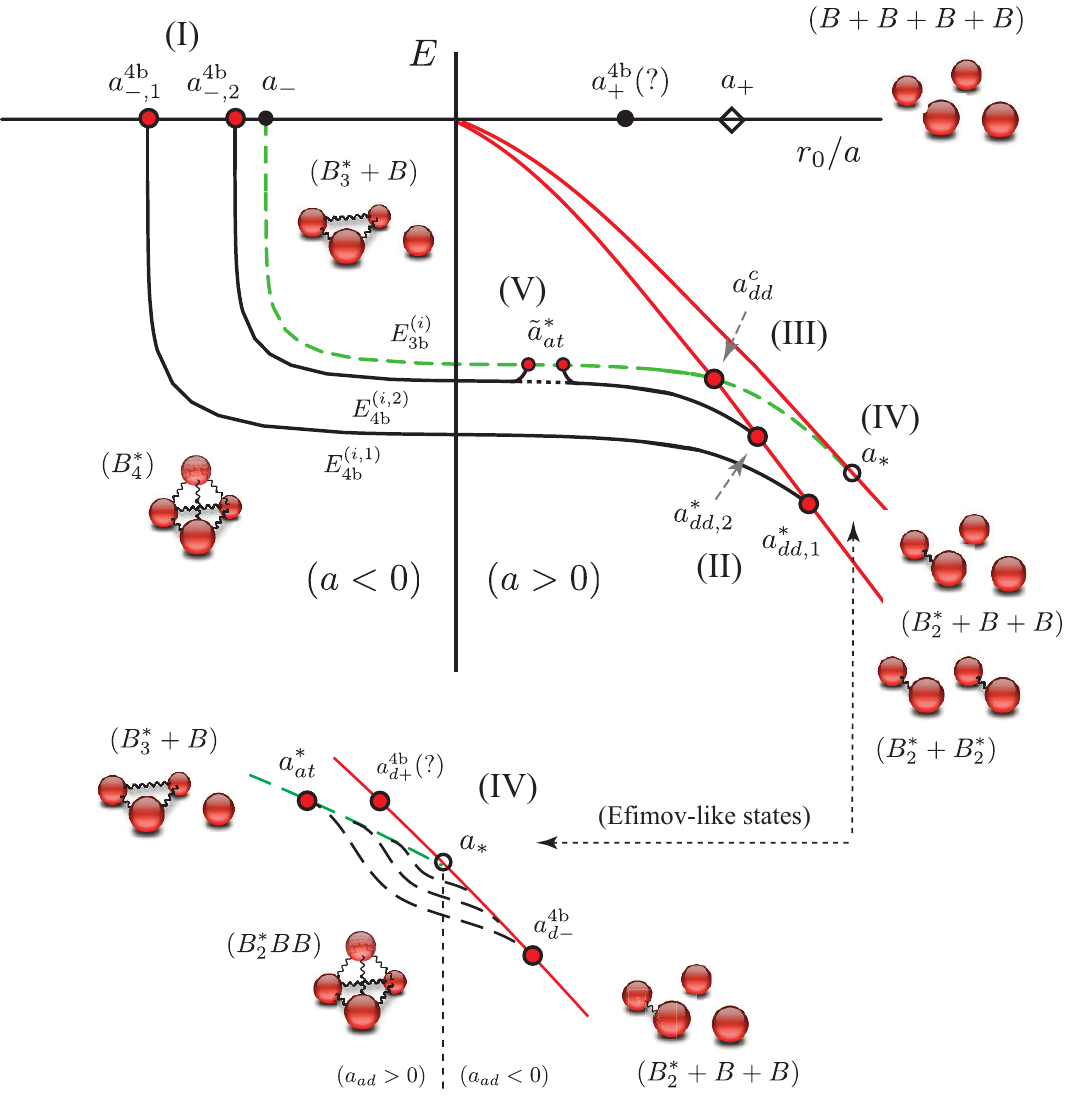}
\caption{Schematic representation of the four-boson energies associated with a single Efimov state as a function of $r_0/a$, 
indicating the four-body parameters relevant for scattering observables and their corresponding three-body parameters 
($a_+$, $a_-$ and $a_*$). In region (I), $a<0$, tetramer states are manifested as resonances in four-body recombination, 
$B+B+B+B$ (see $a_-^{\rm 4b}$ in the figure), while in region (II), $a>0$, tetramer states lead to resonances in dimer-dimer scattering, 
$B^*_2+B^*_2$ ($a_{dd}^{*}$). In region (III) the manifestation of four-boson physics is through the enhancement of 
rearrangement reactions, $B^*_2+B^*_2\rightarrow B^*_3+B$ ($a_{dd}^{c}$), allowing for the formation of
Efimov trimers. In region (IV), i.e., near an atom-dimer resonance ($a_*$), an infinity of Efimov-like states can be formed (see lower panel) 
leading to various scattering features similar to those found for three-bodies, but now for atom-trimer, $B^*_3+B$ ($a_{at}^{*}$), 
and atom-atom-dimer, $B^*_2+B+B$ ($a_{d+}^{\rm 4b}$ and $a_{d-}^{\rm 4b}$), collisions. In region (V), four-bosons states are manifested as 
atom-trimer resonances ($\tilde{a}_{at}^{*}$) because of the unbinding of the excited tetramer state.} \label{4BP}
\end{figure}

The general structure of four-boson states and their relation to Efimov states is shown in Fig.~\ref{4BE},
along with the corresponding three-body parameters $a_+$, $a_-$ and $a_{*}$ (see also Fig.~\ref{EfimovSpectrum}). 
Note that, now, on the four-body level, Efimov states appear as an atom-trimer threshold ($B_3^*+B$), while dimers
appear forming the atom-atom-dimer ($B_2^*+B+B$) and dimer-dimer ($B_2^*+B_2^*$) thresholds.
For $a<0$, four-boson states are formed in the collision threshold for four free atoms ($B+B+B+B$), while 
for $a>0$ these states merge into the dimer-dimer threshold. 
(A more detailed description of the four-boson energy spectrum (including states not shown in Fig.~\ref{4BE}) is given below and 
illustrated in Fig.~\ref{4BP}).
As shown in Fig.~\ref{4BE}, for each Efimov state (dashed-green curves), there exist two universal four-boson states 
(solid back curves). Although a four-boson Efimov effect does not exist, one can see that different families of four-boson states 
are related to each other by the Efimov geometric factor, $e^{\pi/s_0}$, thus defining their connection to the Efimov effect.

Evidently, the same universality that leads to the relationship between four-boson and Efimov energies [Eq.~(\ref{Energies4B})] 
is also reflected in four-boson scattering observables. This universality allows for the determination of various (universal) relations
between different three- and four-body parameters, in the same spirit as those derived for three bodies 
(see Section~\ref{Universality}). Such universal relations can be visualized and understood from the four-body energy spectrum.
In Fig.~\ref{4BP} we singled out a single family of four-boson states and the corresponding three-body parameters ($a_+$, $a_-$ and $a_*$) 
to better analyze the relevant scattering observables associated with four-body scattering. This structure, evidently, repeats for all
families of four-boson states, rescaled by the geometric factor $e^{\pi/s_0}$. 

In region (I) of the four-boson energy spectrum in Fig.~\ref{4BP}, where $a<0$, the four-boson states are manifested as resonances in four-body recombination, 
i.e., the collision process involving four free atoms, $B+B+B+B$ \cite{mehta2009PRL}. 
The values of $a$ in which such resonances occur, namely $a^{\rm 4b}_{-,j}$ ($j=1$ and $2$), are universally connected to the value $a=a_-$ 
in which the trimer becomes bound [see region (I) in Fig.~\ref{4BP}]. In Ref.~\cite{stecher2009NTP}
this relationship was found to be
\begin{align}
a^{\rm 4b}_{-,1}\approx0.43a_-~~~\mbox{and}~~~a^{\rm 4b}_{-,2}\approx0.90a_-,\label{4bp1}
\end{align}
and later refined in Ref.~\cite{deltuva2012PRA} to $a^{\rm 4b}_{-,1}\approx0.4254a_-$ and $a^{\rm 4b}_{-,2}\approx0.9125a_-$. 
Note that an earlier work (Ref.~\cite{hanna2006PRA}) also calculated $a^{\rm 4b}_{-,1}$ associated with the lowest tetramer, 
obtaining $a^{\rm 4b}_{-,1}\approx0.49a_-$ (see also Ref.~\cite{gattobigio2012PRA}). 
Such resonances, associated with the ground Efimov state, have been observed experimentally 
in ultracold quantum gases \cite{ferlaino2009PRL,zaccanti2009NTP,pollack2009Sci,ferlaino2011FBS,dyke2013PRA}
and represent a milestone in the studies of universal properties of few-body systems. [Note, however, that $^{39}$K features originally labeled 
as tetramers \citep{zaccanti2009NTP} are currently being reassigned to trimer resonances.]

In contrast, the other four-boson scattering features illustrated in Fig.~\ref{4BP} remain unverified experimentally and,
in some cases, not even theoretically identified. Such is the case for four-body recombination for $a>0$, where 
interference effects presumably could lead to specific features for specific values of $a=a_{+}^{\rm 4b}$. 
In region (II), for $a>0$, resonant features associated with the two four-boson states have been predicted  \cite{dincao2009PRLb} 
for dimer-dimer collisions, $B^*_2+B^*_2$, for values of $a$ expressed in terms of the three-body atom-dimer resonance, $a_*$,
\begin{align}
a^{*}_{dd,1}\approx2.37a_*~~~\mbox{and}~~~a^{*}_{dd,2}\approx6.60a_*.\label{4bp2}
\end{align}
Such values were further refined by calculating them in the limit in which finite-range effects are negligible, 
in Ref.~\cite{deltuva2011PRA}: $a^{*}_{dd,1}\approx2.1962a_*$ and $a^{*}_{dd,2}\approx6.7854a_*$.

The region (III) of Fig.~\ref{4BP} illustrates the regime in which rearrangement reactions capable of producing Efimov states, 
$B^*_2+B^*_2\rightarrow B^*_3+B$, should take place. This reaction was predicted in Ref.~\cite{dincao2009PRLb} to occur for 
$a>a_{dd}^c$, where 
\begin{align}
a_{dd}^c=6.73a_*\label{4bp3},
\end{align}
with great efficiency, thus providing an exciting possibility of forming Efimov states. Such a rearrangement reaction can
also, in part, explain some of the experimental data in Ref.~\cite{ferlaino2008PRL} (see Ref.~\cite{dincao2009PRLb}). 
The value for $a_{dd}^{c}$ in Eq.~(\ref{4bp3}) is comparable to $a_{dd}^c\approx6.789a_*$ from Ref.~\cite{deltuva2011PRA}.
In the region (IV), near the 
atom-dimer resonance (i.e., when $a=a_*$, and $a_{ad}=\pm\infty$), 
an infinity of four-boson states are formed where a weakly bound dimer binds with two other atoms, $B_2^*BB$, in close analogy
to the three-body Efimov states with one of the bodies having twice the mass of the other two \cite{braaten2006PRep,dincao2009PRLb}.
Many of the four-boson scattering properties in this region resemble those from the three-body problem. 
Such four-boson states, whose geometric scaling is $e^{\pi/s_0}\approx2.01649\times10^5$ ($s_0\approx0.25721$) were studied in
Ref.~\cite{deltuva2012PRAb}. There, it was found that four-boson states cause resonant enhancement in atom-trimer, $B^*_3+B$, collisions
($a_{ad}>0$) and atom-atom-dimer, $B^*_2+B+B$, collisions ($a_{ad}<0$), for values of $a$ given by
\begin{align}
a^*_{at}\approx 1.608a_*~~~\mbox{and}~~~0.9999a_*<a_{d-}^{\rm 4b}<a_*,\label{4bp4}
\end{align}
respectively, in analogy to the $a_*$ ($a>0$) and $a_-$ ($a<0$) three-body parameters. Although, interference effects
can be expected for atom-atom-dimer collisions ($a_{ad}>0$) for values of $a=a_{d+}^{\rm 4b}$, currently there are no 
predictions for such a four-body parameter. 

Similarly, in the region (V), other atom-trimer resonances can also occur \cite{deltuva2010PRAb}. 
Such a case was studied in Ref.~\cite{deltuva2011EPL} where resonant effects in atom-trimer collisions occurred because of the 
unbinding and subsequent binding of the excited four-boson state at the values
\begin{align}
\tilde{a}_{at(l)}^{*}\approx 93.128a_*~~~\mbox{and}~~~\tilde{a}_{at(r)}^{*}\approx6.8002a_*.\label{4bp5}
\end{align}
In Ref.~\cite{deltuva2012PRAb}, however, it was noticed that such atom-trimer resonances occur only for 
the excited families of four-body states. For the lowest family of four-boson states, the excited tetramer always remained below the atom-trimer
threshold, most likely due to the existence of the variational principle that prevents the energy of the bound state of a $N$-body system
from exceeding that of the ground state of the $N-1$ system \cite{bruch1973PRL,lee2007PRA}. 
These results are consistent with the analysis of the numerical calculations performed in 
Refs.~\cite{stecher2009NTP,dincao2009PRLb}.

With the universal relations for the four-body parameters presented in Eqs.~(\ref{4bp1})-(\ref{4bp5}) one can derive other 
relations with respect to the other three-body parameters $a_-$ and $a_+$ (see Section \ref{UniversalRelations}). Such relations
can provide various ways to identify universal features in four-boson scattering. There are also other systems in which the notion
of universality has been explored such as heteronuclear four-body systems 
\cite{castin2010PRL,zhang2012PRA,wang2012PRL,blume2014PRL,schmickler2017FBS,bazac2017PRL} and systems with more than four atoms.
For a more comprehensive review on the theoretical and experimental progress on the four-boson problem,
see Ref.~\cite{wang2013Adv}. In many cases, the adiabatic hyperspherical representation has proved to be instrumental,
offering a simple and conceptually clear physical picture of the problem. The increase of complexity, however, demands novel
numerical and/or analytical tools to be developed.

% !TEX root = ./TutorialJPB.tex

%%%%%%%%%%%%%%%%%%%%%%%%%%%%%%%%%%%%%%%%%%%%%%%%%%%%
\section{Summary and Acknowledgments} \label{Summary}

We provided a general discussion on the importance of Efimov physics for ultracold quantum
gases in the strongly interacting regime. 
The richness of few-body systems along with the experimental ability to control and probe the 
associated universal phenomena make ultracold few-body systems a dynamic field of 
research with much still to be explored.
From the theoretical viewpoint, we present an analysis of the few-body universal phenomena from the 
adiabatic hyperspherical representation perspective. By using a zero-range potential model for the interatomic interactions, 
the convenience of the hyperspherical representation allowed us to discuss general properties Efimov physics, 
including its influence in allowing for a general classification of three-body systems and their bound and scattering 
properties. In fact, one of the major strengths of the adiabatic hyperspherical representation is that it provides a 
simple and conceptually clear physical picture of the few-body problem, while still offering numerically 
exact solutions. Using a simple pathway analysis, we showed that the energy and scattering length dependence for all 
scattering observables can be obtained and related to Efimov physics in an elegant way. 
From this analysis we also provided a general discussion of atomic and molecular losses in ultracold gases, 
highlighting the importance of finite-range corrections as well as finite temperatures effects. We conclude this manuscript 
with a discussion on universal properties of few-body systems and the origin of the universal behavior in both three- 
and four-body systems.

We gratefully acknowledge many years of stimulating and substantive discussions with our
many collaborators (Brett D. Esry and Chris H. Greene, in particular) as well as with others in the field. 
We appreciate various discussions with Victor E. Colussi and Julie Phillips, who guided us toward a better presentation
and elaboration of certain parts of the present material, H.-W. Hammer for sharing his data
presented in figures in the text and Pascal Naidon for stimulating discussions. 
We acknowledge partial support from the U.S. National Science Foundation (NSF), Grant PHY-1607204,
and by the National Aeronautics and Space Administration (NASA).

%\appendix
%\input{SectionAppA}

%\bibliographystyle{unsrt}
%\bibliographystyle{abbrv}
\bibliographystyle{ieeetr}

%\bibliography{ReviewRefs}

\end{document}